\documentclass[11pt]{article}
\usepackage[T1]{fontenc}
\usepackage[letterpaper,margin=1in]{geometry}
\usepackage{amsmath,amssymb,amsthm,mathtools}
\usepackage{microtype}
\usepackage{enumitem}
\usepackage{booktabs,needspace,placeins,url}
\usepackage{tikz}
\usetikzlibrary{calc}
\usetikzlibrary{arrows.meta,decorations.pathreplacing}
\definecolor{acyBlue}{RGB}{35,85,140}
\definecolor{acyInk}{RGB}{65,70,76}
\definecolor{acyGreen}{RGB}{55,110,70}
\definecolor{acyRed}{RGB}{145,65,45}
\definecolor{acyEntry}{RGB}{105,80,145}
\tikzset{
 acy/edge/.style={->,>=Stealth,line width=.8pt,draw=acyInk},
 acy/coreedge/.style={acy/edge,draw=acyBlue},
 acy/rededge/.style={acy/edge,draw=acyRed},
 acy/note/.style={font=\normalsize,align=center},
 acy/title/.style={font=\normalsize\bfseries,align=center},
 acy/box/.style={draw=acyInk,rounded corners=2pt,line width=.65pt,
                fill=white,align=center,font=\normalsize,inner sep=5pt},
 acy/vertex/.style={circle,draw=acyInk,line width=.7pt,minimum size=6mm,
                   inner sep=1pt,font=\normalsize},
 acy/dot/.style={circle,line width=.7pt,minimum size=3.4mm,inner sep=0pt},
 acy/bluedot/.style={acy/dot,draw=acyBlue,fill=acyBlue!12},
 acy/reddot/.style={acy/dot,draw=acyRed,fill=white},
 acy/bluearea/.style={draw=acyBlue!65,fill=acyBlue!4,line width=.6pt},
 acy/redarea/.style={draw=acyRed!45,fill=acyRed!3,line width=.6pt,
                    rounded corners=2pt}
}
\usepackage[hidelinks]{hyperref}
\hypersetup{
 pdftitle={Near-Linear Lower Bounds for Testing Directed Acyclicity},
 pdfsubject={Adaptive property testing; two-sided query lower bounds},
 pdfkeywords={directed acyclicity, property testing, query complexity, information theory}
}
\setlist{itemsep=3pt,topsep=5pt}
\numberwithin{equation}{section}
\newtheorem{theorem}{Theorem}[section]
\newtheorem{lemma}[theorem]{Lemma}
\newtheorem{proposition}[theorem]{Proposition}

\theoremstyle{definition}
\newtheorem{definition}[theorem]{Definition}

\newtheorem*{inftheorem}{Informal Theorem}
\theoremstyle{plain}
\newcommand{\E}{\mathbb E}
\newcommand{\ind}{\mathbf 1}
\newcommand{\fas}{\operatorname{fas}}
\newcommand{\TV}{\operatorname{TV}}
\newcommand{\Dec}{\operatorname{Dec}}
\newcommand{\Dist}{\mathcal{L}}
\newcommand{\Unif}{\operatorname{Unif}}
\newcommand{\Reach}{\operatorname{Reach}}
\newcommand{\eps}{\ensuremath{}\varepsilon}
\newcommand{\poly}{\ensuremath{\textrm{poly}}}
\newcommand{\polylog}{\ensuremath{\textrm{polylog}}}
\newcommand{\Emb}{\operatorname{Emb}}
\newcommand{\DY}{\mathcal D_{\mathrm Y}}
\newcommand{\DN}{\mathcal D_{\mathrm N}}
\newcommand{\hDY}{\widehat{\mathcal D}_{\mathrm Y}}
\newcommand{\hDN}{\widehat{\mathcal D}_{\mathrm N}}
\newcommand{\Sf}{s_{\mathrm f}}
\newcommand{\Sc}{s_{\mathrm c}}

\title{Near-Linear Lower Bounds for Testing Directed Acyclicity}
\author{
    Artur Czumaj\thanks{University of Warwick. Email: A.Czumaj@warwick.ac.uk}
        \and 
    Pan Peng\thanks{University of Science and Technology of China. Email: ppeng@ustc.edu.cn} 
        \and
    Yuyang Wang\thanks{University of Science and Technology of China. Email: wangyvyang@mail.ustc.edu.cn}
}
\date{}

\begin{document}
\hypersetup{pageanchor=false}
\begin{titlepage}
\maketitle
\thispagestyle{empty}
\begin{abstract}
We study testing acyclicity of directed graphs in the bounded-degree incidence-lists model, where a tester can query the outneighbors of a vertex but not its inneighbors. Despite significant effort, no tester with $o(n)$ queries is known, while the best established lower bounds are $\widetilde\Omega(n^{5/9})$ for one-sided testers (Chen, Randolph, Servedio, and Sun, 2022) and $\Omega(n^{1/3})$ for two-sided testers (Bender and Ron, 2002). We show that a query exponent arbitrarily close to one is necessary: for every fixed $\eta\in(0,1)$, there exists a constant $\varepsilon_\eta>0$ such that, for every constant $d\ge2$, every randomized adaptive two-sided $\varepsilon_\eta$-tester for acyclicity of simple $n$-vertex directed graphs whose indegrees and outdegrees are at most $d$ makes $\Omega(n^{1-\eta})$ queries.

Our proof overcomes two barriers. First, in previous constructions, the acyclic part of the graph is easy to recognize, which helps an algorithm stay inside the cyclic part. We hide a high-girth cyclic core inside a hierarchy of acyclic masking regions, in which each region imitates the levels below it and can be told apart from them only through rare events, and an iterated information-theoretic argument shows that every exploration observes only small, low-diameter forest components of the core. Second, a two-sided tester may exploit the statistics of collisions without ever seeing a cycle, and at a near-linear budget collisions are unavoidable. We compare complete transcripts with those of an acyclic input built on a fixed core by counting injective embeddings of the observed forest: in high-girth cores, all connected collision patterns of bounded order contribute identically, which makes the two transcript distributions statistically indistinguishable.
\end{abstract}

\end{titlepage}

\pagenumbering{roman}
\hypersetup{pageanchor=true}
\setcounter{tocdepth}{2}
\tableofcontents
\clearpage
\pagenumbering{arabic}

\section{Introduction}
\label{sec:introduction}
\emph{Property testing} is a relaxation of the classical decision problem of recognizing a property: given query access to an object, the goal is to distinguish objects satisfying a specified property $\Pi$ from objects that are far from every object with $\Pi$. The notion of being ``far'' is problem-dependent; 
the algorithm rejects objects that are $\varepsilon$-far from having $\Pi$, where an object is $\varepsilon$-far from property $\Pi$ if one has to modify at least an $\varepsilon$-fraction of its representation to obtain an object with property $\Pi$.

The notion of property testing was introduced by Rubinfeld and Sudan \cite{RS96} in the context of program checking; Goldreich, Goldwasser, and Ron \cite{GGR98} studied its connections to learning and approximation and initiated the study of testing graph properties and other combinatorial objects.
Since then, a substantial body of work has developed algorithms for testing a wide range of graph properties. 
While property testing in undirected graphs has been studied extensively (see, e.g., surveys \cite{BY22,CS06,Goldreich17,Ron10} and Section~\ref{sec:intro-related-works-model} on related work), our understanding of the \emph{directed} setting remains much more limited. The goal of this work is to advance our understanding of testing properties of \emph{bounded-degree directed graphs} in the \emph{incidence-lists representation}, focusing on one of the most fundamental problems in the area: testing whether a directed graph is acyclic.

Property testing for directed graphs was formally introduced by Bender and Ron \cite{BR02}, who considered two main models: the \emph{adjacency-matrix model}, which is a natural extension of the classical dense graph model for undirected graphs (cf. \cite{GGR98,BY22,Goldreich17} and also \cite{AS04}), and the \emph{bounded-degree incidence-lists model}, which is the natural model for sparse directed graphs.
The focus of this paper is on the latter model: in the (bounded-degree) incidence-lists model, each vertex has a list of length $d$, where $d$ is a bound on the outdegree of the graph. The tester can query a vertex $u \in \{1,\dots,n\}$ and an index $i \in \{1,\dots,d\}$ for the $i$-th outneighbor of vertex $u$ (the head of the $i$-th edge leaving $u$); if no such neighbor exists, then the answer is the symbol $\bot$. (Observe that the tester has access only to outgoing edges, which is why this model is often called the \emph{unidirectional model}.) In this model, the objective is to accept any directed graph $G$ satisfying a given property $\Pi$ and reject any $G$ that is $\eps$-far from $\Pi$, that is, 
any $G$ requiring at least $\eps dn$ edge modifications to obtain $\Pi$ (while still maintaining the degree bound).

In this paper, we consider the fundamental problem of \emph{testing acyclicity}, which in the property testing setting asks whether the input directed graph is acyclic or requires many (at least $\varepsilon d n$) edge deletions to become acyclic. The aim is to distinguish between these cases while inspecting only a small part of the graph. A \emph{two-sided} $\eps$-tester must accept every acyclic input and reject every $\eps$-far input, each with probability at least $\frac23$. A \emph{one-sided tester} must always accept acyclic inputs, so it can reject only after revealing a directed cycle. This ties one-sided testing to cycle finding. A two-sided tester has more freedom: it may reject on statistical evidence even when the observed graph is acyclic. 

\paragraph{Previous bounds for testing acyclicity.}
Bender and Ron \cite{BR02} call acyclicity ``the most basic property of directed graphs'', but despite significant effort, testing acyclicity remains poorly understood: no sublinear-query tester is known, and the main results to date are lower bounds. 
(All lower bounds in this paragraph except that of \cite{CRSS22} hold even if the indegrees are also bounded by $d$; the bounds claimed in \cite{Yoshida26}, discussed below, assume only an outdegree bound.)
Bender and Ron \cite{BR02} proved an $\Omega(n^{1/3})$ lower bound for two-sided testing. Their constructions also yield a birthday-paradox $\Omega(\sqrt{n})$ lower bound for cycle finding (one-sided error testing), as explained by Chen, Randolph, Servedio, and Sun~\cite{CRSS22}; a similar result also follows from the undirected $\Omega(\sqrt{n})$ cycle-freeness testing lower bound due to Goldreich and Ron \cite{GR02}. Chen et al.\ improved this bound by proving an $\widetilde\Omega(n^{5/9})$ cycle-finding lower bound. They hide a cyclic region among acyclic layers 
so that an algorithm making $n^{5/9}/\polylog(n)$ queries sees many collisions yet cannot assemble a directed cycle.
Thus collisions alone do not let a tester find a cycle. A two-sided tester, however, need not find a cycle at all: it can use the statistical information in its answers to decide whether the input is acyclic. The challenge we address is to push the query lower-bound exponent arbitrarily close to one while allowing this more permissive error guarantee.

A recent preprint \cite{Yoshida26} claims an $\widetilde\Omega(n^{2/3})$ one-sided lower bound and an $\Omega(\sqrt{n})$ two-sided lower bound for testing acyclicity. We could not verify the proofs of these bounds; in Appendix~\ref{sec:Yuichi} we briefly show that the distribution used for the one-sided bound admits a simple one-sided algorithm with $O(\sqrt{n})$ queries.

\begin{table}[t]
\centering \begin{minipage}{\linewidth}
\begin{tabular*}{\textwidth}{@{\extracolsep{\fill}}lccc@{}}
\toprule
    Work & Error guarantee & Lower bound \\
    \midrule
    Bender and Ron~\cite{BR02}& one-sided & $\Omega(n^{1/2})$ \\
        & two-sided & $\Omega(n^{1/3})$ \\
\addlinespace
    Chen, Randolph, Servedio, Sun~\cite{CRSS22}
        & one-sided & $\widetilde\Omega(n^{5/9})$ \\
\addlinespace
    Yoshida~\cite{Yoshida26} \footnote{Claimed; see Appendix~\ref{sec:Yuichi}.} & one-sided & $\widetilde\Omega(n^{2/3})$ \\
        & two-sided & $\Omega(n^{1/2})$ \\
\addlinespace
    \textbf{This work} & two-sided & $\Omega(n^{1-\eta})$ \\
\bottomrule
\end{tabular*}
\end{minipage}
\caption{Selected query lower bounds for directed acyclicity (in the unidirectional model) with constant proximity parameter $\eps$ and constant maximum outdegree. The bounds of \cite{BR02} and of this work hold even when indegrees are also bounded; our bound holds for every fixed degree bound $d\ge2$, with a proximity parameter $\eps_\eta$ that depends only on~$\eta$.
The tilde suppresses polylogarithms.}
\label{tab:results}
\end{table} 

\subsection{Our result: a near-linear two-sided lower bound}
\label{sec:intro-result}

We prove that, for every fixed $\eta\in(0,1)$, testing directed acyclicity requires $\Omega(n^{1-\eta})$ queries at every constant degree bound $d\ge2$, even for fully adaptive randomized testers with two-sided error. The proximity parameter depends only on~$\eta$.

\begin{theorem}
\label{thm:main}
For every fixed $\eta\in(0,1)$, there exists a constant $\eps_\eta>0$ such that, for every constant integer $d\ge2$ and every sufficiently large $n$, every randomized adaptive two-sided $\eps_\eta$-tester for directed acyclicity on simple $n$-vertex directed graphs with maximum indegree and maximum outdegree at most $d$ requires $\Omega(n^{1-\eta})$ queries.
\end{theorem}

The input graph has bounded indegrees and outdegrees, but the tester can access only outgoing edges.
The indegree bound is an additional promise on the input; since such a promise can only make testing easier, a lower bound under it is stronger. Moreover, following Chen et al.~\cite{CRSS22}, our lower bound in Theorem~\ref{thm:main} also holds for the model with \emph{whole-vertex queries}: a query to $u$ \emph{returns all its outneighbors}. For simple graphs, we view this answer as a set. At constant $d$, the indexed and whole-vertex conventions change query complexity by at most a constant factor.\footnote{For the set formulation, list outneighbors in increasing ID order. One vertex query then answers any indexed query, while at most $d$ indexed queries recover a vertex answer.}
The tester may choose each query using the entire preceding history and may query any vertex in $[n]$, including one it has not seen before.

The distance $\eps_\eta$ in Theorem~\ref{thm:main} depends only on~$\eta$, not on~$d$ or~$n$. The threshold for sufficiently large~$n$ may depend on both $\eta$ and the fixed degree bound~$d$; we have not attempted to optimize the dependence of $\eps_\eta$ on~$\eta$.  The proof first gives a lower bound at a degree depending on~$\eta$, then reduces the degree to~$2$, and finally uses a complete blow-up to obtain every fixed $d\ge2$ with the same proximity parameter (Section~\ref{rt:transfer}).
Thus Theorem~\ref{thm:main} rules out a uniform polynomial saving over the trivial tester that queries all $n$ vertices: for each fixed $d\ge2$, there is no bound $O_{d,\eps}(n^\tau)$ with a common exponent $\tau<1$ for all fixed distances. Indeed, choosing $\eta<1-\tau$ contradicts such a bound at $\eps_\eta$. The result is a family of constant-parameter lower bounds whose exponents approach one; it leaves open whether, for a fixed pair $(d,\eps)$, there is a tester with $o(n)$ queries.

Table~\ref{tab:results} compares our result with previous lower bounds for acyclicity. Our bound applies to one-sided testers as well, and, unlike earlier one-sided bounds, it also applies to testers that reject on statistical evidence without ever exhibiting a directed cycle.

\paragraph{Is there an $n/\mathrm{polylog}(n)$-query tester?}
Theorem~\ref{thm:main} leaves room for testers with $n^{1-o(1)}$ queries, such as $n/\polylog(n)$. Chen, Randolph, Servedio, and Sun \cite[Section~9, item~2]{CRSS22} proposed a natural candidate of this kind: for a sufficiently large constant $C = C(\varepsilon)$, run an outgoing breadth-first search (BFS) from each of $C$ independent uniformly random roots until $C n / \log n$ vertices have been explored, and reject if a directed cycle is found. They asked whether, on every $\varepsilon$-far graph, one of these searches finds a cycle with constant probability. We show that this approach fails, even at a single fixed pair of degree bound and distance, and even with two-sided error: there are absolute constants $D_{\mathrm{BFS}}$ and $\varepsilon_{\mathrm{BFS}}>0$ such that every tester that runs a constant number of outgoing BFS searches, with adaptively chosen roots and an arbitrary decision based on all their transcripts, needs $\Omega(n)$ queries to distinguish acyclic graphs from $\varepsilon_{\mathrm{BFS}}$-far graphs, even when all indegrees and outdegrees are at most $D_{\mathrm{BFS}}$ (Theorem~\ref{thm:bfs-linear}). The proof in Appendix~\ref{app:bfs} is independent of that of Theorem~\ref{thm:main} and uses a much simpler construction: an acyclic gadget forces every search to explore a linear number of vertices before it can look deep into the part of the graph that carries the cycles. The bound is specific to level-by-level search; an unrestricted tester may follow a deep branch before completing a distance level, as the attacks on the construction of \cite{CRSS22} do. Whether some tester makes $o(n)$ queries at a fixed pair $(d,\varepsilon)$ therefore remains open. 

\subsection{Related work and comments about the models}
\label{sec:intro-related-works-model}

\paragraph{Variations of the model.}

Since our hard inputs have both degrees bounded, reversing all edges shows that Theorem~\ref{thm:main} also holds when \emph{only incoming neighbors can be queried}.

Bender and Ron \cite{BR02} (see also, e.g., \cite{CPS16,PW23,PW27}) also considered the \emph{bidirectional model}, which allows the tester to query both outgoing and incoming edges of a vertex. While this model resembles the model of undirected graphs (see, e.g., \cite{Goldreich10:introduction}), and admits nontrivial constant-query testers, cf. \cite{BR02,OR11,YI10}, it is significantly stronger than the unidirectional model considered in this paper. A typical example where the two models differ substantially is the web graph (see, e.g., \cite{CPS16}), in which vertices are webpages and an edge $(u,v)$ is a hyperlink from $u$ to $v$. Such graphs are easy to explore when both outgoing and incoming edges of a vertex are accessible (in the bidirectional model), but in many applications the queries for incoming edges are not allowed or are too expensive. The unidirectional model is more realistic when incoming edges are unavailable, but harder to analyze because a tester cannot discover incoming edges directly.
We study the limitations of testing acyclicity under this form of local access.

We focus solely on the bounded-degree incidence-lists model, as defined above. There is also a vast literature on another model, the so-called \emph{adjacency-matrix model} (also known as the \emph{dense graph model}), in which a property tester is given query access to the adjacency matrix of an $n$-vertex graph and a proximity parameter $\eps > 0$, and the goal is to distinguish between graphs that satisfy the property and graphs that are $\eps$-far from satisfying it, meaning that at least $\eps n^2$ edge insertions/deletions are required to obtain the property. For extensive references, see, e.g., surveys \cite{BY22,Goldreich17,Ron10}.

Yoshida~\cite{Yoshida26} also considered acyclicity testing in the \emph{tolerant testing} setting (see, e.g., \cite{PRR06} and \cite[Chapter~12.1]{Goldreich17}). Tolerant testing is never easier than standard testing, and often much harder. Yoshida proved an $\Omega(n)$ lower bound for tolerant acyclicity testing, for some absolute constant outdegree bound, via a reduction from bounded-degree 3-colorability. This bound has no implications for the (non-tolerant) problem considered here.

\paragraph{Undirected graphs.}

Graph property testing in undirected graphs has been studied extensively; see, e.g., the surveys \cite{BY22, CS06, Goldreich17, Ron10}. The most investigated setting is the dense graph model (see, e.g., \cite{AFNS09, AS08:onesided, AS08:monotone, GGR98, GT03}), where we now have a good understanding of testability with query complexity independent of $n$ (see, e.g., \cite{AFNS09, AS08:onesided, AS08:monotone}). The bounded-degree model has also received considerable attention; highlights include \cite{BSS10, CGRSSS14, CPS15, CSS09, CS07, GR02, IKN20, NS13}.

Acyclicity testing has been studied for undirected graphs since the first papers on graph property testing (in the undirected setting it is called \emph{cycle-freeness}). The problem is ``easy'' for the dense graph model, where Goldreich, Goldwasser, and Ron gave a two-sided tester with query complexity $O(1/\eps^2)$ \cite[Proposition~10.2.1.2]{GGR98} and Bogdanov and Trevisan \cite{BT04} gave a similar one-sided error tester and provided almost matching lower bounds. For undirected graphs in the \emph{bounded-degree incidence-lists model}, the complexity of testing cycle-freeness depends heavily on whether one-sided or two-sided error is allowed. For the two-sided error setting, Goldreich and Ron provided a tester for cycle-freeness with $\poly(1/\varepsilon)$ query complexity \cite[Theorem~4.2]{GR02}. For one-sided error testing, the complexity is significantly worse: Goldreich and Ron proved that cycle-freeness has no one-sided error tester that makes $o(\sqrt{n})$ queries \cite[Proposition~4.3]{GR02}, and Czumaj et al.\ complemented this bound with an algorithm testing cycle-freeness with $\widetilde{O}(\poly(d/\eps) \sqrt{n})$ queries \cite[Theorem~1.1]{CGRSSS14}. The above-mentioned lower bound of $\Omega(\sqrt{n})$ for testing cycle-freeness in undirected graphs in the bounded-degree incidence-lists model \cite[Proposition~4.3]{GR02} easily extends to the lower bound of $\Omega(\sqrt{n})$ for testing acyclicity with one-sided error in \emph{directed graphs} in the bounded-degree incidence-lists model, even in the \emph{bidirectional model}.

\paragraph{Related work on directed graph testing.}
Despite the central role of directed graphs in a wide range of applications, \emph{property testing in directed graphs} remains significantly less studied than in the undirected setting. The study of directed graphs in the context of property testing was initiated by Bender and Ron \cite{BR02}, who introduced the main computational models and studied two basic problems: acyclicity testing (discussed in detail above) and testing strong connectivity.
For the \emph{adjacency-matrix model}, Bender and Ron \cite{BR02} gave an acyclicity tester whose query and time complexity is $\widetilde{O}(1/\eps^2)$, independent of the graph size. In the same model, Alon and Shapira \cite{AS04} showed that for any fixed directed graph $H$, the property of being $H$-free is testable with query complexity depending only on $\eps$.
For the \emph{bounded-degree incidence-lists model}, Bender and Ron \cite{BR02} showed that while strong connectivity can be tested in the bidirectional model with $\widetilde{O}(1/\varepsilon)$ queries, in the model allowing only access to the outgoing edges, one needs $\Omega(\sqrt{n})$ queries, even when allowing \emph{two-sided error}. Goldreich \cite[Appendix A.3]{Goldreich10:introduction} and Hellweg and Sohler \cite{HS12} extended these arguments to obtain a lower bound of $\Omega(n)$ for \emph{one-sided error}, and designed testers for strong connectivity with $n^{1-\Omega_{\varepsilon,d}(1)}$ queries to the outgoing edges. The bidirectional model has also been investigated for some specific graph properties, and besides strong connectivity \cite{BR02}, the properties studied include testing Eulerianity  \cite{OR11}, $k$-edge-connectivity \cite{OR11,YI10}, and $k$-vertex-connectivity \cite{OR11}. All these testers rely heavily on access to both incoming and outgoing edges and achieve constant query complexity. Czumaj, Peng, and Sohler~\cite{CPS16} studied the relationship between the two models and showed that every property with a constant-query bidirectional tester has an $n^{1-\Omega_{\eps,d}(1)}$-query unidirectional tester. Peng and Wang~\cite{PW23} showed that this transformation is essentially tight using subgraph-freeness properties whose unidirectional lower-bound exponents approach one (their lower bounds do not apply to acyclicity). Results in the bidirectional model, as well as \cite{CPS16}, and \cite{PW23} assume both degrees bounded, which is exactly the setting of Theorem~\ref{thm:main}.
Very recently, Apers et al.\ \cite{AMSS25} initiated quantum testing in the unidirectional model, and Peng and Wu \cite{PW27} extended the \cite{CPS16} transformation to the quantum setting.

\section{Overview of the techniques}
\label{sec:intro-techniques}
\label{sec:techniques}

In this section, we give a high-level overview of the proof of Theorem~\ref{thm:main}. Following Yao's minimax principle~\cite{Yao77}, we construct, at a constant degree depending on~$\eta$, a YES distribution supported on acyclic graphs and a NO distribution whose inputs are far from acyclic with high probability. We show that, for every deterministic adaptive algorithm $\mathcal A$ making $q=\Theta(n^{1-\eta})$ queries, the complete transcripts under these distributions are $o(1)$-close in total variation distance. Hence the rejection probabilities of $\mathcal A$ differ by $o(1)$, whereas a two-sided tester would need a difference of at least $\frac13-o(1)$. Lowering the degree to~$2$ and then raising it to any $d\ge2$ preserves the bound, with a proximity parameter independent of~$d$ (Section~\ref{rt:transfer}). We first recall the construction of Chen et al.~\cite{CRSS22} and explain two barriers that stand between it and our goal.

\subsection{The construction of Chen et al.\ and two barriers}
\label{sec:chen-baseline}

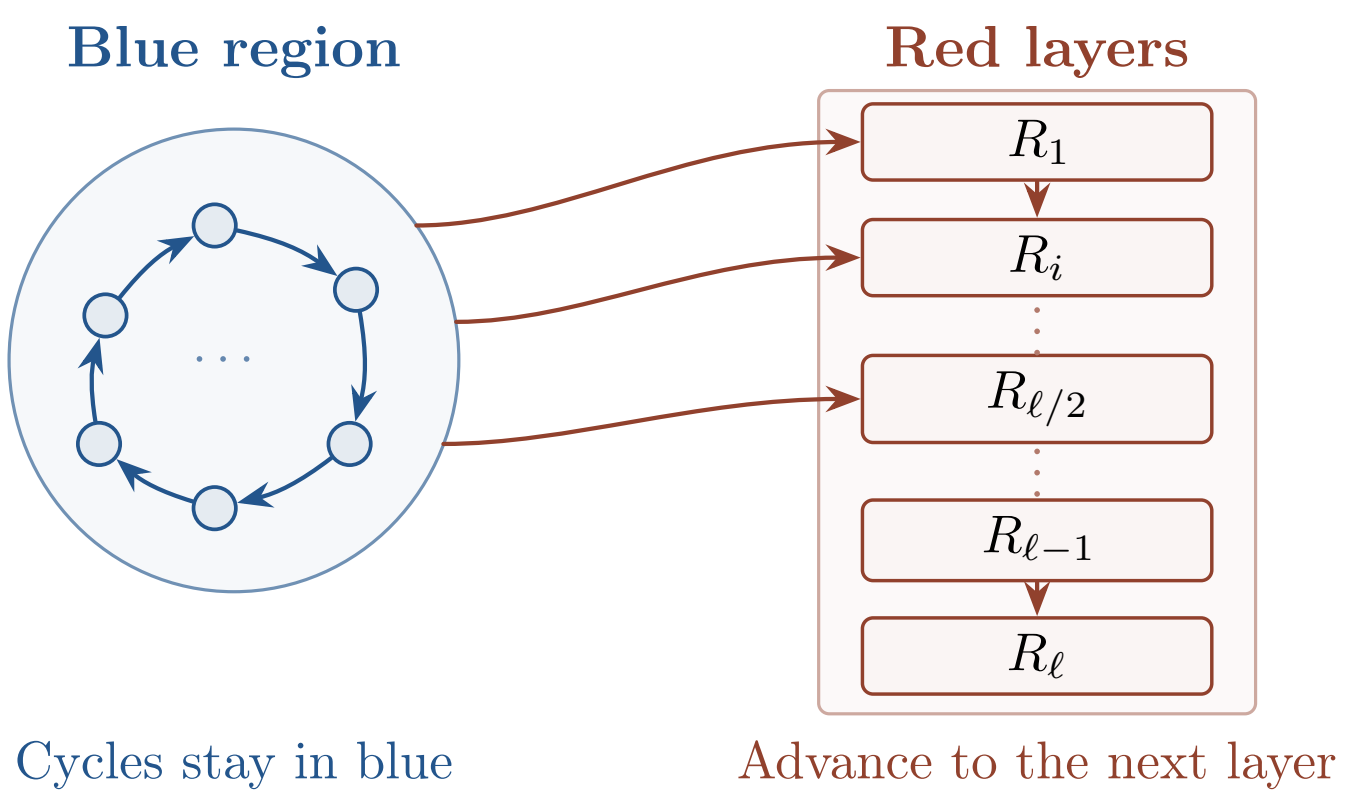
\begin{figure}[!t]
\centering
\begin{tikzpicture}[x=.85cm,y=.85cm,line cap=round,line join=round,
  acy/note/.append style={font=\small},acy/box/.append style={font=\small}]
 \draw[acy/bluearea] (2.0,2.3) ellipse [x radius=1.75,y radius=1.8];
 \node[acy/title,text=acyBlue] at (2.0,4.7) {Blue region};
 \node[acy/title,text=acyRed] at (8.25,4.7) {Red layers};
 \foreach \name/\xx/\yy in {v1/1.0/2.65,v2/1.85/3.35,
       v3/2.95/2.85,v4/2.90/1.65,v5/1.85/1.15,v6/.95/1.65}
   \node[acy/bluedot,minimum size=2.8mm] (\name) at (\xx,\yy) {};
 \foreach \from/\to in {v1/v2,v2/v3,v3/v4,v4/v5,v5/v6,v6/v1}
   \draw[acy/coreedge] (\from) to[bend left=12] (\to);
 \node[acy/note,text=acyBlue!70] at (1.95,2.3) {$\cdots$};
 \draw[acy/redarea] (6.55,-.45) rectangle (9.95,4.4);
 \foreach \tag/\yy/\lab in {a/4.0/1,b/3.1/i,c/2.0/{\ell/2},d/.9/{\ell-1},e/0/\ell}
   \node[acy/box,draw=acyRed,fill=acyRed!5,text width=2.1cm,
         inner sep=3pt,minimum height=.48cm] (\tag) at (8.25,\yy) {$R_{\lab}$};
 \draw[acy/rededge] (3.42,3.35) .. controls (4.5,3.35) and (5.5,4.0) .. (a.west);
 \draw[acy/rededge] (3.73,2.6) .. controls (4.7,2.6) and (5.6,3.1) .. (b.west);
 \draw[acy/rededge] (3.63,1.65) .. controls (4.6,1.65) and (5.6,2.0) .. (c.west);
 \draw[acy/rededge] (a.south)--(b.north);
 \draw[acy/rededge] (d.south)--(e.north);
 \foreach \yy in {2.65,1.55}
   \node[acy/note,text=acyRed!70] at (8.25,\yy) {$\vdots$};
\node[acy/note,text=acyBlue] at (2.0,-.85) {Cycles stay in blue};
 \node[acy/note,text=acyRed] at (8.25,-.85) {Advance to the next layer};
\end{tikzpicture}
\caption{The construction of Chen et al.~\cite{CRSS22}, schematically. Blue edges may form cycles; edges leaving the blue region enter the first half of the red layers and then move forward until they reach the last layer. Since every walk from a red vertex in layer $i$ has exactly $\ell-i$ steps, whereas walks from a blue vertex have varying lengths, a few walks reveal whether a vertex is blue; this is the basis of the attack in Barrier~1.}
\label{fig:chen-baseline}
\end{figure}

The construction of Chen et al.~\cite{CRSS22} contains a blue cyclic region and red acyclic layers (see Figure~\ref{fig:chen-baseline}). In their construction, $m$ \emph{blue} vertices choose random outneighbors among the other blue vertices and the first half of $\ell$ \emph{red} layers of width $w=2m/\ell$, while red vertices choose random outneighbors in the next layer; random IDs hide the partition. Every cycle is blue, and since a random edge of a blue vertex leaves the blue region with constant probability and never returns, it is hard for an algorithm to keep exploring the blue region. A query to $u$ closes a cycle only if it returns an \emph{ancestor} of~$u$, a vertex with a directed path to $u$ in the explored subgraph, so the analysis bounds the number of ancestors of blue vertices. To this end, Chen et al.\ divide the exploration into \emph{epochs}, each of which ends at a \emph{collision}, that is, an answer that repeats a seen ID, or after $\ell/2$ queries, and they show that new blue paths within an epoch are short. Balancing the number of epochs against the lengths of the blue paths yields the exponent $5/9$~\cite[Sections~5--7]{CRSS22}.

\paragraph{Barrier 1: the acyclic part {is easy to recognize}.}
The main challenge in going beyond this bound is that answers from the acyclic region can help an algorithm stay in the blue region. Indeed, walks from a red vertex reach a sink at a distance determined by its layer (see Figure~\ref{fig:chen-baseline}), whereas walks from blue vertices have varying lengths, so a few walks reveal the class of a vertex with about $\ell$ queries. Using this test, the attack sketched in \cite[Section~8]{CRSS22} grows a blue path of length about $\sqrt m$, which closes a cycle by a birthday argument after about $\ell\sqrt m$ queries. If $\ell>m^{1/4}$, an algorithm can first explore almost all of one layer in every $m^{1/4}$ consecutive layers, as in the second attack there; here, each breadth-first search starts at the endpoint of a short random walk, which is red with high probability. Since a layer has $2m/\ell$ vertices, this costs about $m^{3/4}$ queries, and the explored layers can then replace the sinks, so that each test costs only about $m^{1/4}$ queries. One can check that, for every $\ell$, a cycle can be found with $\widetilde O(n^{3/4})$ queries, and no analysis of this construction can give a near-linear lower bound. Thus we need the following.
\begin{itemize}
\item[(R1)] A new construction whose acyclic part {reveals the cyclic part only through rare events}, and an analysis of the information that {these events leak} to an adaptive algorithm, showing that, at a near-linear budget, it sees the cyclic part only in small and shallow pieces.
\end{itemize}

\paragraph{Barrier 2: two-sided testers use statistics.}
Since a two-sided tester may reject on statistical evidence alone, keeping cycles hidden is not enough: we must compare complete transcripts with those of a YES distribution that is acyclic and fixed in advance, including all equalities among IDs. At a near-linear budget, the transcripts contain many collisions, and these cannot simply be discarded. Hence we also need the following.
\begin{itemize}
\item[(R2)] An acyclic YES construction that agrees with the NO construction outside the cyclic part, and a comparison of the complete transcripts under the two constructions that accounts for every collision pattern an adaptive algorithm can create.
\end{itemize}

\noindent The rest of this section explains how we meet these requirements.

\subsection{Our construction}
\label{sec:construction-overview}

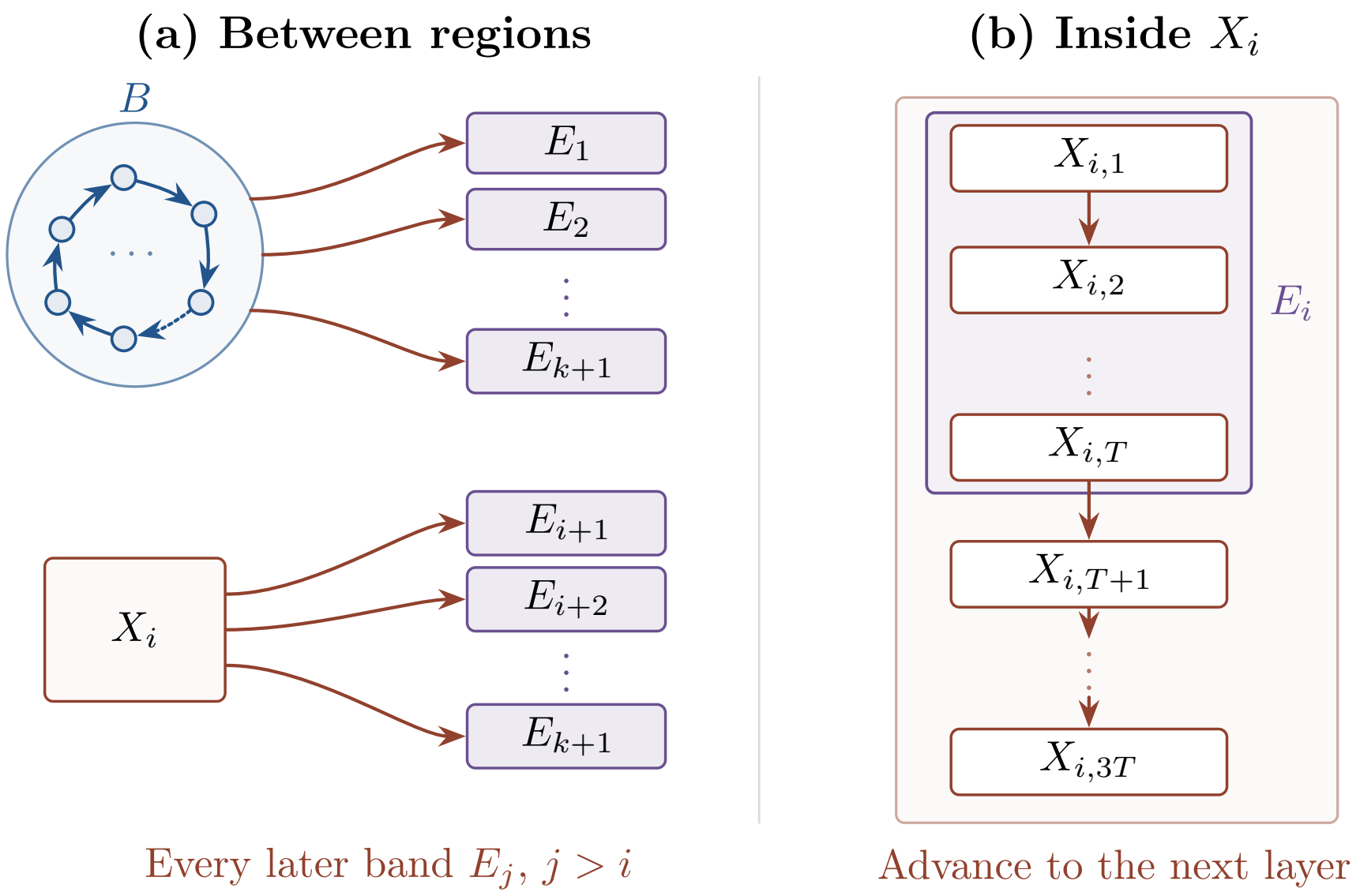
\begin{figure}[!t]
\centering
\begin{tikzpicture}[x=.85cm,y=.85cm,line cap=round,line join=round,
  acy/note/.append style={font=\small},acy/box/.append style={font=\small}]
 \node[acy/title] at (3.15,6.55) {(a) Between regions};
 \node[acy/title] at (10.55,6.55) {(b) Inside $X_i$};
 \draw[draw=acyInk!20,line width=.5pt] (7.05,-1.2)--(7.05,6.15);
 \begin{scope}[yshift=-.17cm]
 \begin{scope}[yshift=.17cm]
 \draw[acy/bluearea] (.9,4.4) ellipse [x radius=1.26,y radius=1.3];
 \node[acy/note,text=acyBlue] at (.9,5.95) {$B$};
 \foreach \name/\xx/\yy in {c1/.18/4.65,c2/.79/5.16,c3/1.58/4.80,
       c4/1.55/3.93,c5/.79/3.57,c6/.14/3.93}
   \node[acy/bluedot,minimum size=2mm] (\name) at (\xx,\yy) {};
 \foreach \from/\to in {c1/c2,c2/c3,c3/c4,c5/c6,c6/c1}
   \draw[acy/coreedge] (\from) to[bend left=10] (\to);
 \draw[acy/coreedge,densely dotted] (c4) to[bend left=12] (c5);
 \node[acy/note,text=acyBlue!70] at (.9,4.4) {$\cdots$};
 \end{scope}
 \foreach \tag/\yy/\lab in {bfirst/5.7/1,bsecond/4.95/2,blast/3.55/{k+1}}
   \node[acy/box,draw=acyEntry,fill=acyEntry!12,text width=1.45cm,
         inner sep=3pt,minimum height=.47cm] (\tag) at (5.15,\yy) {$E_{\lab}$};
 \node[acy/note,text=acyEntry] at (5.15,4.3) {$\vdots$};
 \draw[acy/rededge] (2.04,5.15) .. controls (2.9,5.15) and (3.65,5.7) .. (bfirst.west);
 \draw[acy/rededge] (2.16,4.6) .. controls (3.0,4.6) and (3.65,4.95) .. (bsecond.west);
 \draw[acy/rededge] (2.04,4.05) .. controls (2.9,4.05) and (3.65,3.55) .. (blast.west);
 \end{scope}
 \begin{scope}[yshift=-.38cm]
 \node[acy/box,draw=acyRed,fill=acyRed!3,text width=1.3cm,
       inner sep=3pt,minimum height=1.2cm] (source) at (.9,1.15) {$X_i$};
 \foreach \tag/\yy/\lab in {next/2.2/{i+1},later/1.45/{i+2},last/.1/{k+1}}
   \node[acy/box,draw=acyEntry,fill=acyEntry!12,text width=1.45cm,
         inner sep=3pt,minimum height=.47cm] (\tag) at (5.15,\yy) {$E_{\lab}$};
 \node[acy/note,text=acyEntry] at (5.15,.85) {$\vdots$};
 \draw[acy/rededge] ([yshift=8.5pt]source.east) .. controls (2.65,1.5) and (3.65,2.2) .. (next.west);
 \draw[acy/rededge] (source.east) .. controls (2.65,1.15) and (3.65,1.45) .. (later.west);
 \draw[acy/rededge] ([yshift=-8.5pt]source.east) .. controls (2.65,.8) and (3.65,.1) .. (last.west);
 \end{scope}
 \node[acy/note,text=acyRed] at (3.4,-1.65) {Every later band $E_j$, $j>i$};
 \draw[acy/redarea] (8.4,-1.2) rectangle (12.75,5.95);
 \draw[draw=acyEntry,fill=acyEntry!8,rounded corners=2pt,line width=.7pt]
   (8.7,2.05) rectangle (11.9,5.8);
 \node[acy/note,text=acyEntry] at (12.3,3.925) {$E_i$};
 \node[acy/box,draw=acyRed,fill=white,text width=2.1cm,
       inner sep=3pt,minimum height=.47cm] (entry-first) at (10.3,5.35) {$X_{i,1}$};
 \node[acy/box,draw=acyRed,fill=white,text width=2.1cm,
       inner sep=3pt,minimum height=.47cm] (entry-second) at (10.3,4.15) {$X_{i,2}$};
 \node[acy/box,draw=acyRed,fill=white,text width=2.1cm,
       inner sep=3pt,minimum height=.47cm] (entry-last) at (10.3,2.5) {$X_{i,T}$};
 \draw[acy/rededge] (entry-first.south)--(entry-second.north);
 \node[acy/note,text=acyRed!70] at (10.3,3.325) {$\vdots$};
 \node[acy/box,draw=acyRed,fill=white,text width=2.1cm,
       inner sep=3pt,minimum height=.55cm] (layer) at (10.3,1.25) {$X_{i,T+1}$};
 \node[acy/box,draw=acyRed,fill=white,text width=2.1cm,
       inner sep=3pt,minimum height=.55cm] (terminal) at (10.3,-.6) {$X_{i,3T}$};
 \draw[acy/rededge] (entry-last.south)--(layer.north);
 \draw[acy/rededge] (layer.south)-- ++(0,-.3);
 \draw[acy/rededge] ($(terminal.north)+(0,.3)$)--(terminal.north);
 \node[acy/note,text=acyRed!70] at (10.3,.42) {$\vdots$};
 \node[acy/note,text=acyRed] at (10.55,-1.65) {Advance to the next layer};
\end{tikzpicture}
\caption{The masking hierarchy of a NO instance. (a) Blue vertices can enter every band $E_1,\ldots,E_{k+1}$, and for $i\le k$ the region $X_i$ can enter every later band $E_j$, $j>i$. (b) The region $X_i$ has $3T$ layers $X_{i,1},\ldots,X_{i,3T}$; steps inside $X_i$ advance by one layer, and the entry band $E_i$ consists of the first $T$ layers.}
\label{fig:no-hierarchy}
\end{figure}

{Our} NO {distribution} hides a \emph{core} $B$ of $N=\Theta_\eta(n)$ vertices inside a hierarchy of acyclic \emph{masking regions} $X_1,\ldots,X_{k+1}$ for some appropriately chosen $k$ (see Figure~\ref{fig:no-hierarchy}). {The vertices of $B$ are \emph{blue}, and the core is a random graph given by $32$ permutations $\pi_1,\ldots,\pi_{32}$ of $B$: every blue vertex $u$ has the $32$ core edges $u\to\pi_1(u),\ldots,u\to\pi_{32}(u)$, and the edge to $\pi_a(u)$ has \emph{color}~$a$.} Each region $X_i$ has $3T$ layers of width $W$, where $N=TW$ {and $T=N^{\eta/8+o(1)}$}, and its first $T$ layers form the \emph{entry band} $E_i$. The vertices of $B$ have \emph{level}~$0$, and those of $X_i$ have level~$i$.

The whole input is generated by a single rule. Every adjacency list consists of $D$ entries, each of which has a hidden \emph{category} in $\{0,\ldots,k+1\}$; the categories are arranged in a uniformly random order, with exactly $d_j$ entries of category at most $j$, where $32=d_0<d_1<\cdots<d_{k+1}=D$. An entry of category $c$ at a vertex of level $i$ \emph{steps} within the region of its source if $c\le i$, and \emph{jumps} to a uniformly random vertex of the band $E_c$ if $c>i$. {At a blue vertex $u$, the $a$-th entry of category~$0$ steps to $\pi_a(u)$}, and a step in $X_i$ goes to a uniformly random vertex of the next layer, where the layer after the last one is $E_{i+1}$ {for $i\le k$, and the lists in the last layer of $X_{k+1}$ are empty}. {For instance, if $k=1$ and $(d_0,d_1,d_2)=(32,64,128)$, then every list has $32$ entries of category~$0$, $32$ of category~$1$, and $64$ of category~$2$. A blue vertex follows core edges with its entries of category~$0$ and jumps to $E_1$ and $E_2$ with the others; a vertex of $X_1$ steps with its $64$ entries of category at most~$1$ and jumps to $E_2$ with the others; and a vertex of $X_2$ steps with all $128$ entries.} (A query reveals the endpoints of the edges, but not their categories.) Levels never decrease along an edge, and layers increase inside a region; therefore every cycle lies in $B$.

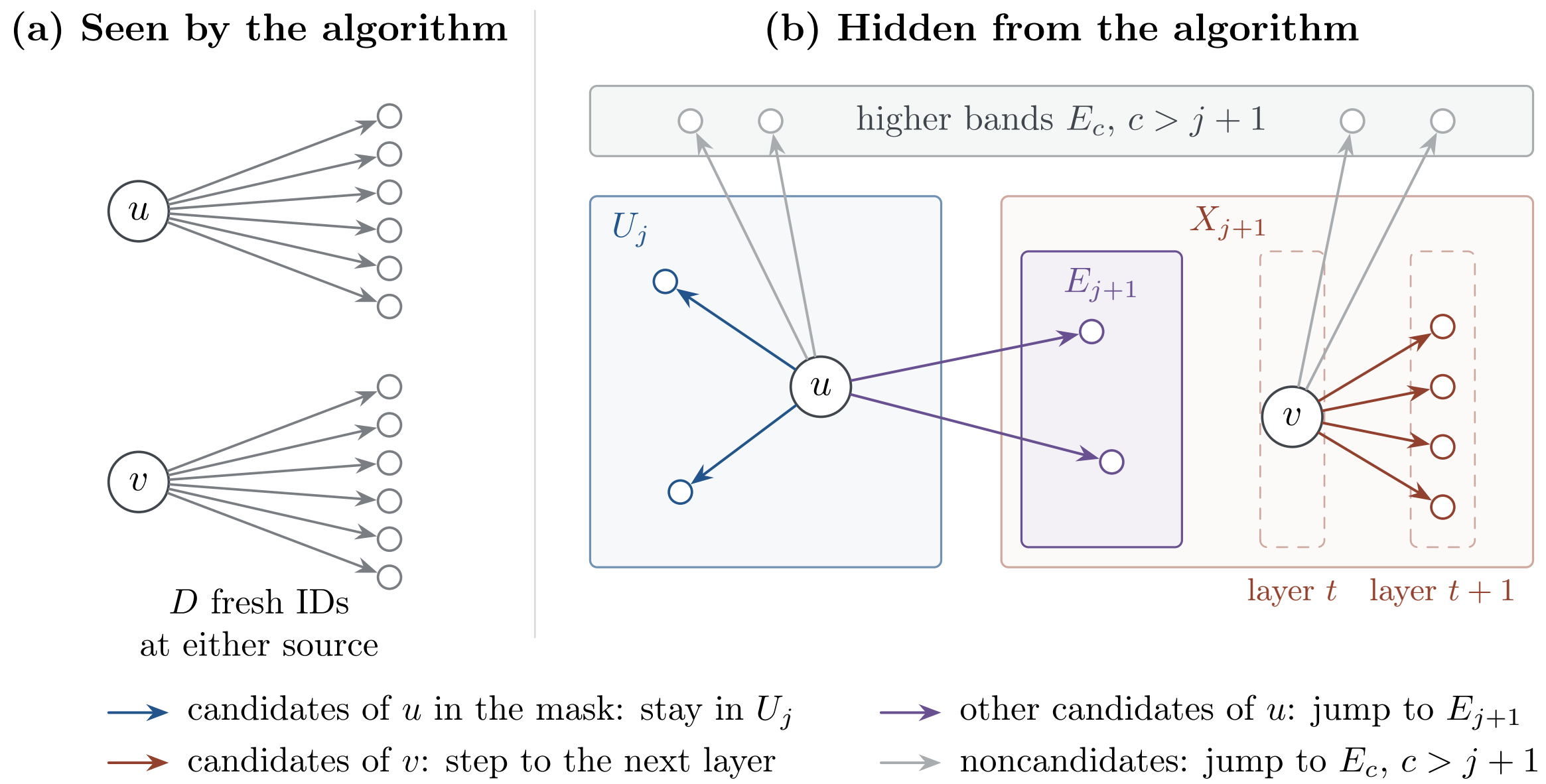
\begin{figure}[!t]
\centering
\begin{tikzpicture}[x=1cm,y=1cm,line cap=round,line join=round,
  tgt/.style={acy/dot,minimum size=2.3mm,draw=#1,fill=white},
  seen/.style={acy/edge,line width=.7pt,draw=acyInk!70},
  sel/.style={acy/edge,line width=.75pt,draw=acyBlue},
  jmp/.style={acy/edge,line width=.75pt,draw=acyEntry},
  stp/.style={acy/edge,line width=.75pt,draw=acyRed},
  hi/.style={acy/edge,line width=.75pt,draw=acyInk!45},
  lab/.style={acy/note,font=\small}]
\node[acy/title] at (1.2,4.75) {(a) Seen by the algorithm};
 \foreach \src/\yy in {u/2.95,v/.25} {
   \node[acy/vertex] (\src) at (0,\yy) {$\src$};
   \foreach \dy [count=\n] in {-.95,-.57,-.19,.19,.57,.95} {
     \node[tgt=acyInk!70] (\src\n) at (2.5,\yy+\dy) {};
     \draw[seen] (\src) -- (\src\n);
   }
 }
 \node[lab] at (1.2,-1.15) {$D$ fresh IDs\\at either source};
 \draw[draw=acyInk!20,line width=.5pt] (3.95,-1.3)--(3.95,4.95);
\begin{scope}[xshift=4.5cm]
  \node[acy/title] at (4.7,4.75) {(b) Hidden from the algorithm};
  \draw[draw=acyInk!45,fill=acyInk!5,rounded corners=2pt,line width=.6pt] (0,3.5) rectangle (9.4,4.2);
  \node[lab,text=acyInk] at (4.7,3.85) {higher bands $E_c$, $c>j+1$};
  \draw[acy/bluearea,rounded corners=2pt] (0,-.6) rectangle (3.5,3.1);
  \node[lab,text=acyBlue,anchor=north west] at (.08,3.07) {$U_j$};
  \draw[acy/redarea] (4.1,-.6) rectangle (9.4,3.1);
  \node[lab,text=acyRed] at (6.37,2.83) {$X_{j+1}$};
  \draw[draw=acyEntry,fill=acyEntry!8,rounded corners=2pt,line width=.6pt] (4.3,-.4) rectangle (5.9,2.55);
  \node[lab,text=acyEntry,anchor=north] at (5.1,2.52) {$E_{j+1}$};
  \foreach \xx/\lab in {7.0/t,8.5/{t+1}} {
    \draw[draw=acyRed!45,dashed,rounded corners=2pt,line width=.5pt] (\xx-.32,-.4) rectangle (\xx+.32,2.55);
    \node[acy/note,font=\footnotesize,text=acyRed] at (\xx,-.85) {layer $\lab$};
  }
  \node[acy/vertex,fill=white] (bu) at (2.3,1.2) {$u$};
  \node[acy/vertex,fill=white] (bv) at (7.0,.9) {$v$};
  \foreach \st/\col/\xx/\yy in {sel/acyBlue/.75/2.25, sel/acyBlue/.9/.15,
       jmp/acyEntry/5.0/1.75, jmp/acyEntry/5.2/.45,
       hi/acyInk!45/1.0/3.85, hi/acyInk!45/1.8/3.85} {
    \node[tgt=\col] (t) at (\xx,\yy) {};
    \draw[\st] (bu) -- (t);
  }
  \foreach \st/\col/\xx/\yy in {stp/acyRed/8.5/0, stp/acyRed/8.5/.6,
       stp/acyRed/8.5/1.2, stp/acyRed/8.5/1.8,
       hi/acyInk!45/7.6/3.85, hi/acyInk!45/8.5/3.85} {
    \node[tgt=\col] (t) at (\xx,\yy) {};
    \draw[\st] (bv) -- (t);
  }
 \end{scope}
\foreach \st/\xx/\yy/\txt in {
    sel/-.3/-2.05/{candidates of $u$ in the mask: stay in $U_j$},
    stp/-.3/-2.55/{candidates of $v$: step to the next layer},
    jmp/7.4/-2.05/{other candidates of $u$: jump to $E_{j+1}$},
    hi/7.4/-2.55/{noncandidates: jump to $E_c$, $c>j+1$}} {
   \draw[\st] (\xx,\yy) -- ++(.6,0);
   \node[lab,anchor=west] at (\xx+.65,\yy) {\txt};
 }
\end{tikzpicture}
\caption{The level-$j$ decoy, drawn for $D=6$, $d_{j+1}=4$, and $d_j=2$. (a) An algorithm sees only the returned IDs, and a typical answer consists of fresh IDs, whether the queried vertex is $u\in U_j$ or $v\in X_{j+1}$. (b) The hidden destinations of the same entries. The \emph{noncandidates}, which have category larger than $j+1$, jump to the same bands from both sources. The other $d_{j+1}$ entries are the \emph{candidates}. At $v$, all of them step to the next layer of $X_{j+1}$. At $u$, only the candidates in the \emph{selector mask}, a uniformly random subset of size $d_j$, can stay in $U_j$, and the others jump to $E_{j+1}$. Hence, to keep exploring $U_j$, an algorithm must learn which candidates are in the mask.}
\label{fig:decoy}
\end{figure}

{Consecutive levels look alike: the region $X_{j+1}$ is a decoy for $U_j=B\cup X_1\cup\cdots\cup X_j$. To see this, let $X_0=B$, and compare a vertex $u\in X_j$ with a vertex $v\in X_{j+1}$. At both, the entries of category larger than $j+1$ jump to the same bands, and the other $d_{j+1}$ entries, the \emph{candidates}, return fresh IDs. At $v$, all candidates step to the next layer; at $u$, only those of category at most $j$ step within $X_j$, and the others jump to $E_{j+1}$. Hence the answers at $u$ and $v$ look the same, and the level of $u$ shows only in which of its candidates stay at level $j$: a uniformly random $d_j$-subset, the level-$j$ \emph{selector mask} (see Figure~\ref{fig:decoy}). The same holds at every vertex of $U_j$, whose masked candidates stay in $U_j$. To keep exploring $U_j$, and in particular the core, an algorithm must therefore guess masked candidates, and each guess succeeds only with probability $d_j/d_{j+1}$, which is small since the category counts are widely separated.}

{Each decoy can still be detected, but only through rare events. Only in the top region $X_{k+1}$ does every entry step, so that walks follow the steps automatically, as in Barrier~1; in a lower region $X_{j+1}$, following the steps requires telling them apart from jumps to higher levels, which is the same problem one level up. Even in the top region, a walk becomes informative only after about $T$ queries, since jumps land in the first $T$ of its $3T$ layers, and exploring a whole layer, as in Barrier~1, is out of reach, since a layer has $W\gg q$ vertices. Finally, with $q$ queries, a new answer is a collision with probability only about $q/W$. Both rates, $1/T$ and $q/W$, are $N^{-\Omega(\eta)}$, so each level passes on only an $N^{-\Omega(\eta)}$ fraction of the exploration above it, and $k+1=O(1/\eta)$ levels suffice. This addresses the first half of~(R1); the second half, the analysis of the information that these events leak, makes this heuristic rigorous and is the subject of Section~\ref{sec:ov-info}.}

{We condition the core on weak girth greater than $\gamma=\Theta_\eta(\log N)$, that is, ignoring directions, it has no cycle of length at most $\gamma$. Then every ball of radius less than $\gamma/2$ is the same colored tree in all such cores, in which every vertex has one outgoing and one incoming edge of each color.} Switching arguments show that conditioning costs only a factor $e^{-o(N)}$ in probability, that unexposed endpoints remain diffuse under adaptive exposure, and that the core remains far from acyclic. {The core has in- and outdegree $32$, whereas the masking degree $D$ is a large constant depending on $\eta$.}

\paragraph{The YES distribution.}
The YES distribution uses the same rule and changes only the core{, which must be acyclic and yet look locally like a NO core}. We fix $32$ permutations $\rho_1,\ldots,\rho_{32}$ of $[T]$ whose union, the \emph{base core}, has weak girth greater than $\gamma$, we identify $B$ with $[W]\times[T]$, and we let the core step of color $a$ go from $(s,x)$ to $(s+1,\rho_a(x))$, except that it jumps into $E_1$ from the last blue layer $s=W$. Thus $B$ becomes one more layered region in front of $X_1$, and every YES instance is acyclic. If the last blue layer wrapped around to the first one instead, we would obtain the \emph{periodic cover} of the base core, which is again a union of $32$ permutations of $B$ with weak girth greater than $\gamma$ (see Figure~\ref{fig:cover}). From now on, girth always means weak girth.

\paragraph{Bounded degrees.}

As described, every vertex has constant expected indegree, but the targets are independent, so the maximum indegree is unbounded. For simplicity, we present the proof for this construction. To bound the indegrees, we give every masking vertex $R=(3k+4)D$ incoming slots for targets chosen in its layer and $R$ for its band, and route the entries to distinct random slots. Then all degrees are at most $2R$; Appendix~\ref{app:indegree} shows that the transcript comparison and exploration bounds remain valid. We next lower the degree to~$2$: each vertex becomes a binary out-tree and a binary in-tree with a common root, and each edge leaves its source at a leaf of the out-tree and enters its target at a random free leaf of the in-tree. Cycles survive, the normalized distance decreases by a factor $O(R)$, and each query costs at most one query to the original graph, since the random leaf can be drawn when its edge is first seen.

To obtain an arbitrary fixed degree bound $d\ge2$, we then replace every vertex of a degree-two graph by $t=\lfloor d/2\rfloor$ copies and every edge by all $t^2$ edges between the corresponding copies. Both degrees are at most $2t\le d$, while the number of vertices is multiplied by~$t$ and the minimum feedback arc set size by~$t^2$. 
Thus the distance, normalized by $dn$, drops by at most an absolute constant factor, and each query still costs at most one query to the degree-two graph.
Section~\ref{rt:transfer} proves both reductions and obtains a proximity parameter independent of~$d$.

\subsection{From transcripts to counting forests}
\label{sec:ov-reduction}

{Since collisions cannot be discarded (Barrier~2), we compute transcript probabilities exactly, which reduces the comparison of transcripts to a counting problem (Section~\ref{rt:section}).} For the analysis, we append to every transcript \emph{marks} that the algorithm does not see: which observed vertices are blue, the coordinates of the other observed vertices, and the category words of the queried lists. Forgetting the marks cannot increase the total variation distance, and given the marks, every answer outside the core is uniform in a known layer or band, under YES and under NO alike, and only the positions of the observed blue vertices in the core remain hidden. {The category word of a queried blue vertex $u$ shows which entries of its list have category~$0$, so every such entry, with color~$a$ and returned ID~$v$, becomes a colored edge $u\xrightarrow{a}v$, meaning that $v=\pi_a(u)$. These edges form the \emph{observed blue graph} $F_\xi$, and a placement of the observed blue vertices in the core is consistent with the transcript exactly if it maps every such edge to a core edge of the same color.} Hence, for a fixed NO core $G$, the probability of a marked transcript $\xi$ factors exactly as
\[
 \Pr[\xi\mid G]=c_{\mathcal A}(\xi)\cdot\Emb_G(F_\xi),
\]
where $\Emb_G(F_\xi)$ is the number of {such placements, that is, of} injective embeddings of $F_\xi$ into $G$, and the factor $c_{\mathcal A}(\xi)$ does not depend on the core. Since the embeddings are injective, every collision among blue vertices is accounted for, and the adaptive choices of queries and the stopping time enter only through $c_{\mathcal A}(\xi)$. {For example, if the algorithm has seen a blue path $u\xrightarrow{1}v\xrightarrow{2}w\xrightarrow{3}x$, then an embedding is determined by the image $y$ of $u$, since it must map $v$, $w$, and $x$ to $\pi_1(y)$, $\pi_2(\pi_1(y))$, and $\pi_3(\pi_2(\pi_1(y)))$. Girth greater than~$3$ makes these four vertices distinct, so there are exactly $N$ embeddings into every NO core, and such a transcript is equally likely for every NO core.}

{The YES distribution agrees with the NO distribution outside the core, as the first half of~(R2) requires, so the same factor appears under YES: an embedding into the periodic cover that avoids its last layer uses only core steps that every YES instance also has, and so the YES probability of $\xi$ is at least $c_{\mathcal A}(\xi)$ times the number of such embeddings. Few embeddings meet the last layer: layer translation is an automorphism of the periodic cover, so every vertex of a uniformly random embedding lies in a uniform layer, and hence at most a fraction $|F|/W$ of the embeddings of a forest $F$ meets the last layer, where $|F|=O(q)$ is much smaller than~$W$. It therefore suffices to show that, for a typical NO transcript, the observed blue graph has almost as many embeddings into the periodic cover as into a random core. We prove this in two steps, one probabilistic and one combinatorial.}

\begin{inftheorem}[Structural bound; Theorem~\ref{os:structural}]
Under the NO distribution, with probability $1-o(1)$, the observed blue graph is a forest whose components have at most $N^{z}$ vertices and diameter at most $b\log N$, where the constants $z,b>0$ can be made as small as needed by the choice of parameters.
\end{inftheorem}

\begin{inftheorem}[Forest comparison; Theorem~\ref{rf:comparison}]
Let $G,G'$ be two cores of the same degree and girth greater than $2K\Lambda+1$, and let $F$ be a forest with $c$ components, each with at most $S$ vertices and diameter at most $\Lambda$. If $cS^2/N$ is sufficiently small, then $\Emb_G(F)$ and $\Emb_{G'}(F)$ agree up to a factor $\exp\bigl(\pm 8e\,c\,(2ecS^2/N)^K\bigr)$.
\end{inftheorem}

{We explain the comparison first, since it shows what the structural bound must deliver.}

\subsection{Matching collision patterns to any fixed order}
\label{sec:ov-counting}

By the reduction above, it suffices to show that every observed forest has almost the same number of embeddings into two high-girth cores. {Both cores in question, the random NO core and the periodic cover, are \emph{eligible}: they are unions of $32$ permutations with girth greater than $\gamma$.} The comparison rests on locality: {every collision pattern that matters fits inside a ball of radius less than half the girth, where all eligible cores look alike.}

\paragraph{Few components: exact equality.}
{As in the example above, a tree component whose diameter is smaller than the girth has exactly $N$ embeddings into any core, since the image of one vertex determines the others. So all dependence on the core comes from the requirement that different components have disjoint images. For instance, two edges of color~$1$ have exactly $N(N-3)$ embeddings into every core of girth greater than~$2$: once the first edge is placed on $(x,\pi_1(x))$, the source of the second edge must avoid $x$, $\pi_1(x)$, and $\pi_1^{-1}(x)$, which are distinct, since the core has no cycles of length $1$ or~$2$; with a cycle of length~$2$ through $x$, the count would change. In general, once one of two components is placed, every placement of the other that overlaps it has its root within distance $2\Lambda$ of the first root, where $\Lambda$ bounds the diameters. If the girth exceeds $4\Lambda+1$, this ball is the same colored tree in every eligible core, so the number of overlapping placements, and with it the embedding count, is the same in all eligible cores.}

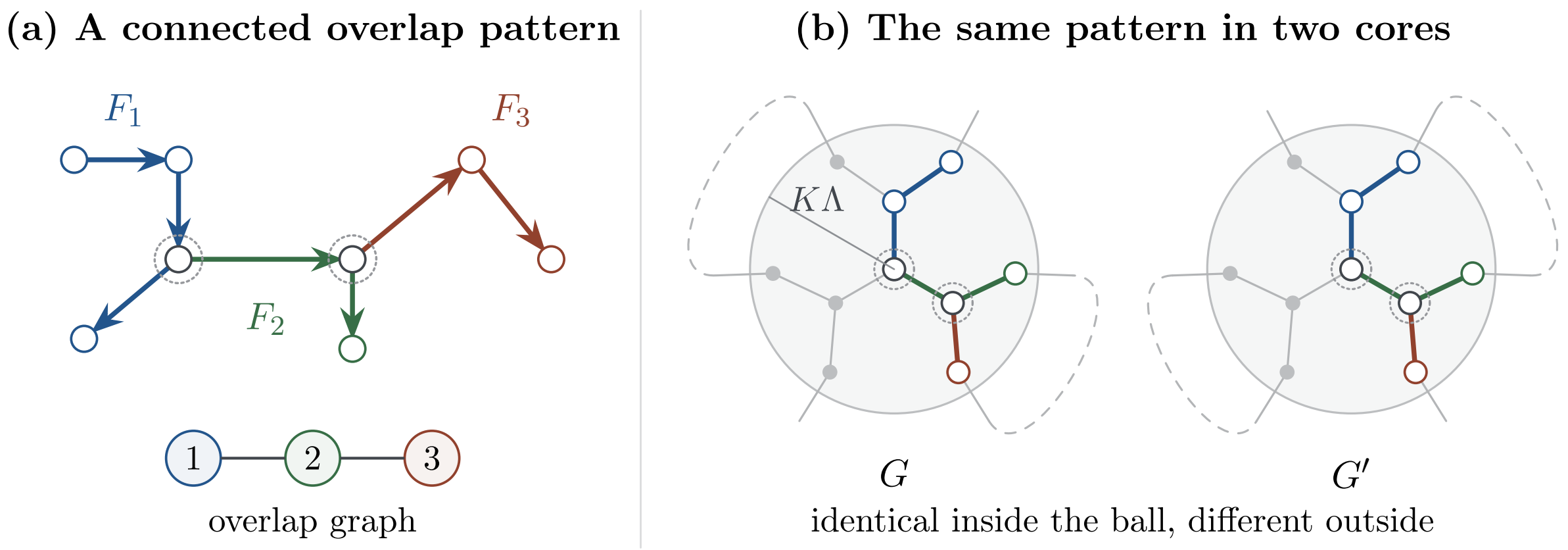
\begin{figure}[!t]
\centering
\begin{tikzpicture}[x=1cm,y=1cm,line cap=round,line join=round,
  comp/.style={line width=1.4pt,shorten >=2.5pt,shorten <=2.5pt},
  tree/.style={draw=acyInk!40,line width=.6pt},
  tv/.style={circle,fill=acyInk!35,inner sep=0pt,minimum size=1.5mm},
  lab/.style={acy/note,font=\small}]
\node[acy/title] at (2.45,1.95) {(a) A connected overlap pattern};
 \foreach \name/\xx/\yy in {f/.05/.65,g/1.1/.65,p/1.1/-.35,h/.15/-1.15,
      q/2.85/-.35,t/2.85/-1.25,r/4.05/.65,s/4.85/-.35}
   \coordinate (\name) at (\xx,\yy);
 \foreach \from/\to in {f/g,g/p,p/h}
   \draw[acy/coreedge,comp] (\from)--(\to);
 \foreach \from/\to in {p/q,q/t}
   \draw[acy/edge,draw=acyGreen,comp] (\from)--(\to);
 \foreach \from/\to in {q/r,r/s}
   \draw[acy/rededge,comp] (\from)--(\to);
 \foreach \name/\color in {f/acyBlue,g/acyBlue,h/acyBlue,t/acyGreen,r/acyRed,s/acyRed}
   \node[acy/dot,draw=\color,fill=white,minimum size=2.6mm] at (\name) {};
 \foreach \name in {p,q} {
   \draw[draw=acyInk!55,densely dotted,line width=.65pt] (\name) circle [radius=.24cm];
   \node[acy/dot,draw=acyInk,fill=white,minimum size=2.6mm] at (\name) {};
 }
 \node[acy/note,text=acyBlue] at (.55,1.15) {$F_1$};
 \node[acy/note,text=acyGreen] at (1.98,-.95) {$F_2$};
 \node[acy/note,text=acyRed] at (4.45,1.15) {$F_3$};
 \foreach \name/\xx/\color/\lab in {o1/1.25/acyBlue/1,o2/2.45/acyGreen/2,o3/3.65/acyRed/3}
   \node[acy/vertex,draw=\color,fill=\color!7,minimum size=5.5mm,font=\small] (\name) at (\xx,-2.35) {$\lab$};
 \draw[draw=acyInk,line width=.8pt] (o1)--(o2)--(o3);
 \node[lab] at (2.45,-3.0) {overlap graph};
 \draw[draw=acyInk!20,line width=.5pt] (5.75,-3.25)--(5.75,2.15);
\node[acy/title] at (10.6,1.95) {(b) The same pattern in two cores};
 \foreach \cx/\nm/\arcs in {8.3/G/{118/182,302/358},12.9/{G'}/{182/238,358/62}} {
  \begin{scope}[shift={(\cx,-.45)}]
   \draw[draw=acyInk!35,fill=acyInk!5,line width=.6pt] (0,0) circle [radius=1.45];
   \coordinate (o) at (0,0);
   \foreach \a in {90,210,330} {
     \coordinate (l\a) at (\a:.68);
     \draw[tree] (o)--(l\a);
     \foreach \d in {-28,28} {
       \pgfmathtruncatemacro\b{\a+\d}
       \coordinate (m\b) at (\b:1.22);
       \draw[tree] (l\a)--(m\b)--(\b:1.8);
     }
   }
   \foreach \x/\y in \arcs
     \draw[tree,dashed] (\x:1.8) to[out=\x,in=\y,looseness=1.15] (\y:1.8);
   \foreach \v in {l90,l210,l330,m62,m118,m182,m238,m302,m358}
     \node[tv] at (\v) {};
   \draw[draw=acyBlue,comp] (m62)--(l90)--(o);
   \draw[draw=acyGreen,comp] (o)--(l330)--(m358);
   \draw[draw=acyRed,comp] (l330)--(m302);
   \foreach \v/\color in {m62/acyBlue,l90/acyBlue,m358/acyGreen,m302/acyRed}
     \node[acy/dot,draw=\color,fill=white,minimum size=2.2mm] at (\v) {};
   \foreach \v in {o,l330} {
     \draw[draw=acyInk!55,densely dotted,line width=.6pt] (\v) circle [radius=.2cm];
     \node[acy/dot,draw=acyInk,fill=white,minimum size=2.2mm] at (\v) {};
   }
   \node[acy/note] at (0,-2.05) {$\nm$};
  \end{scope}
 }
 \draw[draw=acyInk!60,line width=.5pt] (8.3,-.45)--++(150:1.45) node[pos=.62,above,font=\small,text=acyInk] {$K\Lambda$};
 \node[lab] at (10.6,-3.0) {identical inside the ball, different outside};
\end{tikzpicture}
\caption{Matching collision patterns. (a) Independently placed images of the components $F_1,F_2,F_3$ of an observed forest may share vertices (circled), and the overlap graph records which images meet. The cluster expansion assigns an activity to every connected overlap pattern. (b) A connected pattern of at most $K$ images, each of diameter at most $\Lambda$, lies in a ball of radius $K\Lambda$ (shaded). If the girth exceeds $2K\Lambda+1$, this ball is the same tree in every eligible core, although two cores may close different cycles outside it (dashed). Hence the activities of order at most $K$ agree exactly. (Stroke colors distinguish components, not core colors.)}
\label{fig:overlaps}
\end{figure}

\paragraph{Many components: the birthday problem.}
{At a near-linear budget, however, there are many components: for $c$ components of size $S$, the expected number of overlapping pairs is about $c^2S^2/N$, which is large. Collisions are therefore unavoidable, as discussed in Barrier~2, and an argument that treats them one pair at a time stops at the birthday bound.} The right question is not whether components collide, but how many of them collide together. A cluster expansion\footnote{{The term comes from statistical physics and is unrelated to graph expansion.}} makes this precise: it expresses the disjointness probability as a sum over collections of disjoint clusters, in which each cluster contributes its \emph{activity}, a quantity that depends only on the components of the cluster and accounts for the ways in which their images may overlap in a connected pattern (see Figure~\ref{fig:overlaps}(a)). For two components, the expansion is just $\Pr[\text{disjoint}]=1-\Pr[\text{overlap}]$, so the activity of a pair is minus the probability that their images overlap. For the two edges above it is $-3/N$, which recovers their $N^2(1-3/N)=N(N-3)$ embeddings.

\paragraph{Small clusters are local, large clusters are rare.}
A connected cluster of at most $K$ component images, each of diameter at most $\Lambda$, lies in a ball of radius $K\Lambda$. If $\gamma>2K\Lambda+1$, then {this ball is the same colored tree in every eligible core, so the activity of the cluster is the same count in all of them, just as for two components} (see Figure~\ref{fig:overlaps}(b)). Large clusters, on the other hand, are rare: attaching another overlapping component costs a factor $S^2/N$, since a pair of vertices determines at most one position of a root, and there are at most $c$ components to attach. Hence the clusters of more than $K$ components contribute only $O(c\,(cS^2/N)^{K})${; since the activities have signs, turning this into the error bound $8e\,c\,(2ecS^2/N)^K$ needs some care (Section~\ref{rf:section})}. Thus the diameter of the observed components buys locality, and their size buys rarity. This meets the second half of~(R2): no collision is discarded, and only collisions among more than $K$ components at once are charged to the error.

\subsection{Iterated information control}
\label{sec:ov-info}

The main technical contribution of this paper is the proof of the structural bound, which must overcome Barrier~1 at a near-linear budget. The idea is that an algorithm can follow the core only if it knows the selector masks: {blind guesses fail quickly, and the algorithm} learns about the masks only through rare events. We first explain this idea for a single search at a single level $j$, {then how the levels interact, and finally} how the argument covers the whole exploration.

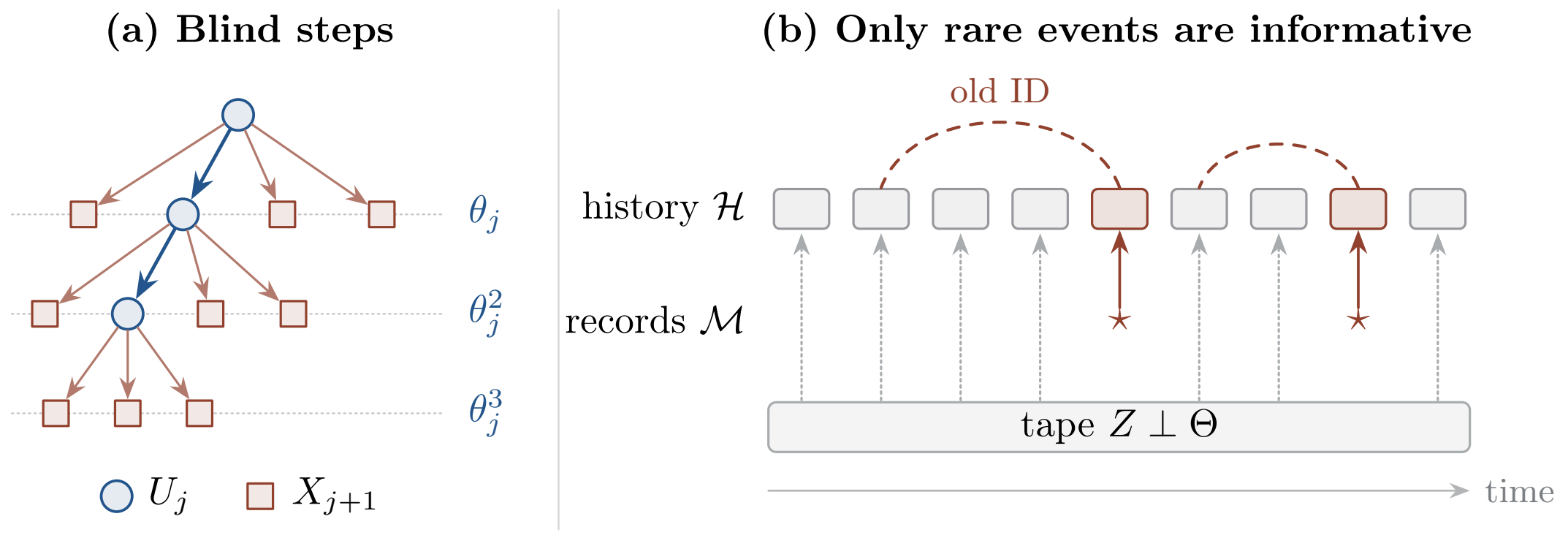
\begin{figure}[!t]
\centering
\begin{tikzpicture}[x=1cm,y=1cm,line cap=round,line join=round,
  pt/.style={acy/title,font=\small\bfseries},
  m/.style={acy/note,font=\small},
  ub/.style={acy/bluedot,minimum size=2.8mm},
  xr/.style={rectangle,draw=acyRed,fill=acyRed!12,line width=.6pt,minimum size=2.3mm,inner sep=0pt},
  keep/.style={acy/edge,draw=acyBlue,line width=.9pt},
  leave/.style={acy/edge,draw=acyRed!70,line width=.6pt},
  contact/.style={draw=acyRed,line width=.8pt,dashed},
  rec/.style={text=acyRed,font=\large,inner sep=0pt},
  ans/.style={rectangle,rounded corners=1.5pt,draw=acyInk!55,fill=acyInk!8,line width=.6pt,minimum width=5mm,minimum height=3.6mm,inner sep=0pt},
  guide/.style={draw=acyInk!30,line width=.45pt,densely dotted},
  piece/.style={draw=acyBlue!60,dashed,rounded corners=6pt,line width=.6pt,fill=acyBlue!3},
  qd/.style={circle,fill=acyInk!55,inner sep=0pt,minimum size=1.3mm}]
\node[pt] at (2.2,4.35) {(a) Blind steps};
 \foreach \yy/\p in {2.7/{\theta_j},1.8/{\theta_j^2},.9/{\theta_j^3}} {
   \draw[guide] (.05,\yy)--(3.95,\yy);
   \node[m,text=acyBlue,anchor=west] at (4.05,\yy) {$\p$};
 }
 \node[ub] (r) at (2.1,3.6) {};
 \foreach \n/\xx in {a1/.7,a3/2.5,a4/3.4} \node[xr] (\n) at (\xx,2.7) {};
 \node[ub] (a2) at (1.6,2.7) {};
 \foreach \n/\xx in {b1/.35,b3/1.85,b4/2.6} \node[xr] (\n) at (\xx,1.8) {};
 \node[ub] (b2) at (1.1,1.8) {};
 \foreach \n/\xx in {c1/.45,c2/1.1,c3/1.75} \node[xr] (\n) at (\xx,.9) {};
 \draw[keep] (r)--(a2); \draw[keep] (a2)--(b2);
 \foreach \n in {a1,a3,a4} \draw[leave] (r)--(\n);
 \foreach \n in {b1,b3,b4} \draw[leave] (a2)--(\n);
 \foreach \n in {c1,c2,c3} \draw[leave] (b2)--(\n);
 \node[ub] at (1.0,.15) {}; \node[m,anchor=west] at (1.15,.15) {$U_j$};
 \node[xr] at (2.3,.15) {}; \node[m,anchor=west] at (2.45,.15) {$X_{j+1}$};
 \draw[draw=acyInk!20,line width=.5pt] (5.0,-.15)--(5.0,4.55);
\begin{scope}[shift={(6.75,0)}]
  \node[pt] at (3.3,4.35) {(b) Only rare events are informative};
  \foreach \i in {1,...,9} {
    \pgfmathsetmacro\xx{.45+.72*(\i-1)}
    \node[ans] (q\i) at (\xx,2.75) {};
  }
  \foreach \i in {5,8} {
    \pgfmathsetmacro\xx{.45+.72*(\i-1)}
    \node[ans,draw=acyRed,fill=acyRed!15] (q\i) at (\xx,2.75) {};
    \node[rec] (s\i) at (\xx,1.75) {$\star$};
    \draw[->,>=Stealth,draw=acyRed,line width=.7pt] (s\i)--(q\i);
  }
  \draw[contact] (q5.north) to[out=110,in=70] node[pos=.5,above=1pt,m,font=\footnotesize,text=acyRed] {old ID} (q2.north);
  \draw[contact] (q8.north) to[out=110,in=70] (q6.north);
  \draw[draw=acyInk!45,fill=acyInk!6,rounded corners=2pt,line width=.6pt] (.15,.55) rectangle (6.5,1.0);
  \node[m] at (3.33,.78) {tape $Z\perp\Theta$};
  \foreach \i in {1,2,3,4,6,7,9} {
    \pgfmathsetmacro\xx{.45+.72*(\i-1)}
    \draw[->,>=Stealth,draw=acyInk!45,densely dotted,line width=.6pt] (\xx,1.02)--(\xx,2.53);
  }
  \node[m,anchor=east] at (.08,2.75) {history $\mathcal H$};
  \node[m,anchor=east] at (.08,1.75) {records $\mathcal M$};
  \draw[->,>=Stealth,draw=acyInk!40,line width=.5pt] (.15,.2)--(6.5,.2) node[right,m,font=\footnotesize,text=acyInk!70] {time};
 \end{scope}
\end{tikzpicture}
\caption{Why the masks stay hidden. (a) Blind steps: if the masks are independent of the history, then every step inside $U_j$ (blue) survives with probability $\theta_j=d_j/d_{j+1}$, and the other candidates leave for $X_{j+1}$ (red); so a vertex at depth $h$ stays in $U_j$ with probability $\theta_j^{h}$. (b) Only rare events are informative: the answers $\mathcal H$ of a search, in order of time. Ordinary answers consist of fresh IDs and are replayed from a tape $Z$ that is independent of the masks $\Theta$ (dotted). A collision (dashed arcs) triggers an exceptional record, and only the records $\mathcal M$ carry information about the masks, $O(\log N)$ bits per record.}
\label{fig:info}
\end{figure}

\paragraph{Warm-up: blind steps.}
Consider a search that starts at a vertex of $U_j$ and only ever receives fresh IDs. Every edge from $U_j$ to $U_j$ is a candidate in the selector mask of its source, and a fixed candidate lies in the mask with probability $\theta_j=d_j/d_{j+1}$, independently at different sources. Hence, if the masks are independent of everything that the algorithm has seen, a fixed path of length $h$ in the discovery tree stays in $U_j$ with probability $\theta_j^h${, which is at most $N^{-80}$ for $h=80\log N/\log(1/\theta_j)$. So a blind search discovers $U_j$ only in pieces of depth less than $h$, each with at most $d_j^{h}$ vertices (see Figure~\ref{fig:info}(a)). By separating the category counts, we make $\theta_j$ so small that $h$ is a small multiple of $\log N$ and $d_j^{h}\le N^{\zeta}$ for a small $\zeta>0$, without changing the core degree $32$, and hence without changing the girth threshold. This is where the small and shallow components required by the comparison come from.}

\paragraph{Only rare events are informative.}
{An adaptive algorithm is not blind: a collision tells it something about where the colliding entry leads, and hence about the mask, and at a near-linear budget, there are many collisions. So we cannot hope that the masks stay uniform given the history; instead, we bound how much the history reveals about them. The decoy limits this: an answer that consists of fresh IDs looks the same whatever the masks are, and so do the later answers at these IDs, since vertices of $U_j$ and of $X_{j+1}$ answer alike. Our key lemma at a single level makes this precise by compression: the local history of a search can be replayed from randomness that is independent of the level-$j$ masks, together with a short \emph{exceptional record} at each rare event where the masks become visible. The main such events are collisions, since whether an entry hits a seen vertex depends on where it really leads. In addition, at a \emph{cutoff}, after a walk has made $T$ queries in $X_{j+1}$, a record reveals its position, before the walk can approach the end of the region, where its answers change, as in Barrier~1. A record names only $O(1)$ hidden identities, so the history reveals at most $O(\log N)$ bits about the masks per record (see Figure~\ref{fig:info}(b)).}

\begin{figure}[!t]
\centering
\begin{tikzpicture}[x=1cm,y=1cm,line cap=round,line join=round,
  pt/.style={acy/title,font=\small\bfseries},
  m/.style={acy/note,font=\small},
  ub/.style={acy/bluedot,minimum size=2.8mm},
  xr/.style={rectangle,draw=acyRed,fill=acyRed!12,line width=.6pt,minimum size=2.3mm,inner sep=0pt},
  keep/.style={acy/edge,draw=acyBlue,line width=.9pt},
  leave/.style={acy/edge,draw=acyRed!70,line width=.6pt},
  contact/.style={draw=acyRed,line width=.8pt,dashed},
  rec/.style={text=acyRed,font=\large,inner sep=0pt},
  ans/.style={rectangle,rounded corners=1.5pt,draw=acyInk!55,fill=acyInk!8,line width=.6pt,minimum width=5mm,minimum height=3.6mm,inner sep=0pt},
  guide/.style={draw=acyInk!30,line width=.45pt,densely dotted},
  piece/.style={draw=acyBlue!60,dashed,rounded corners=6pt,line width=.6pt,fill=acyBlue!3},
  qd/.style={circle,fill=acyInk!55,inner sep=0pt,minimum size=1.3mm}]
\node[pt] at (2.1,4.35) {(a) Records pay for long paths};
 \draw[piece] (.6,2.03) rectangle (3.4,3.88);
 \draw[piece] (.2,.08) rectangle (1.45,1.92);
 \draw[piece] (2.3,.08) rectangle (3.6,1.92);
 \foreach \n/\xx/\yy in {R/2.2/3.6,A1/1.6/2.95,S1/2.9/2.95,A2/1.05/2.3,B1/2.0/2.3,S2/2.9/2.3,
      A3/.75/1.65,B2/2.75/1.65,A4/.5/1.0,A4b/1.15/1.0,B3/2.6/1.0,B3b/3.3/1.0,A5/.5/.35,B4/2.6/.35}
   \node[ub,minimum size=2.5mm] (\n) at (\xx,\yy) {};
 \foreach \u/\v in {R/A1,R/S1,A1/A2,A1/B1,S1/S2,A2/A3,B1/B2,A3/A4,A3/A4b,B2/B3,B2/B3b,A4/A5,B3/B4}
   \draw[keep,line width=.8pt] (\u)--(\v);
 \draw[draw=acyRed,line width=1.8pt] (.746,2.046)--(1.054,1.904);
 \draw[draw=acyRed,line width=1.8pt] (2.264,1.847)--(2.486,2.103);
 \node[rec] at (.82,.32) {$\star$};
 \node[rec] at (2.92,.32) {$\star$};
 \node[m,font=\footnotesize,text=acyRed,anchor=north] at (.82,.02) {record};
 \node[m,font=\footnotesize,text=acyRed,anchor=east] at (.15,1.98) {cut};
 \node[m,font=\footnotesize,text=acyBlue,anchor=north west] at (.68,3.84) {piece};
 \draw[decorate,decoration={brace,amplitude=4pt},draw=acyInk,line width=.6pt] (3.85,3.6)--(3.85,1.65) node[midway,right=4pt,m] {$h$};
 \draw[draw=acyInk!20,line width=.5pt] (5.2,-.15)--(5.2,4.55);
\begin{scope}[shift={(5.85,0)},
   work/.style={draw=acyInk!55,fill=acyInk!14,line width=.6pt},
   own/.style={draw=acyBlue!60,dashed,fill=acyBlue!10,line width=.6pt},
   shrink/.style={->,>=Stealth,draw=acyInk!60,line width=.6pt}]
  \node[pt] at (2.75,4.35) {(b) Discount across levels};
  \foreach \yy/\lab in {3.5/{U_{k+1}},2.6/{U_k},1.7/{U_{k-1}},.5/{U_1}}
    \node[m,anchor=east] at (.62,\yy) {$\lab$};
  \node[m] at (.38,1.18) {$\vdots$};
  \draw[work] (.75,3.33) rectangle (5.0,3.67);
  \foreach \yy/\xx in {2.6/2.97,1.7/2.07,.5/1.29} {
    \draw[work] (1.17,\yy-.17) rectangle (\xx,\yy+.17);
    \draw[own] (.75,\yy-.17) rectangle (1.17,\yy+.17);
  }
  \node[m,anchor=west] at (5.05,3.5) {$q$};
  \draw[shrink] (4.9,3.28)--(3.07,2.8) node[pos=.45,anchor=north west,inner sep=1.5pt,m,font=\footnotesize] {$\times N^{-\Omega(\eta)}$};
  \draw[shrink] (2.9,2.38)--(2.15,1.9) node[pos=.45,anchor=north west,inner sep=1.5pt,m,font=\footnotesize] {$\times N^{-\Omega(\eta)}$};
  \draw[shrink,densely dotted] (2.0,1.48)--(1.37,.7);
  \node[m,font=\footnotesize,text=acyBlue,anchor=north] at (.96,.27) {pieces};
 \end{scope}
\end{tikzpicture}
\caption{From information to work. (a) Records pay for long paths: by a union bound, each record (star) buys $O(1)$ long paths of the real $U_j$ discovery tree, so cutting $O(h)$ edges per record (red) leaves pieces (dashed) of depth less than $h$, each with at most $N^{\zeta}$ vertices; here $h=3$. (b) Discount across levels: each level passes on only an $N^{-\Omega(\eta)}$ fraction of the work above it (gray), since records at level $j$ arise only about once in $N^{\Omega(\eta)}$ queries to $U_{j+1}$. Besides, every level has the small pieces around the roots of the searches (dashed).}
\label{fig:info-paths}
\end{figure}
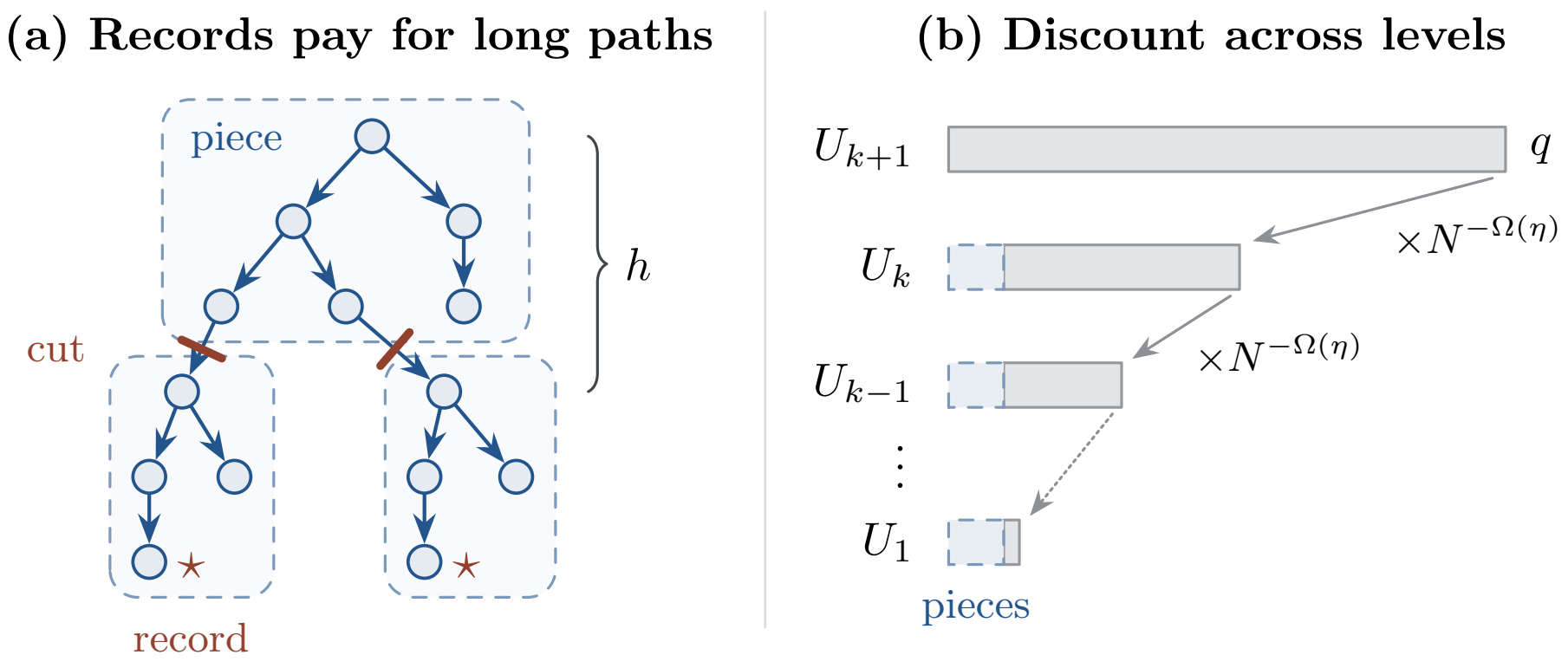

\paragraph{From bits to short paths.}
{It remains to bound the exploration that these bits can pay for. By the warm-up, pieces of the discovery tree inside $U_j$ of depth less than $h$ need no information about the masks, and each has at most $N^{\zeta}$ vertices. A path of length $h$, by contrast, stays in $U_j$ only with probability $N^{-80}$ if the masks are independent of the history, so following it requires $\Omega(\log N)$ bits of information about the masks, and edge-disjoint paths require this much each. A record supplies only $O(\log N)$ bits, so it pays for $O(1)$ such paths, and cutting $O(h)$ edges per record breaks every long path; a union bound makes this precise. Hence the \emph{work} in $U_j$, that is, the number of queried vertices of $U_j$, is at most $N^{\zeta}$ times the number of searches and records (see Figure~\ref{fig:info-paths}(a)).}

\paragraph{Discounting across levels.}
Since the replay cannot tell $U_j$ from its decoy $X_{j+1}$, the records at level~$j$ arise from collisions and cutoffs anywhere in $U_{j+1}$, and these events are rare: as both $q/W$ and $1/T$ are $N^{-\Omega(\eta)}$, they occur only about once in $N^{\Omega(\eta)}$ queries to $U_{j+1}$. Each record buys only about $N^{\zeta}$ queries in $U_j$, so each level passes on only an $N^{-\Omega(\eta)}$ fraction of the work above it, apart from the small pieces around the roots of the searches. The total work is at most $q<N$, so nothing is passed on after $k=O(1/\eta)$ levels, and $U_1$ is explored only in small pieces (see Figure~\ref{fig:info-paths}(b)). In contrast with the epoch analysis of~\cite{CRSS22}, where every collision ends an epoch of the whole exploration, here the information from a collision is charged only to the local search that it affects.

\paragraph{From one search to the whole exploration.}
An adaptive algorithm may revisit IDs and query arbitrary IDs, so its exploration is not a single rooted search. We therefore split the exploration into local searches, called \emph{fragments}, and apply the argument above to each of them. The main difficulty is that one fragment may continue from vertices that another fragment has discovered, so informed searches could form long chains; we show that such chains are short and that the resulting pieces of the core are small, shallow, and acyclic (Section~\ref{os:section}). This yields the structural bound: every piece of information about the masks is charged to a record, and records are rare.

\paragraph{Organization.}
Section~\ref{sec:preliminaries} fixes the notation and states the facts about random high-girth cores that we need. In Section~\ref{sec:distributions}, we fix all parameters and define the two input distributions. Section~\ref{rt:section} proves Theorem~\ref{thm:main}, assuming the structural bound and the forest comparison; these two theorems are proved in Sections~\ref{os:section} and~\ref{rf:section}, respectively. The appendices contain the proofs about random cores (Appendix~\ref{app:core-estimates}), the height estimate for fragments (Appendix~\ref{app:blue-erasure}), the details of the local reconstruction (Appendix~\ref{app:local-reconstruction}), the modification that bounds the indegrees (Appendix~\ref{app:indegree}), the lower bound for the BFS approach (Appendix~\ref{app:bfs}), and our comments on a recent preprint (Appendix~\ref{sec:Yuichi}).

\section{Preliminaries}
\label{sec:preliminaries}

\subsection{Model and notation}
\label{subsec:model}

We consider simple directed graphs $G=([n],E)$ with maximum outdegree at most $d$, using the unidirectional bounded-degree incidence-lists model of Section~\ref{sec:introduction} with the whole-vertex queries of Section~\ref{sec:intro-result}. In Theorem~\ref{thm:main}, the maximum indegree is also at most $d$, while queries still return only outneighbors.
We write $\fas(G)$ for the size of a \emph{minimum feedback arc set}, that is, the minimum number of edges whose deletion makes $G$ acyclic, and we say, as in Section~\ref{sec:introduction}, that $G$ is \emph{$\eps$-far from acyclic} if $\fas(G)\ge\eps dn$. Two-sided and one-sided $\eps$-testers are as defined in Section~\ref{sec:introduction}; their queries may be fully adaptive.

\paragraph{Conventions.}
A vertex's public identifier is its \emph{ID}. Our auxiliary random constructions assign IDs by a uniformly random bijection, independently of all other randomness, and this bijection is hidden from the algorithm. The reductions in Section~\ref{rt:transfer} use explicitly encoded gadget and copy IDs, and fixed IDs for padding vertices. An ID is \emph{seen} once it has been queried or returned; its first occurrence is \emph{fresh}, and later occurrences are \emph{old}. A query is a \emph{first query} if its ID has not been queried before, and a \emph{repeated query} otherwise. Our constructions are multigraphs whose edges, called \emph{edge occurrences}, are identified by their source and their position in the adjacency list of the source (for permutation cores, by their source and color). \emph{Weak} components, paths, cycles, forests, distance, diameter, and girth refer to the underlying undirected multigraph, which forgets directions and retains every occurrence; thus a loop is a weak cycle of length one, and two occurrences on the same unordered pair form a weak cycle of length two, regardless of their directions. For the auxiliary construction, asymptotic notation refers to $N\to\infty$, where $N=\Theta_\eta(n)$ is the number of core vertices, and hides constants depending only on~$\eta$. In the final reduction to a prescribed degree bound~$d$, size thresholds and intermediate comparison constants may also depend on~$d$, which is fixed as $n\to\infty$. Throughout, ``with high probability'' means with probability $1-o(1)$. A subscript, as in $O_{d,\eps}$ or $O_{a,r}$, names the parameters on which the hidden constant may depend; the subscript $\eta$, as in $O_\eta$ or $\Theta_\eta$, only stresses the default dependence. We write $C$ for an absolute constant and $C_\eta$ for a constant that depends only on $\eta$; both may change from line to line. Capital Latin letters with an index usually denote sets, graphs, maps, or random variables, such as the regions $X_i$, the bands $E_i$, and the sets $U_j$; indexed numbers, such as the category counts $d_j$ and the path lengths $h_j$, are denoted by lowercase or Greek letters. Logarithms are natural, and $(x)_s=x(x-1)\cdots(x-s+1)$.

\paragraph{Distributions.}
All distributions in this paper are on finite sets. We write $\Dist(X)$ for the distribution of $X$ and $\Unif(S)$ for the uniform distribution on $S$, and we condition only on events of positive probability. The total variation distance is $\TV(\mu,\nu)=\frac12\sum_x|\mu(x)-\nu(x)|=1-\sum_x\min\{\mu(x),\nu(x)\}$; it bounds the difference between the rejection probabilities of every (possibly randomized) decision rule.

\paragraph{Lower bound strategy.}
Fix an algorithm. Its \emph{transcript} records all queries, all answers, the stopping time, and the output. If a YES distribution $\DY$, supported on acyclic graphs, and a NO distribution $\DN$, which is $\eps$-far from acyclic with probability $1-o(1)$, induce transcript distributions at total variation distance $o(1)$, then the algorithm is not a two-sided $\eps$-tester. Indeed, since the output is part of the transcript, its rejection probabilities under $\DY$ and $\DN$ differ by $o(1)$, whereas a two-sided $\eps$-tester rejects inputs of $\DY$ with probability at most $\frac13$ and inputs of $\DN$ with probability at least $\frac23(1-o(1))$. By averaging over the random seed, it suffices to prove such a bound uniformly over deterministic algorithms. Until Section~\ref{rt:transfer}, we work with an auxiliary oracle that returns ordered adjacency lists, possibly with repeated targets; such answers reveal at least as much as sets of outneighbors.

\subsection{Random high-girth permutation cores}
\label{rc:section}

Both hard distributions use a core whose edges come from $32$ permutations. Let $N$ be even and let $\Pi=(\pi_1,\ldots,\pi_{32})$ be independent uniform permutations of $[N]$. The \emph{occurrence graph} $\Gamma(\Pi)$ has an occurrence $v\xrightarrow{a}\pi_a(v)$ for every $v\in[N]$ and every color $a\in[32]$, so its underlying multigraph is $64$-regular. For an integer $\gamma=\gamma(N)\ge2$ with
\begin{equation}
\label{rc:girth-scale}
 64^{\gamma}\log N=o(N),
\end{equation}
let $\mathcal G_\gamma$ be the event that $\Gamma(\Pi)$ has weak girth greater than $\gamma$. On $\mathcal G_\gamma$, the graph $\Gamma(\Pi)$ is simple, and every vertex has exactly $32$ inneighbors and $32$ outneighbors. A set of colored matches $\pi_a(v)=w$ is \emph{compatible} if some tuple in $\mathcal G_\gamma$ extends it; given such a set and $\mathcal G_\gamma$, the tuple $\Pi$ is uniform on the extensions in $\mathcal G_\gamma$. We need three facts about these cores: conditioning on $\mathcal G_\gamma$ costs only a factor $e^{-o(N)}$ in probability, unexposed endpoints remain diffuse under adaptive exposure, and the conditioned core is far from acyclic. We prove them in Appendix~\ref{app:core-estimates} by switching arguments and a cut estimate.

\begin{proposition}[The cost of girth conditioning]
\label{rc:existence}
Under \eqref{rc:girth-scale}, $\Pr[\mathcal G_\gamma]\ge\exp(-C\,64^{\gamma}\log N)=e^{-o(N)}$.
\end{proposition}

\begin{lemma}[An adaptive atom bound]
\label{rc:atom}
Fix a compatible set of exposed matches and an unexposed source--color pair $(v,a)$, and let $m_a$ be the number of exposed matches of color $a$. Then for every $w\in[N]$,
\[
 \Pr[\pi_a(v)=w\mid\text{exposed matches},\mathcal G_\gamma]
 \le\frac{1}{N-m_a-C\,64^{\gamma}},
\]
provided the denominator is positive. In particular, if the matches are exposed sequentially by a rule that chooses each next pair from the exposed matches and independent auxiliary randomness, and at most $q=o(N)$ matches per color have been exposed, then for large $N$ the next target lies in any specified set of $s$ vertices with conditional probability at most $2s/N$.
\end{lemma}

\begin{proposition}[Farness of the conditioned core]
\label{rc:farness}
Under \eqref{rc:girth-scale}, with probability $1-e^{-\Omega(N)}$ conditional on $\mathcal G_\gamma$, every set of occurrences whose deletion destroys all directed cycles of $\Gamma(\Pi)$ has at least $4N$ elements; in particular, $\fas(\Gamma(\Pi))\ge4N$.
\end{proposition}

\section{The hard input distributions}
\label{sec:distributions}

In this section, we formally define the two input distributions that were described informally in Section~\ref{sec:construction-overview}. They share the masking regions and a single rule that generates all adjacency lists, and they differ only in the core: the NO core is random, whereas the YES core is a fixed core unrolled into layers. We first define both distributions (Section~\ref{os:sampler}), then we fix the parameters (Section~\ref{os:params}), and finally we collect the properties of the NO distribution that the exploration analysis uses (Section~\ref{os:props-sec}). Both distributions are sampled before the algorithm runs. Until Section~\ref{rt:transfer}, a query returns an ordered adjacency list, which may contain repeated targets.

\subsection{The construction}
\label{os:sampler}
\label{rt:yes-sampler}

The construction depends on a few integer parameters, whose values are fixed in Section~\ref{os:params}: the number $k+1$ of masking regions, the band depth $T$ and the layer width $W$, where $N=TW$, the category counts $32=d_0<d_1<\cdots<d_{k+1}=D$, and the girth threshold $\gamma$.

\paragraph{Objects.}
We call a vertex, described by its hidden coordinates, an \emph{object}. The \emph{blue} objects form a set $B$ of size $N$. For $1\le i\le k+1$, the \emph{masking region} $X_i=\{i\}\times[3T]\times[W]$ consists of the $3T$ \emph{layers} $X_{i,t}=\{i\}\times\{t\}\times[W]$, and its first $T$ layers form the \emph{entry band} $E_i$; thus $|E_i|=|B|=N$. The \emph{level} of an object is $0$ on $B$ and $i$ on $X_i$, and $U_j=B\cup X_1\cup\cdots\cup X_j$ is the set of objects of level at most $j$. There are $n_0=(3k+4)N$ objects, and IDs are assigned by a uniformly random bijection from the objects to $[n_0]$, independently of all other randomness.

\paragraph{The two cores.}
The \emph{NO core} $\Pi=(\pi_1,\ldots,\pi_{32})$ consists of independent uniform permutations of $B$ conditioned on $\mathcal G_\gamma$, which is well defined by Proposition~\ref{rc:existence}. The YES core is fixed. Since $64^\gamma\log T=o(T)$ (see Section~\ref{os:params}), Proposition~\ref{rc:existence} at size $T$ shows that there are permutations $\rho_1,\ldots,\rho_{32}$ of $[T]$ whose occurrence graph $G_{\rm base}$, the \emph{base core}, has weak girth greater than $\gamma$. We fix one such tuple, say the first one in a fixed ordering, and we identify $B$ with $[W]\times[T]$ by a fixed bijection; the first coordinate is the \emph{blue layer}. (Since blue coordinates are hidden, this identification changes no public information.)

\paragraph{Adjacency lists.}
Every adjacency list has $D$ ordered \emph{entries}, whose positions are called \emph{list indices}, except that the lists of the last layer of $X_{k+1}$ are empty. Each entry is an edge occurrence, even if two entries have the same target. Every nonempty list carries a \emph{category word}, which is a uniformly random word in $\{0,\ldots,k+1\}^D$ in which exactly $d_j$ letters are at most $j$, for every $j$; thus exactly $d_i-d_{i-1}$ letters equal $i$, where $d_{-1}=0$. The category words are independent of each other and of all other randomness. At a blue object, the $a$-th entry of category $0$ in list order has \emph{color} $a$.

\begin{definition}[The NO and YES distributions]
\label{def:no}\label{def:yes}
Consider an entry of category $c$ in the list of an object $v$ of level $i$.
\begin{enumerate}
\item If $c>i$, the entry \emph{jumps}: it targets a uniform object of $E_c$.
\item If $c\le i$ and $v\in X_{i,t}$, the entry \emph{steps}: it targets a uniform object of the next layer $X_{i,t+1}$, where, for $i\le k$, the next layer after $X_{i,3T}$ is $E_{i+1}$.
\item If $c=0$ and $v\in B$, the entry steps along the core according to its color $a$. In $\DN$, it targets $\pi_a(v)$. In $\DY$, where $v=(s,x)$, it targets $(s+1,\rho_a(x))$ if $s<W$, and a uniform object of $E_1$ if $s=W$.
\end{enumerate}
Given the core and the category words, all targets are independent.
\end{definition}

Thus the two distributions differ only in the core steps. In $\DY$, the blue core is one more layered region, whose last layer steps into $E_1$, just as the last layer of $X_i$ steps into $E_{i+1}$. Hence every input of $\DY$ is acyclic: if we order the blue layers $1,\ldots,W$, followed by the layers of $X_1$, then those of $X_2$, and so on, then every occurrence points forward in this order. Observe that the NO distribution is not conditioned on farness or on the behavior of the exploration. Since the targets in (1) and (2) are independent, the maximum indegree is not bounded by a constant. Appendix~\ref{app:indegree} replaces these targets by uniform injections into a bounded number of incoming slots per object, and shows that the whole analysis carries over.

\subsection{Parameters}
\label{os:params}
\label{rt:parameters}

Fix $\eta\in(0,1)$, and let
\begin{equation}
\label{os:parameters}
 \alpha=1-\frac\eta2,\qquad
 \beta=\frac\eta8,\qquad
 k=\Bigl\lceil\frac4\beta\Bigr\rceil.
\end{equation}
Let $A$ be a sufficiently large integer that depends only on $\eta$; the two lower bounds on $A$ that we need are stated in \eqref{rt:A-choice}, where they are used. The category counts and the degree are
\begin{equation}
\label{os:probabilities}
 d_j=2^{5A^j}\quad(0\le j\le k+1),\qquad
 D=d_{k+1}.
\end{equation}
Finally, let $T$ be even and $N=TW$, where
\begin{equation}
\label{os:scales}
 T=N^{\beta+o(1)},\qquad W=N^{1-\beta+o(1)},\qquad
 \gamma=\Bigl\lfloor\frac14\log_{64}T\Bigr\rfloor,\qquad
 Q=\lfloor N^\alpha\rfloor.
\end{equation}
Here $\gamma=\Theta(\log N)$ is the girth threshold of both cores; since $64^\gamma\le T^{1/4}$, condition \eqref{rc:girth-scale} holds both at size $N$ and at size $T$. The number $Q$ is the query budget against which we prove indistinguishability; since $\alpha>1-\eta$, it exceeds the bound $n^{1-\eta}$ of Theorem~\ref{rt:bounded-degree} for large $n$ (see Section~\ref{rt:transfer}).

These choices balance three effects. Each target hits a specified object with probability at most $1/W$, so an algorithm with $Q$ queries causes collisions only at rate $Q/W=N^{-3\beta+o(1)}$; the band depth $T$, on the other hand, is the length of the cutoff, which is charged at rate $1/T=N^{-\beta+o(1)}$. Moreover, the category counts are widely separated, so at each level a prescribed set of $s$ entries is selected with probability at most $(d_j/d_{j+1})^s$, which makes long selected paths rare. Finally, the core degree stays $d_0=32$, which alone determines the base $64$ in the girth scale, while the degree $D$ is a large constant depending only on $\eta$.

\subsection{Properties used in the exploration}
\label{os:props-sec}

We note that no input of $\DN$ has a self-loop, and that replacing each list by its set of targets yields a simple directed graph of outdegree at most $D$ that contains the blue core unchanged (see Section~\ref{rt:transfer}). By Proposition~\ref{rc:farness}, with probability $1-o(1)$, every feedback arc set contains at least $4N$ occurrences. The following lemma collects the properties of $\DN$ that the exploration analysis uses.

\begin{lemma}[Properties used in exploration]
\label{os:properties}
Every input of $\DN$ has the following properties.
\begin{enumerate}[label=(\roman*)]
\item \emph{Monotonicity.} The level never decreases along an occurrence, and the layer strictly increases along an occurrence inside some $X_i$. In particular, every directed cycle lies in $B$, and the target in $X_{j+1}$ of an occurrence with source in $U_j$ lies in $E_{j+1}$.
\item \emph{Upper degree.} An object of $U_j$ has at most $d_j$ outgoing occurrences with targets in $U_j$, namely among its entries of category at most $j$.
\item \emph{Diffuse endpoints.} Fix an exposure state in which at most $Q$ core matches of each color are exposed, and an unexposed entry. Its conditional probability of targeting any specified object is at most $1/W$ (and at most $2/N$ for a core entry), for all large $N$.
\end{enumerate}
\end{lemma}

\begin{proof}
Properties (i) and (ii) follow directly from Definition~\ref{def:no}: a step stays in the region of its source or enters $E_{i+1}$ from the last layer of $X_i$, and a jump from level $i$ enters $E_c$ with $c>i$. For (iii), core entries are covered by Lemma~\ref{rc:atom}, since $1/(N-Q-C64^\gamma)\le2/N\le1/W$, and nonblue targets are uniform over a layer or a band.
\end{proof}

\section{Proof of the main theorem}
\label{rt:section}

In this section, we prove Theorem~\ref{thm:main}, assuming two technical theorems: the structural bound for exploration under the NO distribution, and the comparison of forest embeddings in high-girth cores. This section formalizes the reduction outlined in Section~\ref{sec:ov-reduction}. We first state the two theorems, and then we derive a lower bound at a larger constant degree (Theorem~\ref{rt:bounded-degree}) from them by means of an exact formula for transcript probabilities; the boundary argument of Section~\ref{sec:ov-reduction} appears as Lemma~\ref{rf:boundary}. Finally, we reduce the degree to~$2$ and then use a complete blow-up to prove Theorem~\ref{thm:main} for every fixed $d\ge2$ with a proximity parameter independent of~$d$. The two theorems themselves are proved in Sections~\ref{os:section} and~\ref{rf:section}, respectively.

\subsection{The two main estimates}
\label{rt:key}

\paragraph{The structural bound.}
Our first estimate concerns the part of the core that an algorithm observes. The \emph{observed blue graph} of an execution under $\DN$ consists of all seen blue vertices (including queried isolated vertices and unqueried targets) and of all exposed occurrences whose two endpoints are blue. Recall that a weak forest has no loops, no opposite occurrences, and no repeated occurrences on one pair.

\begin{theorem}[Structural bound]
\label{os:structural}\label{os:forest}
Under $\DN$, uniformly over all deterministic adaptive algorithms that make at most $Q$ queries, even with access to the disclosures of Section~\ref{os:fragments}, the following holds with probability $1-o(1)$: the observed blue graph is a weak forest whose components have at most $S=N^z$ vertices and weak diameter at most $\Lambda$. Here $z$ and $\Lambda$ are defined in Section~\ref{os:analysis-params}, and they satisfy $z<\eta/8$ and $2K\Lambda+1<\gamma$, where $K=\lceil8/\eta\rceil$.
\end{theorem}

\paragraph{The forest comparison.}
Our second estimate compares the numbers of embeddings of such a forest into two different cores. A \emph{permutation core} on an $N$-set $V$ is given by bijections $P_1,\ldots,P_{32}:V\to V$, and its color-$a$ occurrences are $x\to P_a(x)$. We say that a directed edge-colored forest is \emph{proper} if no vertex has two outgoing edges of one color or two incoming edges of one color. Let $\Emb_G(F)$ be the number of injective maps $V(F)\to V$ that preserve the direction and the color of every edge. (Observe that the maps need not be induced: further edges of $G$ impose no constraint.)

\begin{theorem}[Forest comparison]\label{rf:comparison}
Let $K,\Lambda\ge1$ be integers, and let $G,G'$ be permutation cores on $N$ vertices with weak girth greater than $2K\Lambda+1$. Let $F$ be a proper colored forest with $c$ components, each with at most $S$ vertices and weak diameter at most $\Lambda$. If $\kappa=cS^2/N\le1/(16e^2)$, then
\begin{equation}\label{rf:log-comparison}
 \bigl|\log\Emb_G(F)-\log\Emb_{G'}(F)\bigr|
 \le 8e\,c(2e\kappa)^K,
\end{equation}
and $\Emb_G(F)=\Emb_{G'}(F)$ if $c\le K$.
\end{theorem}

\begin{figure}[!htbp]
\centering
\begin{tikzpicture}[x=1cm,y=1cm,line cap=round,line join=round]
 \foreach \tag/\yy in {p/2.2,a/-.7} {
   \foreach \col/\xx/\lab in {1/1.7/1,2/4.0/2,3/7.1/{W-1},4/9.4/W} {
     \draw[acy/bluearea,rounded corners=2pt] (\xx-.57,\yy-.72) rectangle (\xx+.57,\yy+.72);
     \node[acy/note,text=acyBlue] at (\xx,\yy-.99) {\(\lab\)};
     \foreach \lane/\dy in {u/.4,m/0,l/-.4}
       \node[acy/bluedot] (\tag\col\lane) at (\xx,\yy+\dy) {};
   }
   \foreach \left/\right in {1/2,3/4}
     \foreach \lane in {u,m,l}
       \draw[acy/coreedge] (\tag\left\lane)--(\tag\right\lane);
   \foreach \lane in {u,m,l}
     \draw[acy/coreedge,densely dotted] (\tag2\lane)--(\tag3\lane);
   \node[acy/note,text=acyBlue] at (5.55,\yy+.55) {\(\cdots\)};
 }
 \node[acy/title,anchor=east,text width=1.9cm,text=acyInk] at (.65,2.2) {Periodic\\cover};
 \node[acy/title,anchor=east,text width=1.9cm,text=acyInk] at (.65,-.7) {YES\\instance};
 \draw[acy/coreedge] (p4u) .. controls (9.4,4.1) and (1.7,4.1) .. (p1u);
 \node[acy/note,fill=white,inner sep=3pt] at (5.55,3.57) {Wrap around};
 \node[acy/box,draw=acyRed,fill=acyRed!6,text width=1.85cm,minimum height=1.4cm]
   (mask) at (12.05,-.7) {Band\\$E_1$};
 \draw[acy/rededge] (a4m)--(mask.west);
 \node[acy/note,text=acyRed] at (10.33,-.42) {step};
 \node[acy/note,text=acyBlue] at (5.55,-2.55) {Same transitions away from the last layer};
\end{tikzpicture}
\caption{The periodic cover $G_\circ$ and an acyclic YES instance. Both have the same transitions between consecutive blue layers, each a copy of the base core on $T$ vertices. At layer $W$ the periodic cover wraps around, whereas a YES instance steps into the band $E_1$. Embeddings avoiding layer $W$ are valid in both graphs; Lemma~\ref{rf:boundary} bounds the fraction lost.}
\label{fig:cover}
\end{figure}
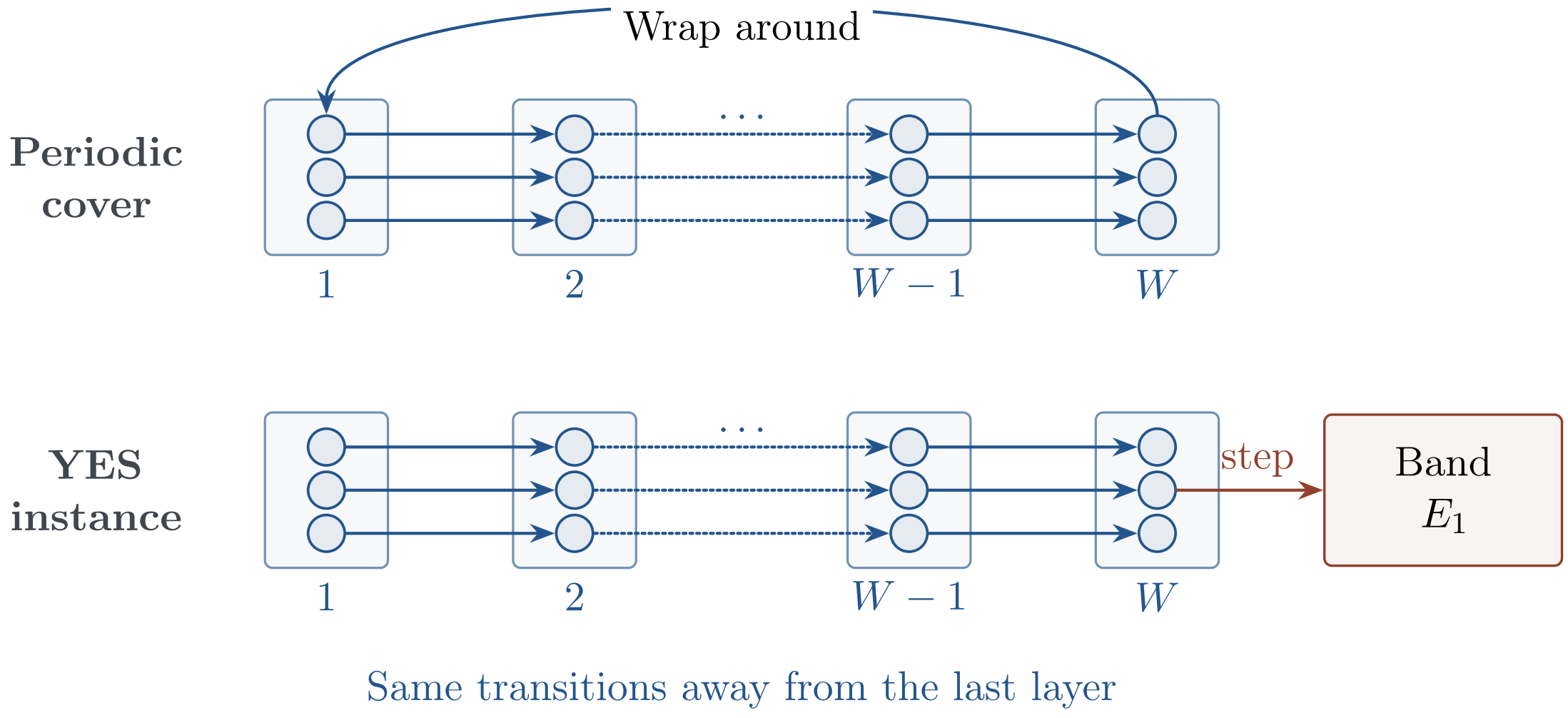

\paragraph{The periodic cover.}
The YES core is not a permutation core, since its last blue layer steps out of $B$. For counting, we close it up: let $G_\circ$ be the permutation core on $B$ with transitions
\begin{equation}\label{rt:periodic-core}
 P_a(s,x)=(s+1,\rho_a(x)),
 \qquad a\in[32],
\end{equation}
where blue layers are read modulo $W$, so that layer $W$ is residue $0$ and its successor is layer $1$. Thus every YES instance agrees with $G_\circ$, except that the wrap-around edges from layer $W$ step into $E_1$ (see Figure~\ref{fig:cover}). By the following lemma, $G_\circ$ has weak girth greater than $\gamma$.

\begin{lemma}[Periodic cover]\label{rf:periodic}
Let $G_{\rm base}$ be a permutation core on $T$ vertices with maps $\rho_1,\ldots,\rho_{32}$ and weak girth greater than $\gamma$, and let $W\ge1$. The core $G_\circ$ on $\mathbb Z/W\mathbb Z\times V(G_{\rm base})$ with $P_a(i,x)=(i+1,\rho_a(x))$ is a permutation core on $TW$ vertices with weak girth greater than $\gamma$, and the layer translation $(i,x)\mapsto(i+1,x)$ is a color-preserving automorphism.
\end{lemma}
\begin{proof}
Each $P_a$ is a bijection, with inverse $(i,x)\mapsto(i-1,\rho_a^{-1}(x))$. The projection to $G_{\rm base}$ maps the incoming and the outgoing occurrence of each color at a vertex bijectively to those at its image, and so it is a covering map. A weak cycle in $G_\circ$ projects to a nonbacktracking closed walk of the same length; indeed, an immediate reversal in the projection would lift, by the uniqueness of lifts, to a reversal in $G_\circ$. Since a nonbacktracking closed walk of length $\ell$ contains a weak cycle of length at most $\ell$, the weak girth of $G_\circ$ is at least that of $G_{\rm base}$. Finally, the translation commutes with every $P_a$.
\end{proof}

We will apply Theorem~\ref{rf:comparison} to the NO core and to $G_\circ$. Only the embeddings into $G_\circ$ that avoid the last blue layer are also valid in every YES instance, and translation symmetry shows that they are almost all of them.

\begin{lemma}[Boundary loss after injectivity]\label{rf:boundary}
Let $F$ be a colored forest on $v$ vertices, and let $\Emb_{G_\circ,\mathrm{avoid}}(F)$ count the injective embeddings of $F$ into $G_\circ$ whose images avoid layer $W$ (residue $0$). Then
\begin{equation}\label{rf:boundary-factor}
 \Emb_{G_\circ,\mathrm{avoid}}(F)\ge(1-v/W)\Emb_{G_\circ}(F).
\end{equation}
\end{lemma}
\begin{proof}
We may assume that $\Emb_{G_\circ}(F)>0$, and we take a uniformly random injective embedding. Composing with the layer translation is a bijection of the set of injective embeddings; hence the layer of the image of each fixed vertex of $F$ is uniform on the $W$ layers, even though global injectivity is imposed. A union bound over the $v$ vertices completes the proof.
\end{proof}

\subsection[Exact probabilities of marked transcripts]{Exact probabilities of marked transcripts}
\label{rt:likelihood}

We now relate transcript probabilities to embedding counts. Fix a deterministic algorithm $\mathcal A$ that makes at most $Q$ queries. The algorithm may query arbitrary IDs, repeat queries, and stop early. A transcript $\xi$ records every query, its ordered answer (also for repeated queries), the stopping time, and the output. Let $V_\xi$ be the set of IDs that occur in $\xi$ as queried sources or as targets; then $|V_\xi|\le(D+1)Q$. After the execution, we append to the transcript the following \emph{marks}, which the algorithm does not see:
\begin{enumerate}[label=(\roman*)]
\item for every ID in $V_\xi$, whether it is blue;
\item for every nonblue ID in $V_\xi$, its object (region, layer, coordinate);
\item for every queried nonempty adjacency list, its category word.
\end{enumerate}
Observe that no blue object and no blue coordinate is marked. Both distributions induce laws on the same finite space of marked transcripts, and forgetting the marks recovers the ordinary transcript.

\paragraph{The blue graph of a transcript.}
Let $\xi$ have positive NO probability, let $B_\xi\subseteq V_\xi$ be the set of its blue IDs, and let $v=|B_\xi|$. The colored directed graph $F_\xi$ on $B_\xi$ has, for every queried blue source $u$ and every color $a$, an occurrence of color $a$ from $u$ to the ID returned by the $a$-th entry of category~$0$ in the list of $u$. In $\DN$, only these entries of blue sources target $B$, so $F_\xi$ is exactly the observed blue graph, including its isolated vertices. The graph $F_\xi$ is proper, since each color is injective on sources and on targets, but it need not be a forest. For a permutation core $G$ on $B$, let $\Emb_G(F_\xi)$ be the number of injective maps $B_\xi\to B$ that preserve every colored occurrence.

\paragraph{The common factor.}
Let $\ell(\xi)$ be the number of queried nonempty lists. By Definition~\ref{def:no}, every nonblue target is uniform in a layer of some $X_i$ or in an entry band, which we call its \emph{pool}. Among the exposed occurrences with nonblue targets, let $\ell_W(\xi)$ be the number of those whose pool is a layer, and let $\ell_N(\xi)$ be the number of those whose pool is an entry band. Observe that these counts, as well as all pool memberships, are determined by the marks. For $\xi$ of positive NO probability, let
\begin{equation}\label{rt:common-factor}
 c_{\mathcal A}(\xi)=
       \binom{D}{d_0,\,d_1-d_0,\,\ldots,\,d_{k+1}-d_k}^{-\ell(\xi)}\frac{W^{-\ell_W(\xi)}N^{-\ell_N(\xi)}}{(n_0)_{|V_\xi|}}.
\end{equation}

\begin{lemma}[Exact likelihood and a YES subcount]
\label{rt:exact-likelihood}
Let $\mu_{\rm core}$ be the law of the NO core. For every marked transcript $\xi$ with positive NO probability and every core $G$ in the support of $\mu_{\rm core}$,
\begin{equation}\label{rt:no-fixed-core}
 \Pr_{\DN}[\xi\mid G]=c_{\mathcal A}(\xi)\Emb_G(F_\xi),
 \qquad\text{hence}\qquad
 \Pr_{\DN}[\xi]=c_{\mathcal A}(\xi)\,\E_{G\sim\mu_{\rm core}}\Emb_G(F_\xi).
\end{equation}
Moreover,
\begin{equation}\label{rt:yes-subcount}
 \Pr_{\DY}[\xi]\ge c_{\mathcal A}(\xi)\Emb_{G_\circ,\mathrm{avoid}}(F_\xi),
\end{equation}
where $\Emb_{G_\circ,\mathrm{avoid}}$ counts the embeddings into $G_\circ$ whose images avoid blue layer $W$.
\end{lemma}
\begin{proof}
Fix an injective map $f:B_\xi\to B$. Together with the marked nonblue objects, $f$ specifies the object of each of the $|V_\xi|$ IDs, and the uniform ID bijection agrees with this partial assignment with probability $1/(n_0)_{|V_\xi|}$. (This holds for arbitrary adaptive choices of IDs, since we evaluate a completed transcript.) Each of the $\ell(\xi)$ category words read for the first time contributes the inverse of the multinomial coefficient in \eqref{rt:common-factor}, and repeated queries contribute nothing. Given the words, every nonblue target is uniform in its pool and contributes $1/W$ or $1/N$, independently of $f$. Two occurrences with the same target contribute separately, and their equality is enforced by the partial bijection. The lists of the last layer of $X_{k+1}$ are deterministically empty. An entry of category~$0$ and color~$a$ at a blue source $u$ with returned ID $w$ contributes the indicator $\ind\{f(w)=P_a(f(u))\}$, and the product of these indicators is the event that $f$ embeds $F_\xi$. Finally, the queries, the stopping time, and the output are determined by the answers, so they contribute only a consistency indicator, which equals $1$; all unread variables integrate out. Summing over $f$ gives the first identity in \eqref{rt:no-fixed-core}, and averaging over $G$ gives the second.

For \eqref{rt:yes-subcount}, we restrict the sum to embeddings into $G_\circ$ that avoid layer $W$. Then every queried blue source lies in a layer $i<W$ and uses the same colored transition in $\DY$ as in $G_\circ$, while all other factors are unchanged. (We note that this bounds a part of the YES mass from below; it does not condition $\DY$ on any event.)
\end{proof}

\begin{lemma}[Pointwise domination on forest transcripts]
\label{rt:pointwise}
Suppose that the NO cores and $G_\circ$ have weak girth greater than $2K\Lambda+1$. Let $\xi$ have positive NO probability, and suppose that $F_\xi$ is a forest with $c$ components, each with at most $S$ vertices and weak diameter at most $\Lambda$. If $cS^2/N\le1/(16e^2)$ and $v<W$, then
\begin{equation}\label{rt:pointwise-bound}
 \Pr_{\DY}[\xi]
 \ge
 \left(1-\frac vW\right)
 \exp\left[-8e\,c\left(\frac{2e\,cS^2}{N}\right)^K\right]
 \Pr_{\DN}[\xi].
\end{equation}
\end{lemma}
\begin{proof}
For every $G$ in the support of $\mu_{\rm core}$, Lemma~\ref{rf:boundary} and Theorem~\ref{rf:comparison} (applied to $G_\circ$ and $G$) give $\Emb_{G_\circ,\mathrm{avoid}}(F_\xi)\ge(1-v/W)e^{-8ec(2ecS^2/N)^K}\Emb_G(F_\xi)$. Averaging over $G$ and applying Lemma~\ref{rt:exact-likelihood} completes the proof.
\end{proof}

\subsection[Indistinguishability of transcripts]{Indistinguishability of transcripts}
\label{rt:tv}

It remains to sum the pointwise bound over typical transcripts. Since $z<\eta/8$, we have $\alpha+2z-1<-\eta/4$, and so the choice $K=\lceil8/\eta\rceil$ gives
\begin{equation}\label{rt:order-choice}
 \alpha+K(\alpha+2z-1)<0.
\end{equation}
For every algorithm with at most $Q$ queries, the number $v$ of seen blue vertices and the number $c$ of blue components are at most $(D+1)Q$, and $S=N^z$. Hence
\begin{align}
 \frac{(D+1)Q}{W}&=N^{-3\eta/8+o(1)}=o(1),
      \label{rt:boundary-small}\\
 \frac{(D+1)QS^2}{N}&=O\bigl(N^{\alpha+2z-1}\bigr)=o(1),
      \label{rt:activity-small}\\
 8e(D+1)Q\left(\frac{2e(D+1)QS^2}{N}\right)^K&=O\bigl(N^{\alpha+K(\alpha+2z-1)}\bigr)=o(1),
      \label{rt:tail-small}
\end{align}
where the last bound follows from \eqref{rt:order-choice}. Together with \eqref{rt:girth-diameter}, these estimates verify the hypotheses of Lemma~\ref{rt:pointwise} for every transcript described by Theorem~\ref{os:structural}.

\begin{proposition}[Distance between complete transcripts]\label{rt:transcript-tv}
For the distributions and parameters above, the transcript distributions under $\DY$ and $\DN$ of every deterministic adaptive algorithm with at most $Q$ queries have total variation distance $o(1)$, uniformly over the algorithm. The same holds for the joint law of the random seed and the transcript of every randomized algorithm with at most $Q$ queries.
\end{proposition}
\begin{proof}
Consider first a deterministic algorithm, and let $\mathcal E$ be the set of marked transcripts whose observed blue graph $F_\xi$ is a weak forest with components of size at most $S=N^z$ and weak diameter at most $\Lambda$. By Theorem~\ref{os:structural}, we have $\Pr_{\DN}[\mathcal E]\ge1-\delta_N$, where $\delta_N=o(1)$ is given by \eqref{os:structural-failure}. Let $u_N=(D+1)Q/W$ and $t_N=8e(D+1)Q(2e(D+1)QN^{2z}/N)^K$; both are $o(1)$ by \eqref{rt:boundary-small}--\eqref{rt:tail-small}. By Lemma~\ref{rt:pointwise}, we have $\Pr_{\DY}[\xi]\ge(1-u_N)e^{-t_N}\Pr_{\DN}[\xi]$ for all $\xi\in\mathcal E$, and therefore
\[
 \TV\bigl(\Dist_{\DY}(\xi),\Dist_{\DN}(\xi)\bigr)
 =1-\sum_\xi\min\{\Pr_{\DY}[\xi],\Pr_{\DN}[\xi]\}
 \le1-(1-u_N)e^{-t_N}(1-\delta_N)
 \le\delta_N+u_N+t_N.
\]
Observe that this argument uses only the NO probability of $\mathcal E$. Forgetting the marks does not increase the total variation distance. For a randomized algorithm, we fix its random seed, which is independent of the input; the resulting deterministic algorithm obeys the same query bound, and averaging the uniform bound over the seeds bounds the distance between the joint laws.
\end{proof}

\subsection{Simple graphs and arbitrary constant degree}
\label{rt:transfer}

We first prove a lower bound at a degree depending on~$\eta$, then reduce the degree to~$2$, and finally extend the bound to every fixed $d\ge2$ with a proximity parameter independent of $d$.

\begin{theorem}[Degree depending on~$\eta$]
\label{rt:bounded-degree}
For every fixed $\eta\in(0,1)$, there exist an integer $d\ge2$ and a constant $\eps>0$ such that, for every sufficiently large $n$, every randomized adaptive two-sided $\eps$-tester for directed acyclicity on simple $n$-vertex directed graphs with maximum indegree and maximum outdegree at most $d$ requires at least $n^{1-\eta}$ queries.
\end{theorem}

\begin{proof}[Proof of Theorem~\ref{rt:bounded-degree}]
We first realize the construction at every sufficiently large input size, then project it to simple graphs, and finally derive the query bound from Proposition~\ref{rt:transcript-tv}. Fix $\eta$ and the constants of Section~\ref{rt:parameters}, and let $d=2(3k+4)D$. We use the capacity-constrained distributions $\hDY$ and $\hDN$ of Appendix~\ref{app:indegree}, which have the same objects, cores, and parameters as $\DY$ and $\DN$, and for which Proposition~\ref{rt:transcript-tv} holds by Theorem~\ref{in:main}. (The same argument applied to $\DY$ and $\DN$, with $D$ in place of $d$, proves the theorem for graphs whose outdegrees alone are bounded.)

\emph{All large input sizes.} Given $n$, put
\begin{equation}\label{rt:all-n}
 n'=\left\lfloor\frac n{3k+4}\right\rfloor,
 \qquad T=2\left\lfloor\frac{(n')^\beta}{2}\right\rfloor,
 \qquad W=\left\lfloor\frac{n'}T\right\rfloor,
 \qquad N=TW.
\end{equation}
Then $T$ and $N$ are even, $n_0=(3k+4)N\le n$, and $0\le n'-N<T=o(n')$. Thus $T=N^{\beta+o(1)}$ and $W=N^{1-\beta+o(1)}$, and all preceding estimates apply.

\emph{Simple graphs and farness.} For an auxiliary input $G$, let $G^\flat$ be the graph obtained by replacing each ordered list by its set of targets. Then $G^\flat$ has $n_0$ vertices, outdegree at most $D$, and indegree at most $2R=d$ by Lemma~\ref{in:degrees}, and it has no self-loops. Indeed, in $\hDN$ the core has weak girth greater than $\gamma\ge2$, and every other occurrence leaves $B$ or advances a layer or a region; in $\hDY$, the whole graph is acyclic. We pad both distributions with $n-n_0$ isolated vertices with fixed IDs, which keeps YES instances acyclic. In $\hDN$, every directed cycle lies in the simple core, which the projection leaves unchanged, so $\fas(G^\flat)=\fas(\Gamma(\Pi))$, and this is at least $4N$ with probability $1-o(1)$ by Proposition~\ref{rc:farness}. With
\begin{equation}\label{rt:distance}
 \eps=\frac1{10(3k+4)d},
\end{equation}
and since $n\le2n_0$ for large $n$, we have $\eps dn\le N/5<4N$. Thus a simple NO instance is $\eps$-far from acyclic with probability $1-o(1)$.

\emph{Simulating ordinary queries.} An ordinary query to a vertex that is not a padding vertex is answered by querying its complete ordered list in the auxiliary model and returning the set of targets, with caching; queries to padding vertices return $\emptyset$ without an auxiliary query. This simulates every ordinary algorithm with no more auxiliary queries, on every input and every random seed. Since the ordinary transcript is a function of the auxiliary transcript and the seed, the total variation distance can only decrease.

\emph{The query bound.} A two-sided $\eps$-tester with at most $Q$ queries would reject the simple YES distribution with probability at most $\frac13$, and the simple NO distribution with probability at least $\frac23(1-o(1))$; this contradicts Proposition~\ref{rt:transcript-tv}. Hence every two-sided $\eps$-tester makes more than $Q=\lfloor N^\alpha\rfloor$ queries. For large $n$, we have $N\ge n/(6k+8)$ and $Q\ge N^\alpha/2$, and since $\alpha>1-\eta$, we obtain $Q\ge n^{1-\eta}$. All constants depend only on $\eta$, and both distributions are fixed independently of the algorithm.
\end{proof}

Next, we lower the degree to~$2$ by replacing every vertex with two binary trees, at the cost of a smaller distance.

\begin{lemma}[Degree two]\label{rt:degree-two}
For every fixed $\eta\in(0,1)$, there exist constants $\delta_\eta,a_\eta>0$ such that, for every sufficiently large $n$, every randomized adaptive two-sided $\delta_\eta$-tester for directed acyclicity on simple $n$-vertex directed graphs with maximum indegree and maximum outdegree at most~$2$ requires at least $a_\eta n^{1-\eta}$ queries, even with whole-vertex queries.
\end{lemma}
\begin{proof}
Let $b$ and $\eps$ denote the degree bound and proximity parameter supplied by Theorem~\ref{rt:bounded-degree}. Set $\delta_\eta=\eps/(10b)$ and $a_\eta=1/(8b)$.

\emph{Double trees.} Fix a rooted tree with $b$ leaves, labeled $1,\ldots,b$, in which every internal vertex has two children. Orienting its edges away from the root gives the \emph{out-tree}, and orienting them toward the root gives the \emph{in-tree}. Let $G$ be a simple directed graph on $[m]$ whose indegrees and outdegrees are at most $b$, and fix, for every vertex $v$, an injection $\sigma_v$ from the edges entering $v$ into $[b]$. The graph $H=H(G,\sigma)$ replaces each vertex $u$ by a \emph{double tree}, a copy of the out-tree and a copy of the in-tree with a common root $r_u$, and each edge $e$ from $u$ to its $i$-th outneighbor $v$ by an edge from leaf $i$ of the out-tree of $u$ to leaf $\sigma_v(e)$ of the in-tree of $v$. Every leaf is incident with at most one edge between double trees, so $H$ is a simple graph on $(4b-3)m$ vertices whose indegrees and outdegrees are at most $2$.

\emph{Cycles and farness.} Each edge $e$ of $G$ from $u$ to $v$ becomes a path from $r_u$ to $r_v$ in $H$, the \emph{path of $e$}. Every leaf lies on at most one of these paths, and so every edge of $H$ lies on at most $b$ of them. Since the double trees are acyclic, every directed cycle of $H$ leaves some double tree, and contracting the double trees maps it to a closed walk in $G$; thus $H$ is acyclic if $G$ is. Conversely, if $F$ is a feedback arc set of $H$, then the edges of $G$ whose paths meet $F$ form a feedback arc set of $G$, as a directed cycle avoiding them would give a closed walk in $H-F$. Hence $\fas(G)\le b\fas(H)$.

\emph{Simulating queries.} Let $\mathcal A'$ be a two-sided $\delta_\eta$-tester for simple $n$-vertex graphs whose indegrees and outdegrees are at most~$2$, and put $m=\lfloor n/(4b-3)\rfloor$. On an $m$-vertex input $G$ with both degrees at most $b$, an algorithm $\mathcal A$ runs $\mathcal A'$ on $H(G,\sigma)$, padded with $n-(4b-3)m$ isolated vertices. IDs encode the double tree and position of each vertex, and the injections $\sigma_v$ are uniformly random and independent of each other and of the tester's seed. Every query is answered without querying $G$, except a query to leaf $i$ of the out-tree of $u$. Such a query costs at most one whole-vertex query to $u$, whose outneighbors are ordered by increasing ID. If $u$ has fewer than $i$ outneighbors, the leaf has no outgoing edge. Otherwise, when the edge $e=(u,v)$ first appears, the simulator draws $\sigma_v(e)$ uniformly from the unused leaves of the in-tree of $v$ and caches it. This lazy sampling is exact: conditional on its values at any set of arguments, a uniform injection maps a new argument uniformly to an unused value. Queries within in-trees reveal only fixed parent edges, so no incoming-neighbor queries are needed. Thus $\mathcal A$ runs $\mathcal A'$ on $H(G,\sigma)$ for a uniformly random~$\sigma$ and makes at most as many whole-vertex queries as $\mathcal A'$.

\emph{The query bound.} If $G$ is acyclic, then so is $H$. If $G$ is $\eps$-far, then $\fas(H)\ge\fas(G)/b\ge\eps bm/b=\eps m\ge2\delta_\eta n$ for large $n$, since $n<4b(m+1)$. Hence $H$ is $\delta_\eta$-far with respect to degree bound~$2$, for every choice of~$\sigma$. Theorem~\ref{rt:bounded-degree} therefore forces $\mathcal A$, and consequently $\mathcal A'$, to make at least $m^{1-\eta}\ge a_\eta n^{1-\eta}$ queries for large $n$.
\end{proof}

We next amplify both the degree and the feedback arc number so that the normalization of distance by $dn$ does not introduce a loss depending on~$d$.

\begin{lemma}[Complete blow-up]\label{rt:blowup}
Let $G$ be a simple directed graph on $m$ vertices and let $t\ge1$ be an integer. Form $G^{(t)}$ by replacing each vertex $u$ by copies $(u,1),\ldots,(u,t)$ and each edge $(u,v)$ by all $t^2$ edges $((u,i),(v,j))$, $i,j\in[t]$. Then $G^{(t)}$ is simple, has $tm$ vertices, and each directional degree is multiplied by~$t$. Moreover,
\begin{equation}\label{rt:blowup-fas}
 \fas(G^{(t)})=t^2\fas(G).
\end{equation}
One whole-vertex query to $G$ suffices to answer any whole-vertex query to $G^{(t)}$.
\end{lemma}
\begin{proof}
Simplicity and the degree claims follow from the construction. Deleting all $t^2$ copies of every edge in a feedback arc set of $G$ makes $G^{(t)}$ acyclic: a topological ordering of the remaining base graph extends by placing the copies of each vertex consecutively. This proves the upper bound in \eqref{rt:blowup-fas}.

For the lower bound, let $F$ be a feedback arc set of $G^{(t)}$, and choose one copy of each vertex of~$G$, independently and uniformly at random. The chosen copies induce a copy of~$G$ in $G^{(t)}$, and $F$ contains at least $\fas(G)$ of its edges. Since $G$ has no self-loops, each edge of $F$ joins two chosen copies with probability exactly $1/t^2$. Taking expectations gives $|F|/t^2\ge\fas(G)$, proving the lower bound. Finally, a query to $(u,i)$ is answered by querying $u$ in $G$ and returning all copies of each of its outneighbors; repeated queries may be cached.
\end{proof}

\begin{proof}[Proof of Theorem~\ref{thm:main}]
Fix $\eta\in(0,1)$ and apply Lemma~\ref{rt:degree-two} with $\eta/2$ in place of~$\eta$. Write $\delta=\delta_{\eta/2}$ and $a=a_{\eta/2}$, and set $\eps_\eta=\delta/3$, independently of~$d$. Fix an integer $d\ge2$, put $t=\lfloor d/2\rfloor$, and let $m=\lfloor n/t\rfloor$. For an $m$-vertex graph $G$ of maximum indegree and outdegree at most~$2$, form $G^{(t)}$ and pad it with $n-tm$ isolated vertices to obtain an $n$-vertex graph~$H$. Both degrees of $H$ are at most $2t\le d$, and $H$ is acyclic whenever $G$ is.

If $G$ is $\delta$-far with respect to degree bound~$2$, Lemma~\ref{rt:blowup} gives
\[
 \frac{\fas(H)}{dn}
 =\frac{t^2\fas(G)}{dn}
 \ge \delta\,\frac{2t}{d}\,\frac{tm}{n}
 \ge \frac{\delta}{3}
 =\eps_\eta
\]
for $n\ge2t$, since $2t/d\ge2/3$ and $tm/n\ge1/2$. Thus $H$ is $\eps_\eta$-far from acyclic with respect to degree bound~$d$.

Any randomized adaptive two-sided $\eps_\eta$-tester on these $n$-vertex graphs induces a two-sided $\delta$-tester on the $m$-vertex degree-two graphs. Encode the nonpadding IDs as pairs $(u,i)$, answer their queries using Lemma~\ref{rt:blowup}, and return the empty list for padding vertices. This simulation makes at most one whole-vertex query to~$G$ per query to~$H$. Lemma~\ref{rt:degree-two} therefore gives the lower bound
\[
 a m^{1-\eta/2}
 \ge a\left(\frac{n}{2t}\right)^{1-\eta/2}
 \ge n^{1-\eta}
\]
for all sufficiently large $n$, where the last inequality holds because $t$ is fixed and $n^{\eta/2}\to\infty$. The size threshold may depend on $\eta$ and~$d$, but $\eps_\eta$ depends only on~$\eta$. This proves Theorem~\ref{thm:main}, even with whole-vertex queries and hence also with indexed queries.
\end{proof}

\Needspace{12\baselineskip}

\section{Counting forests in high-girth cores}
\label{rf:section}

In this section, we prove Theorem~\ref{rf:comparison}, following the outline of Section~\ref{sec:ov-counting}. Observe that each component has exactly $N$ placements in either core, so all dependence on the core comes from the requirement that the images of different components be disjoint. We organize these exclusions by connected overlap patterns, using a cluster expansion (see, e.g.,~\cite{KP86}). The proof rests on a dichotomy. Patterns of at most $K$ components are local, so the girth condition makes their activities identical in both cores (Lemma~\ref{rf:small-activities}); patterns of more components are rare, since each attached component costs a factor $S^2/N$ (Lemma~\ref{rf:tree-bound}). An estimate for signed partition functions turns these two facts into the logarithmic comparison (Lemma~\ref{rf:partition}).

\subsection[Independent roots and the cluster expansion]{Independent roots and the cluster expansion}

Let $F_1,\ldots,F_c$ be the components of $F$, let $s_i=|V(F_i)|\le S$, and fix a root $r_i\in V(F_i)$ in each component. Once the image $x_i$ of $r_i$ is chosen, the map of $F_i$ is determined: one follows $P_a$ along a color-$a$ edge and $P_a^{-1}$ against it. Since $F$ is proper, every path in $F_i$ is a reduced colored word; hence if two vertices of $F_i$ had the same image, we would obtain a reduced closed walk of length at most $\Lambda$, and thus a weak cycle of length at most $\Lambda$. Therefore the map of $F_i$ is injective for every choice of $x_i$. Moreover, the image of each fixed vertex of $F_i$ is a bijective function of $x_i$, being a composition of permutations.

It remains to impose injectivity between different components. Choose $x_1,\ldots,x_c$ independently and uniformly at random, let $O_{ij}$ indicate that the images of $F_i$ and $F_j$ overlap, and let $\Phi_G(F)$ be the probability that all images are pairwise disjoint. Then
\begin{equation}\label{rf:embedding-partition}
 \Emb_G(F)=N^c\Phi_G(F),\qquad
 \Phi_G(F)=\E\prod_{1\le i<j\le c}(1-O_{ij}).
\end{equation}
For $I\subseteq[c]$ with $|I|\ge2$ define the \emph{activity}
\begin{equation}\label{rf:activity}
 a_I(G)=\E\sum_{\substack{\Gamma\text{ connected}\\V(\Gamma)=I}}
                 \prod_{ij\in E(\Gamma)}(-O_{ij}),
\end{equation}
which depends only on the components indexed by $I$ (see Figure~\ref{fig:overlaps}(a)). If we expand the product in \eqref{rf:embedding-partition} and group the edge sets into their connected components, we obtain the exact cluster expansion
\begin{equation}\label{rf:polymer-identity}
 \Phi_G(F)=
 \sum_{\mathcal C}\prod_{I\in\mathcal C}a_I(G),
\end{equation}
where $\mathcal C$ ranges over all collections of pairwise disjoint subsets of $[c]$ of size at least two, and the empty collection contributes $1$. (The factorization uses the independence of the roots in disjoint index sets.)

\begin{lemma}[Tree bound for the activities]\label{rf:tree-bound}
For $m=|I|\ge2$,
\begin{equation}\label{rf:activity-bound}
 |a_I(G)|\le m^{m-2}\left(\frac{S^2}{N}\right)^{m-1}.
\end{equation}
\end{lemma}
\begin{proof}
Fix the roots, and let $\mathcal O$ be the overlap graph on $I$. The integrand of \eqref{rf:activity} is $\sum_\Gamma(-1)^{|E(\Gamma)|}$, where the sum ranges over the connected spanning subgraphs $\Gamma$ of $\mathcal O$. We bound this alternating sum by grouping the connected edge sets according to a spanning tree. Fix a total order on the pairs in $I$, and assign to each such $\Gamma$ its minimum spanning tree $\mathcal T$, obtained by Kruskal's algorithm. For a spanning tree $\mathcal T\subseteq\mathcal O$, let $\operatorname{Ext}(\mathcal T)$ be the set of edges of $\mathcal O\setminus\mathcal T$ that are later than every edge of their $\mathcal T$-path. Then the subgraphs $\Gamma$ with tree $\mathcal T$ are exactly those with $E(\mathcal T)\subseteq E(\Gamma)\subseteq E(\mathcal T)\cup\operatorname{Ext}(\mathcal T)$. Indeed, an edge of $\operatorname{Ext}(\mathcal T)$ joins vertices that are already connected by earlier tree edges, so it does not change the output of Kruskal's algorithm; conversely, a non-tree edge that is earlier than the latest edge of its $\mathcal T$-path could replace that edge, which contradicts the minimality of $\mathcal T$. The signed sum over this interval is $(-1)^{m-1}(1-1)^{|\operatorname{Ext}(\mathcal T)|}\in\{0,\pm1\}$. Hence the integrand is at most the number of spanning trees of $\mathcal O$ in absolute value.

We now bound the probability that the overlaps prescribed by a fixed labeled tree on $I$ occur. Expose the roots starting from an arbitrary vertex, each child after its parent. Given the image of the parent, the child component overlaps it with probability at most $s_is_j/N\le S^2/N$, since each of the $s_is_j$ vertex pairs determines at most one child root. Thus all tree overlaps occur with probability at most $(S^2/N)^{m-1}$, and by Cayley's formula there are $m^{m-2}$ labeled trees.
\end{proof}

\subsection{Equality of the small activities}

\begin{lemma}[Universality of local counts]\label{rf:small-activities}
Under the hypotheses of Theorem~\ref{rf:comparison}, $a_I(G)=a_I(G')$ whenever $2\le|I|\le K$.
\end{lemma}

\begin{proof}
Let $m=|I|\le K$. The integrand of \eqref{rf:activity} vanishes unless the overlap graph on $I$ is connected, and in that case the union of the $m$ images is connected and has diameter at most $m\Lambda$. Fix the image $x$ of one root; then the whole union lies in the weak ball of radius $m\Lambda$ around $x$. Since the weak girth exceeds $2m\Lambda+1$, this ball is isomorphic, as a rooted colored directed graph, to the ball of the same radius in the universal $32$-colored tree. Indeed, two distinct reduced walks of length at most $m\Lambda$ from $x$ that end at the same vertex, or an extra edge inside the ball, would create a weak cycle of length at most $2m\Lambda+1$. Hence the placements of the other $m-1$ roots with connected overlap correspond bijectively, with the same overlap graphs, to placements in the universal ball, and the sum of the integrand over them is an integer that depends only on the colored shapes of the components. Since there are $N^m$ equally likely root tuples and $N$ choices of $x$, the activity $a_I(G)$ is $N^{1-m}$ times this integer, and so $a_I(G)=a_I(G')$. We note that this argument compares the whole activity, including multiple overlaps, and not only the probabilities of pairwise overlaps.
\end{proof}

\subsection[Signed partition functions]{Signed partition functions}

It remains to show that matching the small activities controls the whole partition function. Since the activities have signs, and an interpolation between two cores need not describe a probability, we prove positivity directly. For real numbers $w_I$ indexed by subsets $I\subseteq[c]$ with $|I|\ge2$, and for $J\subseteq[c]$, let
\[
 \Phi_J(w)=\sum_{\mathcal C}\prod_{I\in\mathcal C}w_I,
\]
where $\mathcal C$ ranges over all collections of pairwise disjoint subsets of $J$ of size at least two; in particular, $\Phi_\varnothing(w)=1$.

\begin{lemma}[Positivity and logarithmic derivatives]\label{rf:partition}
Let $p=S^2/N$ and $\kappa=cp\le1/(16e^2)$, and suppose that $|w_I|\le|I|^{|I|-2}p^{|I|-1}$ for all $I$. Then every $\Phi_J(w)$ is positive, $|\Phi_J(w)/\Phi_{J\setminus\{x\}}(w)-1|\le\frac14$ for $x\in J$, and for every $I$,
\begin{equation}\label{rf:derivative}
 0<\frac{\partial\log\Phi_{[c]}(w)}{\partial w_I}
       =\frac{\Phi_{[c]\setminus I}(w)}{\Phi_{[c]}(w)}
       \le2^{|I|}.
\end{equation}
\end{lemma}
\begin{proof}
Since $\binom{c-1}{m-1}\le c^{m-1}/(m-1)!$ and $m^{m-1}/m!\le e^m$, for every $x$,
\begin{equation}\label{rf:incident-sum}
 \sum_{I\ni x}|w_I|\,2^{|I|-1}
 \le\sum_{m\ge2}\binom{c-1}{m-1}m^{m-2}p^{m-1}2^{m-1}
 \le e\sum_{m\ge2}(2e\kappa)^{m-1}
 \le4e^2\kappa\le\tfrac14.
\end{equation}
We proceed by induction on $|J|$; sets of size at most one have $\Phi_J=1$. If we separate the collections according to the set containing $x$, if any, we obtain
\[
 \Phi_J(w)=\Phi_{J\setminus\{x\}}(w)+\sum_{I\subseteq J,\,I\ni x}w_I\Phi_{J\setminus I}(w).
\]
By the induction hypothesis, deleting the other $|I|-1$ elements of $I$ from $J\setminus\{x\}$ one at a time multiplies $\Phi$ by a factor in $[\frac45,\frac43]$ each time, and so $|\Phi_{J\setminus I}/\Phi_{J\setminus\{x\}}|\le(4/3)^{|I|-1}\le2^{|I|-1}$. Now \eqref{rf:incident-sum} gives the ratio bound, and positivity follows. Deleting the elements of $I$ from $[c]$ one at a time gives $0<\Phi_{[c]\setminus I}/\Phi_{[c]}\le2^{|I|}$. Finally, $\partial \Phi_{[c]}/\partial w_I=\Phi_{[c]\setminus I}$, because the collections containing $I$ correspond to the collections on $[c]\setminus I$.
\end{proof}

\begin{proof}[Proof of Theorem~\ref{rf:comparison}]
For $0\le t\le1$, let $w_I(t)=(1-t)a_I(G)+ta_I(G')$. By Lemma~\ref{rf:tree-bound}, Lemma~\ref{rf:partition} applies along the whole interpolation, and integrating \eqref{rf:derivative} gives
\[
 |\log\Phi_G(F)-\log\Phi_{G'}(F)|\le\sum_I2^{|I|}|a_I(G)-a_I(G')|.
\]
By Lemma~\ref{rf:small-activities}, only sets with $|I|>K$ contribute. By Lemma~\ref{rf:tree-bound} and the bounds $\binom cm\le c^m/m!$ and $m^{m-2}/m!\le e^m$, we have
\[
 \sum_{|I|>K}2^{|I|}|a_I(G)-a_I(G')|
 \le2\sum_{m>K}\binom cm2^mm^{m-2}p^{m-1}
 \le4e\,c\sum_{m>K}(2e\kappa)^{m-1}
 \le8e\,c(2e\kappa)^K,
\]
where we used $2e\kappa\le\frac12$. The factor $N^c$ in \eqref{rf:embedding-partition} cancels, which proves \eqref{rf:log-comparison}. If $c\le K$, then all activities agree, and so the counts are equal.
\end{proof}

Thus only connected overlaps of more than $K$ components can distinguish two cores. Observe that the geometric hypothesis of the theorem concerns the diameter, while its tail bound concerns the size of the components; Theorem~\ref{os:structural} provides both.

\section{Adaptive exploration}\label{os:section}

In this section, we prove Theorem~\ref{os:structural}, which is the main technical part of the paper; an informal outline is given in Section~\ref{sec:ov-info}. Section~\ref{os:analysis-params} collects the constants used in the analysis. In Section~\ref{os:fragments}, we split the exploration into fragments and fix the probabilistic conventions. Section~\ref{os:reconstruction} proves the work estimate at one level, Section~\ref{os:generations} iterates this estimate over levels and generations, and Section~\ref{os:chronology} combines clans into components. We use two different auxiliary experiments: an erased process with a \emph{multiplicative} bound at every birth, which controls the height of fragments (Lemma~\ref{os:blue-size} and Appendix~\ref{app:blue-erasure}), and an allocated experiment with masks independent of all other randomness, which controls the \emph{expected} work (Lemma~\ref{os:local}). Keeping the two experiments separate is important, since an additive error at a possibly rare birth would not give the height bound.

\subsection{Parameters of the analysis}
\label{os:analysis-params}

The analysis uses a few further constants and scales, which do not affect the construction. We set
\begin{equation}
\label{os:analysis-constants}
 \zeta=\frac{\beta}{1000(k+1)^2},\qquad
 z=(k+3)\zeta,\qquad
 r=\Bigl\lceil\frac2\beta\Bigr\rceil+2,\qquad
 K=\Bigl\lceil\frac8\eta\Bigr\rceil,
\end{equation}
and we require the integer $A$ of Section~\ref{os:params} to satisfy
\begin{equation}\label{rt:A-choice}
 A-1>\max\Bigl\{\frac{320}{\zeta},\ \frac{4000Kr(k+1)}{\beta}\Bigr\}.
\end{equation}
Here $\zeta$ and $z$ are small slack exponents (observe that $z<\eta/8$), $r$ bounds the number of clans in an observed component, and $K$ is the comparison order of Section~\ref{rt:section}. For $0\le j\le k$, let $\theta_j=d_j/d_{j+1}=2^{-5(A-1)A^j}\le\frac12$ be the selection ratio at level $j$, and let
\begin{equation}
\label{os:path-lengths}
 h_j=\Bigl\lceil\frac{80\log N}{\log(1/\theta_j)}\Bigr\rceil,\qquad
 \Lambda=2r(k+1)(h_0+1)+(r-1),\qquad
 \Delta=C_\eta\Bigl(\frac QW+\frac1T\Bigr).
\end{equation}
Thus $\theta_j^{h_j}\le N^{-80}$, the quantity $\Lambda$ is the diameter bound of Theorem~\ref{os:structural}, and $\Delta$ bounds the rate of the exceptional events discussed in Section~\ref{os:params}. By \eqref{os:scales},
\begin{equation}
\label{os:small-scales}
 \frac QW=\frac{QT}{N}=N^{-3\beta+o(1)},\qquad
 \Delta=N^{-\beta+o(1)},\qquad N^\zeta\Delta=o(1).
\end{equation}
Moreover, $h_0$, $\Lambda$, and $\gamma$ are all of order $\log N$, and the choice of $A$ makes the coefficient of $\log N$ in $\Lambda$ small compared with that in $\gamma$: we have $\Lambda=\frac{32r(k+1)}{(A-1)\log2}\log N+O(1)$ and $\gamma=\frac{\beta}{24\log2}\log N+o(\log N)$, so that, for large $N$,
\begin{equation}\label{rt:girth-diameter}
 2K\Lambda+1<\gamma.
\end{equation}
Finally, the first condition in \eqref{rt:A-choice} controls the growth of short paths.

\begin{lemma}[Deterministic growth]
\label{os:growth}
For all large $N$, every input of $\DN$, all $0\le j\le k$, and all $v\in U_j$, the set $\Reach_{U_j}(v,\le h_j+3)$ of vertices reachable from $v$ by paths of length at most $h_j+3$ inside $U_j$ has at most $N^{\zeta/2}$ elements.
\end{lemma}

\begin{proof}
By Lemma~\ref{os:properties}(ii), the set has at most $\sum_{i\le h_j+3}d_j^i\le2d_j^{h_j+4}$ elements, and $d_j^{h_j}\le N^{80\log d_j/\log(1/\theta_j)+o(1)}=N^{80/(A-1)+o(1)}\le N^{\zeta/4+o(1)}$ by \eqref{rt:A-choice}.
\end{proof}

\subsection{Fragments, disclosures, and clans}
\label{os:fragments}

Since an algorithm may revisit IDs or query arbitrary IDs, its exploration need not be a single rooted search. We therefore divide fresh discoveries into fragments and charge each first query to a local search. To simplify the analysis, an analytical oracle makes additional \emph{disclosures}, called announcements; these disclosures can only help the algorithm, so a bound proved in their presence also holds without them.

\begin{definition}[Fragment protocol]
\label{def:protocol}
A \emph{fragment} consists of a root, a set of owned vertices, and a status, which is either active or retired. Each seen vertex has at most one active fragment, called its \emph{owner}, and a vertex without an active owner is called \emph{inactive}. A \emph{contact} is a newly exposed occurrence whose target has already been seen, that is, a collision; it is an \emph{upper contact} if its source and target lie in $U_1$.
\begin{enumerate}
\item \emph{Births and ownership.} A first query to an unseen ID starts an \emph{initial} fragment, which owns its root. A fresh target inherits the active owner of its source, if any; an old target keeps its owner.
\item \emph{Retirement.} After a complete answer, every active owner incident to an upper contact in that answer retires, and so does an owner that has just been charged with its $T$-th first query with source in $U_1$. Retirement clears the active ownership of the fragment; the oracle then announces the fragment, its root ID, and the IDs of all blue vertices that it owns (possibly none). Observe that region and layer boundaries cause no retirement.
\item \emph{Dependent births.} A first query to an inactive, announced, unqueried blue vertex starts a \emph{dependent} fragment rooted there, whose \emph{parent} is the announcing fragment. A query to an inactive nonblue vertex starts no fragment, and a query to a vertex with an active owner is charged to that owner.
\end{enumerate}
Only first queries expose entries, count toward cutoffs, or create contacts; repeated queries change nothing. The entries of an answer are processed in list order, with no algorithmic decision between them, and an ID introduced by one entry is old at the later entries of the same answer. Seen sets, discovery parents, and ownership are updated entry by entry, whereas announcements are emitted after the complete answer.
\end{definition}

Every fragment has a \emph{discovery tree}, an out-tree rooted at its root, which consists of the occurrences that first introduced its other owned vertices. (The edge through which a parent discovered a dependent root belongs to the tree of the parent.) The \emph{blue discovery tree} of a fragment is the subtree on its blue vertices. For a fragment with a blue root, this subtree is connected, because only blue sources have blue targets, and its height is called the \emph{blue height} of the fragment. A fragment with a nonblue root never discovers a blue vertex, and every first query is charged to at most one owner.

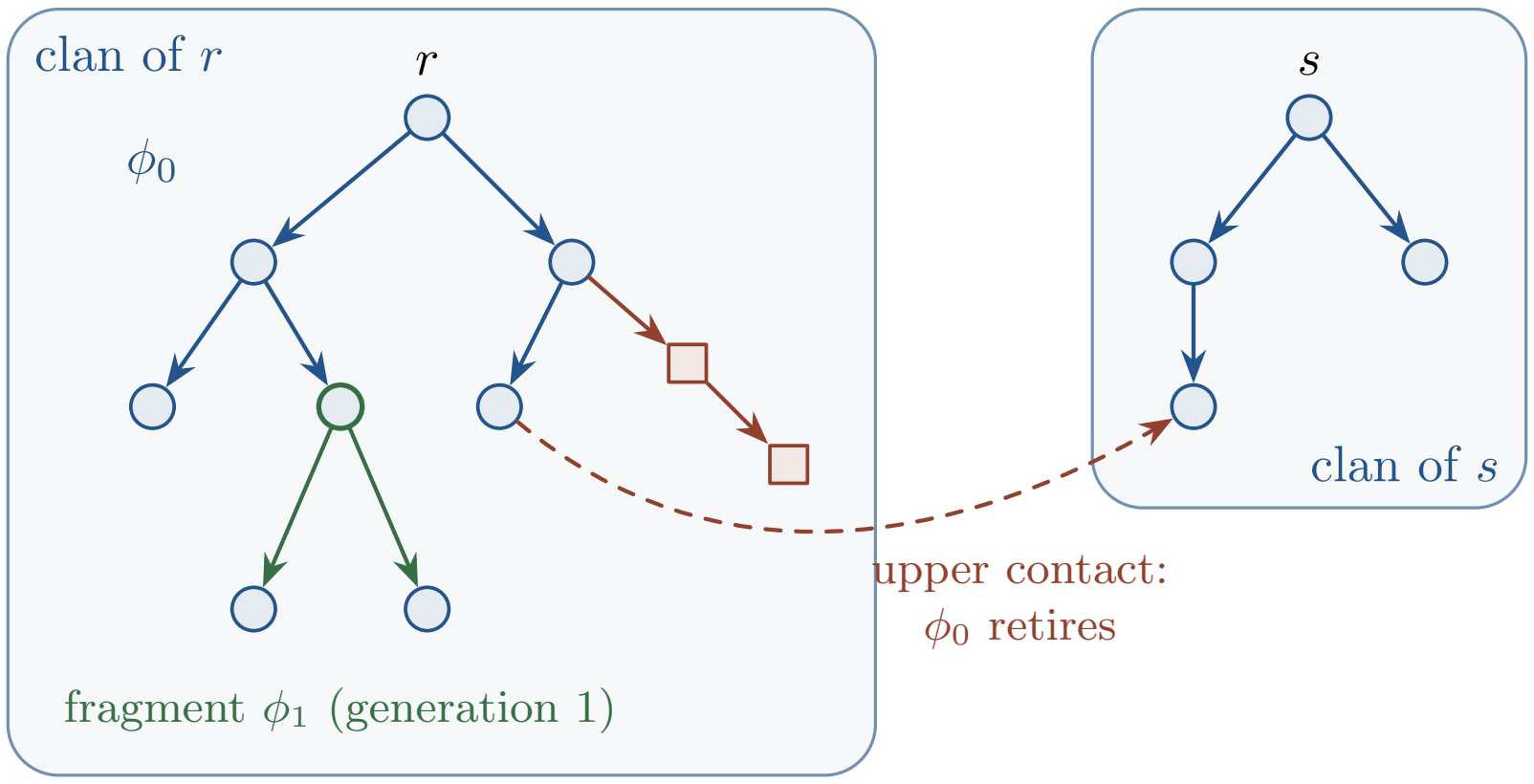
\begin{figure}[!t]
\centering
\begin{tikzpicture}[x=1cm,y=1cm,line cap=round,line join=round,
  bv/.style={acy/bluedot,minimum size=3mm},
  rv/.style={rectangle,draw=acyRed,fill=acyRed!12,line width=.7pt,minimum size=2.6mm,inner sep=0pt},
  e0/.style={acy/edge,draw=acyBlue,line width=.8pt},
  e1/.style={acy/edge,draw=acyGreen,line width=.8pt},
  er/.style={acy/edge,draw=acyRed,line width=.8pt},
  lab/.style={acy/note,font=\small}]
 \draw[acy/bluearea,rounded corners=10pt] (-2.9,-1.55) rectangle (3.1,3.75);
 \node[lab,text=acyBlue,anchor=north west] at (-2.85,3.7) {clan of $r$};
 \node[bv,label={[lab]above:$r$}] (r) at (0,3.0) {};
 \node[bv] (a) at (-1.2,2.0) {};
 \node[bv] (b) at (1.0,2.0) {};
 \node[bv] (c) at (-1.9,1.0) {};
 \node[bv,draw=acyGreen,line width=1pt] (d) at (-0.6,1.0) {};
 \node[bv] (e) at (0.5,1.0) {};
 \node[rv] (f) at (1.8,1.3) {};
 \node[rv] (g) at (2.5,0.6) {};
 \foreach \u/\v in {r/a,r/b,a/c,a/d,b/e} \draw[e0] (\u)--(\v);
 \draw[er] (b)--(f); \draw[er] (f)--(g);
 \node[bv] (h) at (-1.2,-0.4) {};
 \node[bv] (i) at (0.0,-0.4) {};
 \draw[e1] (d)--(h); \draw[e1] (d)--(i);
 \node[lab,text=acyBlue] at (-1.9,2.7) {$\phi_0$};
 \node[lab,text=acyGreen,align=center,font=\footnotesize] at (-0.6,-1.1) {fragment $\phi_1$ (generation $1$)};
 \draw[acy/bluearea,rounded corners=10pt] (4.6,0.3) rectangle (7.6,3.75);
 \node[lab,text=acyBlue,anchor=south east] at (7.55,0.35) {clan of $s$};
 \node[bv,label={[lab]above:$s$}] (s) at (6.1,3.0) {};
 \node[bv] (u) at (5.3,2.0) {};
 \node[bv] (w) at (6.9,2.0) {};
 \node[bv] (x) at (5.3,1.0) {};
 \foreach \u/\v in {s/u,s/w,u/x} \draw[e0] (\u)--(\v);
 \draw[er,dashed] (e) to[out=-40,in=-150] (x);
 \node[lab,text=acyRed,align=center,font=\footnotesize] at (4.1,-0.35) {upper contact:\\$\phi_0$ retires};
\end{tikzpicture}
\caption{Fragments and clans. The initial fragment $\phi_0$, started by a query to the unseen vertex $r$, owns the vertices that it discovers, including nonblue ones (squares). When its exploration exposes an edge to a previously seen vertex of another clan (dashed), $\phi_0$ retires after this answer and announces the blue vertices that it owns. A later first query to an announced, unqueried vertex starts the dependent fragment $\phi_1$ (green), which belongs to the clan of $r$. Edges between different clans always have old targets.}
\label{fig:fragments}
\end{figure}

\paragraph{Generations and clans.}
Initial fragments with blue roots have generation $0$, and a dependent fragment has generation one more than its parent. Every blue vertex carries a persistent \emph{clan}, namely the initial fragment with a blue root from which it descends; a fresh blue target inherits the clan of its source, and a dependent root keeps its clan. We note that every first query to a blue source has an active owner with a blue root. Indeed, an unseen source starts an initial fragment, a blue source that is seen but unqueried either still has its active owner or was announced and starts a dependent fragment, and a repeated query exposes nothing. By induction, every seen blue vertex lies in the discovery tree of its clan, that is, the union of the blue discovery trees of its fragments, which is connected. Moreover, since only blue sources have blue targets, every exposed occurrence between different clans has an old target. Therefore the fragment and clan bounds below cover the whole observed blue graph of Theorem~\ref{os:structural}. Figure~\ref{fig:fragments} illustrates these notions.

\paragraph{The analytical state.}
The exposed ID--object pairs form a partial bijection, and its uniform completion between unseen IDs and unused objects is called the \emph{residual bijection}. We implement it by fresh randomness: a query to an unseen ID scans an independent uniform priority order on objects and takes the first unmatched object, and a fresh target object scans an independent uniform priority order on IDs. The \emph{analytical state} contains the matched identities, the category words read so far, the exposed targets and colored matches of the core, the ownership records and announcements, and the inspected prefixes of priorities. It contains no unread category word, no unexposed core match, and no uninspected suffix of a priority. (Observe that the proof may read the category word of a fresh source before exposing its list, since category words are independent of the core.)

Let $s$ be the analytical state at a birth, after the root has been matched to its object $o$, and before its category word is read and its list is exposed. For a source $v$, let $C_v$ and $Y_v$ be its category word and the samples of its nonblue endpoints (which are constant at the sources in the last layer of $X_{k+1}$); together, they form the \emph{seed} of $v$. Then the conditional law at $s$ is
\begin{equation}\label{os:posterior}
 \psi_s(\Pi)\otimes\Unif(\text{residual bijection})
       \otimes\bigotimes_{v\text{ unqueried}}\Dist(C_v,Y_v),
\end{equation}
where $\psi_s$ is the uniform distribution on the tuples in $\mathcal G_\gamma$ that extend the exposed colored matches. Indeed, given the exposed variables, all past queries and announcements are deterministic, and each core observation imposes one colored equality. Moreover, no operation reads an unqueried word, information arriving at an object through incoming edges constrains the core but not the word of that object, and the skip predicates of priority scans involve only matched identities. Induction over the operations gives \eqref{os:posterior}; Appendix~\ref{app:local-allocation} verifies it directly from the likelihood. We note that unexposed core entries are correlated, but the unused words are independent of the whole posterior core.

\paragraph{Conditioning conventions.}
Three conventions are used throughout.
\begin{enumerate}[label=(C\arabic*)]
\item Every bound on the probability of a contact or a match is a bound on the conditional probability given the analytical state \emph{immediately before} the relevant exposure or scan; it is obtained from Lemma~\ref{os:properties}(iii) or by direct counting. We never condition on future events, on a future work or size budget, or on future randomness.
\item Estimates at a birth are exact statements at \emph{every} birth state $(s,o)$ of positive probability. Therefore they apply to arbitrarily rare dependent births, and they can be summed over any family of fragments whose membership is determined at birth.
\item Different births and levels use separate conditional experiments, which need not coexist on one probability space; we combine their bounds by the tower rule.
\end{enumerate}

\paragraph{Height and size of a single fragment.}

\begin{lemma}[Blue height and size of fragments]
\label{os:blue-size}
Except with probability at most $N^{-20}$, every fragment with a blue root, initial or dependent, has blue height at most $h_0+1$ and owns at most $\Sf=N^\zeta$ blue vertices.
\end{lemma}

\begin{proof}
Lemma~\ref{lem:conditional-blue-height} (Appendix~\ref{app:blue-erasure}) shows that at every birth state $(s,o)$ and for every $h\ge1$,
\[
 \Pr[\text{blue height}\ge h+2\mid s,o]
 \le \exp(C_\eta QT/N)\,(D+1)Q\,\theta_0^h .
\]
By \eqref{os:small-scales}, the exponential factor is at most $2$, and $\theta_0^{h_0}\le N^{-80}$. A union bound over at most $Q$ births gives a failure probability of at most $2(D+1)Q^2N^{-80}\le N^{-20}$. On the complementary event, the blue vertices of every fragment lie within distance $h_0+1$ of its root inside $B$, and hence in a set of size at most $N^{\zeta/2}\le\Sf$ by Lemma~\ref{os:growth}.
\end{proof}

\subsection[The work at one level]{The work at one level}
\label{os:reconstruction}

The goal of this subsection is to bound the work of a family of fragments at each level of the hierarchy. We say that a family $\mathcal F$ of fragments with blue roots is \emph{decided at birth} if the membership of each fragment is determined by its analytical state at birth; for example, the family of all fragments of a fixed generation is decided at birth. For a fragment $\phi$, let $\mathrm{work}_j(\phi)$ be the number of first queries with source in $U_j$ that are charged to $\phi$, and let $\mathrm{work}_j$ be the sum of $\mathrm{work}_j(\phi)$ over $\phi\in\mathcal F$; let $\mathrm{ret}$ be the number of retirements in $\mathcal F$, and let $f=\E|\mathcal F|$.

\begin{lemma}[Local estimate]
\label{os:local}
For every family decided at birth and every $1\le j\le k$,
\begin{align}
 \E\,\mathrm{work}_j
 &\le N^\zeta\bigl(f+\Delta\,\E\,\mathrm{work}_{j+1}+\E\,\mathrm{ret}\bigr)+N^{-18},
       \label{os:local-rec}\\
 \E\,\mathrm{ret}&\le\Delta(\E\,\mathrm{work}_1+f),
       \label{os:retirement-rec}\\
 \mathrm{work}_{k+1}&\le Q.
       \label{os:work-cap}
\end{align}
\end{lemma}

Figures~\ref{fig:info} and~\ref{fig:info-paths} illustrate the idea of the proof. In words, a birth supplies a local root, a retirement costs one record, and work in $U_{j+1}$ creates exceptional records at rate $\Delta$. Since $N^\zeta\Delta=o(1)$, each level passes on only a small fraction of the work above it. Retirements need not be cheap themselves; instead, \eqref{os:retirement-rec} bounds them in terms of the work and the births that cause them, and they are absorbed in Section~\ref{os:generations}. The bound \eqref{os:work-cap} holds since each first query is charged to at most one owner. (We note that the fragments and their retirement times are the same at all levels; the level only affects the analysis.)

\paragraph{Strategy.}
The proof of \eqref{os:local-rec} follows the outline of Section~\ref{sec:ov-info}. We fix a level $j$ and a birth, and we separate the level-$j$ masks $\Theta$ from all other randomness, which forms a tape $Z$. The heart of the argument is the bound \eqref{os:cut-expectation}: in expectation, the number of edges needed to cut all long downward paths inside $U_j$ of the discovery tree is at most $O(\log N)$ per exceptional record. It is proved in two steps. First, the observable history is a deterministic function of $Z$ and of the record word (Lemma~\ref{os:decoder}). Second, for a fixed record word, the decoded tree depends only on $Z$, which is independent of $\Theta$, so long retained paths in it are exponentially rare, and a union bound over the short record words transfers this to the actual tree. Finally, short trees in $U_j$ are small by Lemma~\ref{os:growth}, and records are rare by \eqref{os:record-count}. The conditional experiment and the allocation of the randomness below serve only to make the first step an exact identity in which $\Theta$ is independent of $Z$.

\paragraph{The conditional experiment.}
Fix a level $1\le j\le k$ and a birth state $(s,o)$ of a fragment~$\phi\in\mathcal F$, where $o\in B$; we call $\phi$ the \emph{distinguished} fragment, or simply \emph{the fragment}. In this experiment, \emph{owned} means owned by $\phi$, \emph{exterior} means everything else, \emph{upper} means $U_{j+1}$, and \emph{lower} means the complement of $U_{j+1}$. We follow the entire exploration, including exterior queries, up to the complete answer that retires the distinguished fragment, and we stop before the following announcements; if there is no retirement, we stop when the algorithm stops or at query $Q$, the \emph{horizon}. Let $\mathcal H$ be the observable history, that is, the queries, the answers, the permitted announcements, and a stop tag indicating the reason for stopping. Observe that $\mathcal H$ contains no object identities and no categories.

Three kinds of indices occur in the analysis. Every entry has a \emph{list index} in $[D]$, a \emph{candidate} entry (see Figure~\ref{fig:decoy}) also has a rank among the candidates, and an entry of category~$0$ at a blue source has a color, which is its rank among the entries of category~$0$ in its list. We number the first queries of this experiment by their \emph{query positions} $t=1,2,\ldots$. At every future query position $t$, let $G_t\subseteq[D]$ be a uniform $d_{j+1}$-set of \emph{candidate} indices, and let the \emph{selector mask} $\Theta_t$ be a uniform $d_j$-subset of the candidate ranks $[d_{j+1}]$. At a $U_j$ source of the fragment, the selected candidates receive a uniform arrangement of the categories at most $j$, the other candidates receive category $j+1$, and the noncandidates receive a uniform arrangement of the categories larger than $j+1$ (there are none if $j=k$, since $d_{k+1}=D$); this is exactly a uniform category word. Thus a candidate is selected with probability $\theta_j$, and the masks $\Theta=(\Theta_t)_t$ are independent. An $X_{j+1}$ source of the fragment ignores $\Theta_t$, since all its candidate entries advance to the next layer (or follow the rule of Definition~\ref{def:no} at the last layer).

All other randomness forms a tape $Z$ that is independent of $\Theta$: the candidate sets, the arrangements of the categories at most $j$ and larger than $j+1$, the whole posterior core $\Pi\sim\psi_s$, the samples of nonblue endpoints, and the priorities. All entries of category~$0$ at blue sources read the same core $\Pi$, indexed by the source object and the color. Since the choice of sources depends on earlier answers, $Z$ contains, at every position, separate copies of the randomness for every possible source object and every \emph{mode} (a $U_j$, $X_{j+1}$, or lower source of the fragment, or an exterior source), and the actual source determines the copy that it reads. The only shared randomness is the following: at a noncandidate index, every hypothetical upper source of the fragment has category $c>j+1$ and destination $E_c$, so all of them read one common category and target. The allocation is specified in Appendix~\ref{app:local-allocation}.

\begin{lemma}[Exact law of the continuation]\label{os:continuation-law}
At every birth state $(s,o)$ of positive probability, the allocated experiment, in which $\Theta$ and $Z$ are independent with the laws above, has the same joint law as the original conditional continuation from $(s,o)$, augmented by suitable auxiliary kernels. The identity holds jointly for the posterior core, all variables of the adjacency lists, the completed ID bijection, all allocated tapes and priorities, and the complete private state at the stop (assignments of sources to positions, ownership and discovery records, caches, inspected prefixes of priorities, records, and stop tag).
\end{lemma}

The proof, given in Appendix~\ref{app:local-allocation}, matches the complete seeds of the sources and the completed ID bijection by finite density identities. We use the lemma in two ways. The reconstruction below is a deterministic statement about the allocated experiment, and so it may read all of $Z$. Probability bounds for records, by contrast, follow (C1) and use only the analytical state before an exposure, since conditioning on the whole tape would bias the unused seeds. Observe that $\Theta$ and $Z$ are independent a priori at the birth; what $\mathcal H$ reveals about $\Theta$ is exactly what we must bound.

\paragraph{Records and the decoder.}
The \emph{decoder}, which formalizes the replay of Section~\ref{sec:ov-info}, is an analytical device that recovers $\mathcal H$ from $(s,o)$, from $Z$, and from a word of exceptional records. (It is a tool for the analysis, rather than an operation available to the algorithm.) It maintains a partial map from IDs to objects, and a seen vertex is called \emph{resolved} if its object is known to the decoder. Exterior and lower vertices are always resolved, while upper vertices of the fragment may remain unresolved. Observe that knowing the object of a $U_j$ vertex does not reveal its adjacency list, which depends on $\Theta$. A \emph{component in $X_{j+1}$} is a maximal subtree of the discovery tree of the fragment whose edges lie inside $X_{j+1}$. Its \emph{entry vertex} is the first $X_{j+1}$ vertex on the discovery path from the root, and by monotonicity it lies in $E_{j+1}$. Since $X_{j+1}$ sources ignore $\Theta$, the object of the entry vertex allows the decoder to replay the exposed part of the component, together with the lower vertices below it, from $Z$. The decoder binds each computed endpoint to the ID in the stored \emph{public pattern}, that is, the ordered tuple of returned IDs, including repetitions and empty lists. A component is resolved when all its exposed vertices are resolved.

The decoder needs identity information only at the following exceptional events. Here a \emph{contact in $U_{j+1}$} is a contact whose source and target both lie in $U_{j+1}$.
\begin{enumerate}
\item[(A)] A $U_j$ list of the fragment has a contact in $U_{j+1}$. Record the source's ID and object, its public pattern, its mask $\Theta_t$, and the ID--object pairs of its targets.
\item[(B)] An exterior list hits a seen $U_j$ vertex of the fragment. Record its public pattern and the ID--object pairs of the hit targets.
\item[(C)] A contact in $U_{j+1}$ is first incident to an unresolved component in $X_{j+1}$, or such a component reaches $T$ first queries to its own vertices. Record the public pattern and the pair of the entry vertex (both entry vertices if two unresolved components meet). Contacts that involve only lower descendants do not trigger (C).
\item[(D)] The distinguished fragment retires. Record the retiring answer and the stop tag.
\item[(E)] During the priority scan of an exterior query to an unseen ID, the object of an unresolved vertex is skipped. Record its ID--object pair, and the pair of its component's entry vertex if it lies in an unresolved component in $X_{j+1}$. These records precede the answer.
\end{enumerate}
An auxiliary encoder writes these records into a word $\mathcal M=\mathcal M_{\phi,j}$, where $\phi$ is the distinguished fragment; the records appear in chronological order and are tagged by their query position and by whether they are processed before or after the answer. Each record specifies $O_\eta(1)$ positions, list indices, IDs, and objects, so it lies in an alphabet $\Sigma$ of size $N^{C_\eta}$. One answer produces $O_\eta(1)$ records of types (A)--(D), while a scan may produce several records of type (E).

\begin{lemma}[Exact reconstruction]\label{os:decoder}
There is a deterministic map $\Dec$ such that, on the support of the allocated experiment,
\begin{equation}
\label{os:decoder-identity}
 \mathcal H=\Dec(s,o,Z,\mathcal M).
\end{equation}
\end{lemma}

We now explain why these records suffice. The decoder keeps the public records and leaves fresh upper vertices of the fragment unresolved. For a $U_j$ source of the fragment, or for an unresolved $X_{j+1}$ source of the fragment, an answer without records can be computed without the mask: the noncandidate targets come from the common template in $Z$ (they are lower), and the candidate entries receive fresh IDs from their shared priorities. When a contact makes an identity relevant, records (A)--(C) supply the missing pairs, and resolving an entry vertex lets the decoder replay its component. The cutoff in (C) resolves a component before it can reach the last layer of $X_{j+1}$, since its entry vertex lies in the first $T$ layers and reaching layer $3T$ requires $2T$ further first queries in the component. Records (E) make exterior scans reproducible, and record (D) supplies the final answer. Appendix~\ref{app:decoder} verifies the invariant, including repetitions within an answer and interacting replays; replays read only exposed lists and add no records. Hence all information in $\mathcal H$ beyond $Z$ is carried by $\mathcal M$.

\paragraph{Expected number of records.}
Having established exact reconstruction, we bound how often records are needed; all bounds follow (C1). At most $(D+1)Q$ IDs are ever seen, so by Lemma~\ref{os:properties}(iii), an upper entry of the fragment makes a contact in $U_{j+1}$ with conditional probability at most $C_\eta Q/W$. This bounds the outgoing records of types (A) and (C), summed over $\mathcal F$, by $C_\eta(Q/W)\E\,\mathrm{work}_{j+1}$. For incoming contacts, observe that the upper vertices of an active fragment of $\mathcal F$ consist of its root and of at most $D$ targets per first query to one of them. Summed over all active fragments of $\mathcal F$, the target set at any time has size at most $|\mathcal F|+D\,\mathrm{work}_{j+1}$ (evaluated at the end). This set is predictable, because membership in $\mathcal F$ is decided at birth, and every object in it is hit with conditional probability at most $1/W$ per entry. Summing over at most $DQ$ exposed entries bounds the incoming records by $C_\eta(Q/W)(f+\E\,\mathrm{work}_{j+1})$. The cutoffs in (C) charge disjoint sets of $T$ first queries, which costs at most $\E\,\mathrm{work}_{j+1}/T$, and records of type (D) cost $\E\,\mathrm{ret}$. For (E), let $u=\Omega(N)$ be the number of unmatched objects before an exterior scan and let $b$ be the number of unresolved vertices. Each unresolved object precedes all $u$ unmatched objects with probability $1/(u+1)$, so the expected number of skips is $b/(u+1)$. Since $b\le1+D\,\mathrm{work}_{j+1}(\phi)$ at all times, summing over at most $Q$ scans and over $\mathcal F$ gives $C_\eta(Q/N)(f+\E\,\mathrm{work}_{j+1})$. Finally, each fragment of $\mathcal F$ queries its root, so $f\le\E\,\mathrm{work}_j\le\E\,\mathrm{work}_{j+1}$. After increasing the constant in $\Delta$, we obtain
\begin{equation}
\label{os:record-count}
 \E\sum_{\phi\in\mathcal F}|\mathcal M_{\phi,j}|
 \le C_\eta\bigl(\Delta\,\E\,\mathrm{work}_{j+1}+\E\,\mathrm{ret}\bigr).
\end{equation}
(By the tower rule, the left side equals the sum of the conditional expectations over the births of the fragments of $\mathcal F$.)

\paragraph{From records to a path cut.}
Exact reconstruction lets us bound the expected maximum number of edge-disjoint long retained paths by $O_\eta(1)$ times the expected number of records, up to a negligible additive error. A record word of length $\ell$ has at most $N^{O_\eta(\ell)}$ possible values, whereas a fixed path of length $h_j$ is retained by masks independent of the tape with probability at most $N^{-80}$. The triple $(s,o,\mathcal H)$ determines the discovery tree of the fragment, with all region memberships forgotten. This tree has one root and at most $D$ children per first query charged to the fragment, and hence at most $(D+1)Q$ vertices; distinct edges have distinct (position, list index) addresses, and for an edge whose list index lies in $G_t$, its candidate rank is determined by $Z$. The edges of the tree are directed from parent to child, and every path in it is a \emph{downward} path, that is, a directed path from a vertex to one of its descendants; its length is its number of edges. We call an edge \emph{retained} if its list index lies in $G_t$ and its candidate rank lies in $\Theta_t$. Let $\tau$ be the maximum number of edge-disjoint retained paths of length $h_j$, and let the \emph{path cut} $\widehat\chi$ be the minimum number of edges meeting every retained path of length $h_j$; then $\widehat\chi\le h_j\tau$.

Fix a finite word $w$ over $\Sigma$. Whenever $(Z,w)$ lies in the support of $(Z,\mathcal M)$ conditional on $(s,o)$, let $T_w$ be the discovery tree determined by $(s,o)$ and $\Dec(s,o,Z,w)$, with the candidate ranks given by $Z$; otherwise, let $T_w$ consist only of its root. Thus, for each fixed $w$, the tree $T_w$ is a function of $Z$, has at most $(D+1)Q$ vertices, and has distinct (position, list index) addresses on distinct edges. Let $\tau_w$ be the maximum number of edge-disjoint retained paths of length $h_j$ in $T_w$. Since $\Theta$ is independent of $Z$, it keeps its prior law given $Z$, and so a set of edges of $T_w$ requiring $a_t$ distinct ranks at position $t$ is retained with conditional probability at most $\prod_t(d_j)_{a_t}/(d_{j+1})_{a_t}\le\theta_j^{\sum_ta_t}$, since $(d_j-b)/(d_{j+1}-b)\le\theta_j$. A downward path of length $h_j$ is determined by its last vertex, since it consists of the last $h_j$ edges of the path from the root to that vertex, so there are at most $(D+1)Q$ such paths, and $i$ edge-disjoint paths are all retained with probability at most $\theta_j^{ih_j}$. Hence, since $\theta_j^{h_j}\le N^{-80}$ and $(D+1)Q\le N$,
\[
 \Pr[\tau_w\ge i\mid s,o]\le\bigl((D+1)Q\,\theta_j^{h_j}\bigr)^i\le N^{-79i}.
\]
By Lemma~\ref{os:decoder}, the actual tree is $T_{\mathcal M}$, and so $\tau=\tau_{\mathcal M}$. Since $|\Sigma|=N^{C_\eta}$, there are at most $2N^{C_\eta\ell}$ words of length at most $\ell$. For $\ell=\lfloor39i/C_\eta\rfloor$, a union bound over these words gives
\[
 \Pr[\tau\ge i\mid s,o]
 \le\Pr\bigl[|\mathcal M|>\ell\bigm|s,o\bigr]+2N^{C_\eta\ell-79i}
 \le\Pr\bigl[C_\eta|\mathcal M|\ge39i\bigm|s,o\bigr]+2N^{-40i}.
\]
Summing over $i\ge1$ gives $\E(\tau\mid s,o)\le(C_\eta/39)\,\E(|\mathcal M|\mid s,o)+O(N^{-40})$, and since $\widehat\chi\le h_j\tau$ and $h_j=O(\log N)$, we obtain
\begin{equation}
\label{os:cut-expectation}
 \E(\widehat\chi\mid s,o)\le C_\eta\log N\,\E(|\mathcal M|\mid s,o)+O(N^{-30}).
\end{equation}

Let $\chi_{\phi,j}$ be the minimum number of edges meeting all downward paths of length $h_j$ in the part of the discovery tree of $\phi$ inside $U_j$ in the actual run. Every actual discovery edge from $U_j$ to $U_j$ is retained (steps out of the last layer of $X_j$ only remove such edges), so every downward path of length $h_j$ in this part is a retained path and is met by every retained cut; hence $\chi_{\phi,j}\le\widehat\chi$. By Lemma~\ref{os:continuation-law}, \eqref{os:cut-expectation} also holds for $\chi_{\phi,j}$, and by (C3), we may sum it over the births of the fragments of $\mathcal F$.

\begin{proof}[Proof of Lemma~\ref{os:local}]
A first query with source in $U_j$ that is charged to $\phi$ goes to a vertex owned by $\phi$, so $\mathrm{work}_j(\phi)$ is at most the number of vertices in the part of the discovery tree of $\phi$ inside $U_j$. Removing $\chi_{\phi,j}$ edges from this part (which is connected by monotonicity) leaves at most $1+\chi_{\phi,j}$ rooted subtrees of height less than $h_j$, since every downward path of length $h_j$ loses an edge. Each of them lies inside the $U_j$ out-ball of its root, which has at most $N^{\zeta/2}$ vertices by Lemma~\ref{os:growth}. Hence
\[
 \mathrm{work}_j\le N^{\zeta/2}\Bigl(|\mathcal F|+\sum_{\phi\in\mathcal F}\chi_{\phi,j}\Bigr).
\]
Taking expectations and using \eqref{os:cut-expectation} and \eqref{os:record-count} gives
\[
 \E\,\mathrm{work}_j\le N^{\zeta/2}\Bigl(f+C_\eta\log N\,(\Delta\E\,\mathrm{work}_{j+1}+\E\,\mathrm{ret})+O(QN^{-30})\Bigr),
\]
which implies \eqref{os:local-rec}, since $C_\eta\log N\le N^{\zeta/2}$ for large $N$.

It remains to prove \eqref{os:retirement-rec}. The list of an owned vertex in $U_1$ makes an upper contact with conditional probability at most $C_\eta Q/W$. For incoming contacts, the objects in $U_1$ owned by active fragments of $\mathcal F$ number at most $|\mathcal F|+D\,\mathrm{work}_1$, so summing over at most $DQ$ entries gives at most $C_\eta(Q/W)(f+\E\,\mathrm{work}_1)$. Retirements at cutoffs charge disjoint sets of $T$ first queries in $U_1$. Therefore $\E\,\mathrm{ret}\le C_\eta(Q/W+1/T)(\E\,\mathrm{work}_1+f)\le\Delta(\E\,\mathrm{work}_1+f)$.
\end{proof}

\subsection{Iteration over levels and generations}
\label{os:generations}

\paragraph{Levels.}
The local estimate is applied twice, with a different purpose each time. We first iterate it over the levels $j=1,\ldots,k$, which discounts the remaining query budget, and then we apply the resulting bound across fragment generations. Fix a family decided at birth, and write $x_j=\E\,\mathrm{work}_j$, $y=\E\,\mathrm{ret}$, and $\lambda =N^\zeta\Delta\le\frac14$. Iterating \eqref{os:local-rec} for $j=1,\ldots,k$ and using \eqref{os:work-cap}, we obtain
\[
 x_1\le\sum_{i<k}\lambda^i\bigl(N^\zeta(f+y)+N^{-18}\bigr)+\lambda^kQ
 \le2N^\zeta(f+y)+\lambda^kQ+2N^{-18}.
\]
Substituting $y\le\Delta(x_1+f)$ and moving $2\lambda x_1$ to the left,
\[
 x_1\le\frac{2N^\zeta(1+\Delta)f+N^{k\zeta}\Delta^kQ+2N^{-18}}{1-2\lambda}.
\]
Let $\sigma=(k+2)\zeta$. For large $N$, the coefficient of $f$ satisfies $2N^\zeta(1+\Delta)/(1-2\lambda)+1\le N^\sigma$. Thus this bound, together with its insertion into $y\le\Delta(x_1+f)$, gives
\begin{align}
 x_1&\le N^\sigma\bigl(f+\Delta^kQ\bigr)+4N^{-18},
       \label{os:collapsed-work}\\
 y&\le N^\sigma\bigl(\Delta f+\Delta^{k+1}Q\bigr)+4\Delta N^{-18}.
       \label{os:collapsed-retire}
\end{align}

\paragraph{Generations.}
Retired parents disclose which of their owned vertices are blue, and so they allow dependent fragments to start at these vertices. To control this effect, let $\mathrm{ret}_g$ be the number of retired generation-$g$ fragments, and let $f_g$ be the expected number of generation-$g$ births. Generation $g$ is decided at birth, so \eqref{os:collapsed-retire} applies to it. From \eqref{os:retirement-rec}, $\mathrm{work}_1\le Q$, and $f_0\le Q$, we get $\E\,\mathrm{ret}_0\le2\Delta Q$. On the event of Lemma~\ref{os:blue-size}, a retired parent has at most $\Sf$ dependent births, since each of them uses a distinct announced blue vertex. As there are at most $Q$ births in total, we obtain $f_g\le\Sf\E\,\mathrm{ret}_{g-1}+QN^{-20}$ for $g\ge1$. Therefore
\begin{equation}
\label{os:generation-rec}
 \E\,\mathrm{ret}_g\le\omega\,\E\,\mathrm{ret}_{g-1}+\nu,
 \qquad \omega=N^{\sigma+\zeta}\Delta,\qquad
 \nu =N^\sigma\Delta^{k+1}Q+N^\sigma\Delta QN^{-20}+4\Delta N^{-18},
\end{equation}
and, since $\omega=o(1)$,
\begin{equation}
\label{os:deep-retirements}
 \E\,\mathrm{ret}_k\le2\omega^k\Delta Q+2\nu =o(1).
\end{equation}
Indeed, the exponents of $\omega^k\Delta Q$ and of $N^\sigma\Delta^{k+1}Q$ are at most $\alpha-(k+1)\beta+k(\sigma+\zeta)+o(1)$ and $\alpha-(k+1)\beta+\sigma+o(1)$, respectively, and both are below $-2$, because $\alpha=1-4\beta$, $k\ge4/\beta$, and $k(\sigma+\zeta)<\beta/100$. All other terms are $O(N^{-18})$.

\paragraph{Clans.}
By Markov's inequality and \eqref{os:deep-retirements}, with probability $1-o(1)$ no generation-$k$ fragment retires, and therefore no generation-$(k+1)$ fragment is born. On this event and on the event of Lemma~\ref{os:blue-size}, a clan has at most $\Sf^g$ fragments of generation $g$, each of which owns at most $\Sf$ blue vertices. Thus every clan has at most
\begin{equation}
\label{os:clan-size}
 \Sc=2\Sf^{k+1}
\end{equation}
blue vertices. Moreover, since every dependent root is a vertex of the blue discovery tree of its parent, a path from the root of the discovery tree of a clan passes through at most $k+1$ fragments, each of which contributes at most $h_0+1$ edges; hence this tree has height at most $(k+1)(h_0+1)$.

\subsection[Contacts between clans and cycle closure]{Contacts between clans and cycle closure}
\label{os:chronology}

The generation argument bounds individual clans, and we now control the components that are formed when clans meet. To keep all target sets predictable, we stop the exposure as soon as some clan exceeds $\Sc$ vertices, or some weak component of the observed blue graph meets $r$ clans. Observe that, before the triggering occurrence is exposed, every clan has at most $\Sc$ vertices and every component meets fewer than $r$ clans.

\paragraph{Connected groups of clans.}
The order of clan contacts is adaptive, so we cannot simply multiply the unconditional probabilities of contacts. Instead, we describe each potential contact by birth indices and positions. A weak component meeting at least $r$ clans contains a connected set of exactly $r$ clans, and such a set has a \emph{witness} consisting of the following data: the birth indices of $r$ distinct clans; a labeled spanning tree on them; for each tree edge, an orientation, which names the source clan; and for each tree edge, a pair consisting of the ordinal of a first query to a blue source within the source clan and a list index (there are at most $D\Sc$ choices before the stop). The witness requires that the designated occurrence has its target in the current seen part of the target clan. Since occurrences between different clans have old targets, a witness exists whenever the event occurs. For fixed witness data, each designated target set is predictable, has at most $\Sc$ elements, and is empty if the target clan is not yet born; hence, by Lemma~\ref{os:properties}(iii), each constraint holds with conditional probability at most $p=\Sc/W$. We may assume that the $r-1$ designated positions are distinct, since one occurrence cannot have its target in two clans, and time advances by one exposed entry, so that each step exposes at most one designated position. Let $\ell_t$ be the number of designated positions exposed by time $t$. Then
\[
 p^{-\ell_t}\,\ind\{\text{all designated constraints exposed by time $t$ hold}\}
\]
is a nonnegative supermartingale with initial value $1$. On the event that all $r-1$ designated positions exist and their constraints hold, its final value is $p^{-(r-1)}$, and so, by optional stopping at the bounded horizon, this event has probability at most $p^{r-1}$. There are at most $Q^r$ choices of birth indices, $r^{2(r-1)}$ spanning trees, $2^{r-1}$ orientations, and $(D\Sc)^{r-1}$ query--index pairs; therefore
\begin{equation}
\label{os:clan-witness}
 \Pr[\text{a component meets $r$ clans before the stop}]
 \le r^{2(r-1)}(2D)^{r-1}Q^r\left(\frac{\Sc^2}{W}\right)^{r-1}.
\end{equation}
The exponent of $N$ on the right is
\begin{equation}
\label{os:clan-exponent}
 \alpha+(r-1)\bigl(-3\beta+2(k+1)\zeta\bigr)+o(1)<0,
\end{equation}
since $2(k+1)\zeta<\beta/500$, $(r-1)\beta>2$, and $\alpha<1$.

\paragraph{Cycle closure.}
The contact bound also lets us rule out weak cycles. Before the stop, a component has at most $r\Sc$ vertices. A newly exposed blue occurrence closes a weak cycle (including loops and repeated or opposite occurrences) only if its target lies in the current component of its source, and by (C1) this has probability at most $r\Sc/W$. Summing over at most $DQ$ entries, we obtain
\begin{equation}
\label{os:cycle-closing}
 \Pr[\text{an exposed blue weak cycle before the stop}]
 \le\frac{DQr\Sc}{W}=N^{-3\beta+(k+1)\zeta+o(1)}=o(1).
\end{equation}

\begin{proof}[Proof of Theorem~\ref{os:structural}]
Outside the events bounded in Lemma~\ref{os:blue-size}, \eqref{os:deep-retirements}, \eqref{os:clan-witness}, and \eqref{os:cycle-closing}, no stop occurs, and the observed blue graph is a weak forest in which every component meets fewer than $r$ clans and has at most $r\Sc\le N^{(k+3)\zeta}=N^z$ vertices. Within a component, the forest property implies that each clan induces exactly its discovery tree, which has diameter at most $2(k+1)(h_0+1)$, and contracting the clans gives a tree on fewer than $r$ vertices. Thus every weak path consists of at most $r$ clan subpaths and $r-1$ connecting edges, and so its length is at most $\Lambda$. The failure probability is at most
\begin{equation}\label{os:structural-failure}
 \delta_N=N^{-20}+2\omega^k\Delta Q+2\nu
 +r^{2(r-1)}(2D)^{r-1}Q^r\left(\frac{\Sc^2}{W}\right)^{r-1}
 +\frac{DQr\Sc}{W}=o(1),
\end{equation}
a function of the parameters only.
\end{proof}

\section{Concluding remarks and open problems}
\label{sec:conclusion}

We showed that, for every fixed $\eta\in(0,1)$, there is a proximity parameter $\eps_\eta>0$ at which testing directed acyclicity requires $\Omega(n^{1-\eta})$ queries at every constant degree bound $d\ge2$, even for adaptive testers with two-sided error. Our proof combines two ingredients that may be of independent interest. The first is an information-theoretic accounting in which the information that an algorithm gains from a collision is charged only to the local search that it affects, and which can therefore be discounted through a hierarchy of masking regions. The second is an exact comparison of connected collision patterns of bounded order in high-girth cores, which allows us to compare complete transcripts far beyond the birthday scale. 

The main open question is to determine the tight query complexity of testing directed acyclicity. In particular, is there a tester with $o(n)$ queries for a fixed pair $(d,\eps)$, or are $\Omega(n)$ queries necessary? More broadly, both ingredients of our proof address settings in which an algorithm necessarily encounters many collisions, as is typical in near-linear lower bounds for sparse models. It would be interesting to see whether these techniques, particularly the comparison of collision patterns to high order, can yield stronger two-sided lower bounds for other directed graph properties in the unidirectional model or for testing problems in other sparse models.

\clearpage
\appendix
\addtocontents{toc}{\protect\setcounter{tocdepth}{1}}
\addcontentsline{toc}{section}{Appendix}

\centerline{\huge{\textbf{Appendix}}}

\section{Random high-girth cores: proofs}
\label{app:core-estimates}

In this appendix, we prove the three facts about random high-girth cores stated in Section~\ref{rc:section}. Throughout, all cycles are weak cycles in $\Gamma(\Pi)$. To bound the cost of conditioning on large girth, we first count short cycles in an unconditioned core, and then we show that a controlled sequence of switches removes them. The same switches give the atom bound, and a lower tail for balanced cuts gives farness.

\subsection{Short cycles, safe switches, and the conditioning cost}
\label{app:core-conditioning}

\begin{lemma}[Cycle count]
\label{rc:cycle-count}
Let $\mathrm{cyc}_{\le\gamma}(\Pi)$ be the number of weak cycles of length at most $\gamma$ in $\Gamma(\Pi)$. Under \eqref{rc:girth-scale}, for uniform independent permutations and all large $N$, $\E\,\mathrm{cyc}_{\le\gamma}(\Pi)\le4\cdot64^{\gamma}$.
\end{lemma}

\begin{proof}
We describe a cycle of length $j$ by a cyclic list of vertices together with, for each step, a color and a direction of traversal. There are at most $(64N)^j$ such descriptions, and we keep those that use $j$ distinct occurrences (for $j=2$, this excludes traversing one occurrence back and forth). Every cycle has at least one description. A description imposes, for each color $a$, $s_a$ distinct matches with $\sum_as_a=j$; these hold with probability $0$ if they conflict, and with probability $\prod_a1/(N)_{s_a}\le(N-\gamma)^{-j}$ otherwise. Since $\gamma=O(\log N)$, we have $(N/(N-\gamma))^j\le e^{2\gamma^2/N}\le2$, and so $\E\,\mathrm{cyc}_{\le\gamma}(\Pi)\le2\sum_{j\le\gamma}64^j\le4\cdot64^\gamma$.
\end{proof}

\begin{lemma}[A switch that creates no short cycle]
\label{rc:safe-switch}
Let $v\to w$ and $x\to y$ be occurrences of the same color in a permutation tuple such that (1) both $x$ and $y$ have distance greater than $\gamma$ from $\{v,w\}$, and (2) the occurrence $x\to y$ lies on no cycle of length at most $\gamma$. If we replace them by $v\to y$ and $x\to w$, with the same color, then the result is again a permutation tuple, and the switch creates no new cycle of length at most $\gamma$. This remains true if $v\to w$ is a loop or lies on a cycle of length two.
\end{lemma}

\begin{proof}
By (1), we have $x\ne v$, so exchanging the images of $v$ and $x$ preserves the permutation. By (2), we have $x\ne y$, and by (1), neither new occurrence is a loop. Let $\Gamma'$ be the multigraph obtained by deleting the two old occurrences. A new short cycle that uses only $vy$ would contain a $v$--$y$ path in $\Gamma'$ of length at most $\gamma-1$, which contradicts (1); the same argument applies to $xw$, and it also excludes new parallel pairs. Consider next a cycle that uses both new occurrences. If $v\ne w$, then removing the two new occurrences leaves two paths in $\Gamma'$, which pair the endpoints either as $(v,x),(y,w)$ or as $(v,w),(y,x)$. In the first case, a path joins $\{x,y\}$ to $\{v,w\}$, and so it has length greater than $\gamma$. In the second case, an $x$--$y$ path avoiding $xy$ has length at least $\gamma$ by (2), and so the cycle has length at least $\gamma+2$. Finally, if $v=w$, then the cycle consists of the two new occurrences and an $x$--$y$ path avoiding $xy$, and again its length is at least $\gamma+2$.
\end{proof}

\begin{proof}[Proof of Proposition~\ref{rc:existence}]
Let $\tau_\gamma=\lceil8\cdot64^\gamma\rceil$. By Lemma~\ref{rc:cycle-count} and Markov's inequality, at least half of the $(N!)^{32}$ tuples have at most $\tau_\gamma$ short cycles. We map each such tuple into $\mathcal G_\gamma$ by at most $\tau_\gamma$ switches. If a short cycle exists, we choose one of its occurrences $v\to w$, of color $a$. The radius-$\gamma$ balls around $v$ and $w$ contain at most $C64^\gamma$ vertices, so at most $C64^\gamma$ color-$a$ occurrences have an endpoint in them; moreover, at most $\gamma\tau_\gamma$ occurrences lie on short cycles. Since $C64^\gamma+\gamma\tau_\gamma<N$ by \eqref{rc:girth-scale}, some color-$a$ occurrence $x\to y$ satisfies the conditions of Lemma~\ref{rc:safe-switch}. The switch destroys the chosen cycle and creates no short cycle, and therefore the number of short cycles decreases. Fixing deterministic choices makes this procedure a well-defined map.

To count the preimages of this map, observe that a switch is determined by its color and its two sources, and that it is an involution; hence a resulting tuple, together with a reverse word of at most $\tau_\gamma$ switches, determines the initial tuple. There are at most $(32N^2)^{\tau_\gamma+1}$ such words, and so $|\mathcal G_\gamma|\ge(N!)^{32}/(2(32N^2)^{\tau_\gamma+1})$. Thus $\Pr[\mathcal G_\gamma]\ge\frac12(32N^2)^{-\tau_\gamma-1}\ge\exp(-C\,64^\gamma\log N)$, which is $e^{-o(N)}$ by \eqref{rc:girth-scale}.
\end{proof}

\begin{proof}[Proof of Lemma~\ref{rc:atom}]
We compare the tuples with $\pi_a(v)=w$ with the other tuples by switches: each of the former can be switched in many ways to one of the latter, and each switched tuple arises at most once. Let $\Omega$ be the set of extending tuples in $\mathcal G_\gamma$, and let $\Omega_w\subseteq\Omega$ be the set of those with $\pi_a(v)=w$. Fix a tuple in $\Omega_w$. We call a color-$a$ occurrence $x\to y$ a \emph{partner} if its match is unexposed and both $x$ and $y$ are at distance greater than $\gamma$ from $\{v,w\}$. The two radius-$\gamma$ balls contain at most $C64^\gamma$ vertices, so there are at least $N-m_a-C64^{\gamma}$ partners. Since the tuple lies in $\mathcal G_\gamma$, no occurrence lies on a short cycle, and so, by Lemma~\ref{rc:safe-switch}, the switch to $v\to y$, $x\to w$ yields a tuple in $\mathcal G_\gamma$. The switch changes only unexposed matches, so the new tuple lies in $\Omega$, and in fact in $\Omega\setminus\Omega_w$, because $y\neq w$. Distinct partners give distinct images of $v$. Conversely, a switched tuple determines the partner source $x$ as the color-$a$ preimage of $w$, and exchanging the images of $x$ and $v$ recovers the starting tuple. Double counting now gives $(N-m_a-C64^\gamma)|\Omega_w|\le|\Omega\setminus\Omega_w|$, and hence $|\Omega_w|/|\Omega|\le1/(N-m_a-C64^\gamma+1)$.

For the sequential statement, observe that, given the history, each choice of the next pair is a function of the preceding answers and of the inspected auxiliary randomness. Thus the choices impose no constraint on unexposed matches, and by induction, $\Pi$ is uniform on the compatible extensions in $\mathcal G_\gamma$. The bound then follows from the first part, together with $q=o(N)$ and $64^\gamma=o(N)$, and a union bound over the $s$ specified vertices gives $2s/N$.
\end{proof}

\subsection[Balanced cuts]{Balanced cuts}
\label{app:core-farness}

To prove farness, we first show that, under the unconditioned permutation distribution, every balanced set sends many occurrences to its complement. The following estimate for sampling without replacement gives an exponentially small lower tail, which is strong enough to absorb the cost of the girth conditioning.

\begin{lemma}[A symmetric-mean inequality]
\label{rc:symmetric-mean}
For $z_1,\ldots,z_N\ge0$ and a uniform $s$-subset $I\subseteq[N]$, $\E\prod_{i\in I}z_i\le\bigl(\frac1N\sum_iz_i\bigr)^s$.
\end{lemma}

\begin{proof}
This is a form of Maclaurin's inequality; we include a short proof. The cases $s\le1$ are immediate. For $s\ge2$, let $e_s$ be the elementary symmetric polynomial of degree $s$. As a function of two coordinates $x,y$, it equals $xy\,e_{s-2}'+(x+y)e_{s-1}'+e_s'$, with nonnegative coefficients that depend on the other coordinates; hence replacing $x,y$ by their average does not decrease it. Repeatedly averaging a largest and a smallest coordinate converges to the constant vector (each step decreases the sum of squared deviations by half the squared difference of the pair), and so, by continuity, $e_s(z)\le\binom Ns(\frac1N\sum_iz_i)^s$. Dividing by $\binom Ns$ completes the proof.
\end{proof}

\begin{lemma}[Lower tail of balanced cuts]
\label{rc:cut-tail}
Let $V_1\subseteq[N]$ have size $N/2$, and let $\mathrm{cut}(V_1)$ be the number of occurrences of $\Gamma(\Pi)$ from $V_1$ to its complement, for uniform independent permutations. Then $\Pr[\mathrm{cut}(V_1)\le4N]\le e^{-N}$.
\end{lemma}

\begin{proof}
For one color, $\pi_a(V_1)$ is a uniformly random $N/2$-subset. If we give weight $1$ to the vertices of $V_1$ and weight $e^{-t}$ to the vertices of its complement, then Lemma~\ref{rc:symmetric-mean} yields $\E e^{-t\,\mathrm{cut}_a(V_1)}\le(1-\frac12(1-e^{-t}))^{N/2}$ for the cut count $\mathrm{cut}_a(V_1)$ in color $a$. By the independence of the colors and the inequality $1-u\le e^{-u}$, we obtain $\E e^{-t\,\mathrm{cut}(V_1)}\le\exp(-8N(1-e^{-t}))$. Markov's inequality therefore bounds the lower tail by $\exp(4Nt-8N(1-e^{-t}))$, and the choice $t=\frac12$, together with $e^{-1/2}\le\frac58$, gives the exponent $2N-3N=-N$.
\end{proof}

\begin{proof}[Proof of Proposition~\ref{rc:farness}]
For a set $V_1\subseteq[N]$ of size $N/2$, let $\mathrm{cut}(V_1)$ be the number of occurrences from $V_1$ to its complement. Its mean is $32\cdot\frac N2\cdot\frac12=8N$, and Lemma~\ref{rc:cut-tail} gives $\Pr[\mathrm{cut}(V_1)<4N]\le e^{-N}$ without conditioning. A union bound over the at most $2^N$ such sets shows that, except with probability $e^{-(1-\log2)N}$, every set of size $N/2$ sends at least $4N$ occurrences to its complement. To turn this cut bound into a bound on deletions, suppose that, on this event, deleting a set $F$ of occurrences leaves an acyclic graph. Fix a topological order of the remaining graph, and let $V_1$ be its last $N/2$ vertices. Every occurrence from $V_1$ to its complement points backwards, and so it must lie in $F$; thus $|F|\ge4N$. Finally, dividing the failure probability $e^{-(1-\log2)N}$ by the lower bound $\Pr[\mathcal G_\gamma]\ge e^{-o(N)}$ of Proposition~\ref{rc:existence} shows that the conditional failure probability is $e^{-\Omega(N)}$. Since $\Gamma(\Pi)$ is simple on $\mathcal G_\gamma$, the bound $|F|\ge4N$ also gives $\fas(\Gamma(\Pi))\ge4N$.
\end{proof}

\section[The height of fragments with blue roots]{The height of fragments with blue roots}
\label{app:blue-erasure}

In this appendix, we prove the height estimate used in Lemma~\ref{os:blue-size}. In contrast to the union bound of Section~\ref{os:reconstruction}, which controls the expected work, this estimate bounds a failure probability at \emph{every} birth state. Since a dependent birth may be arbitrarily rare, the comparison must be multiplicative. We obtain such a comparison on execution prefixes, and then we sum it at the first appearance of a long blue path.

Recall that a fragment with a blue root owns its fresh descendants until it retires, which happens after the answer containing its first upper contact (where $U_1=B\cup X_1$) or the answer to its $T$-th first query with source in $U_1$ (the cutoff). The local experiment ends after the retiring answer and before the ensuing announcements; if there is no retirement, it ends when the algorithm stops or at query $Q$, the horizon.

\begin{lemma}[Conditional estimate of the blue height]
\label{lem:conditional-blue-height}
Assume $Q=o(N)$ and $QT/N=o(1)$. Let $s$ be a birth state at which the root has been matched to $o\in B$, but its category word has not been read and its list is not yet exposed. Uniformly over deterministic adaptive algorithms, over such states, and over $h\ge1$,
\begin{equation}\label{eq:appendix-blue-height}
 \Pr[\text{the fragment has blue height at least }h+2\mid s,o]
 \le \exp(C_\eta QT/N)\,(D+1)Q\,\theta_0^h.
\end{equation}
\end{lemma}

\begin{proof}
\emph{Overview.} The proof makes the blind search of the warm-up in Section~\ref{sec:ov-info} exact. We construct an erased execution, called the \emph{skeleton}, in which every candidate entry of an upper list of the fragment returns a fresh vertex. The skeleton therefore never learns which candidates are selected, so its discovery tree is independent of the private selector masks, and long selected paths are rare in it. The real execution agrees with the skeleton until the fragment makes an upper contact or reaches its cutoff, and then the fragment retires; hence a long blue path of the real fragment already appears while the two executions agree. We compare the probabilities of such common prefixes multiplicatively. The only unequal factor comes from exterior matches of unseen IDs, because the skeleton excludes only the objects in its stored matching, and it costs at most a factor $\exp(C_\eta QT/N)$. (We note that agreement is used only to choose which prefixes to compare; we never condition on eventual agreement.)

Write $\theta=\theta_0=32/d_1$. By \eqref{os:posterior}, we may sample the whole posterior core $\Pi\sim\psi_s$ once, and both executions read core entries from this table. In this appendix, \emph{upper} means $U_1$, and \emph{exterior} means not owned by the distinguished fragment. The skeleton below may produce histories that no real input produces; we extend the algorithm's query rule arbitrarily to such histories, keeping its query bound; this changes no real run.

\paragraph{The skeleton.}
We use the terminology of resolved and unresolved vertices from Section~\ref{os:reconstruction} for the stored ID--object matching of the skeleton, which runs the algorithm with the following rules.
\begin{itemize}
\item \emph{Resolved vertices.} The root $o$, all vertices seen at the birth, all exterior vertices, and all owned vertices outside $U_1$ are resolved. For the other owned vertices, the skeleton keeps only their IDs and their discovery and ownership records. The root counts as an upper source of the fragment.
\item \emph{Upper lists of the fragment.} At the first exposure of such a list, the skeleton chooses a uniform $d_1$-set of candidate indices. The other indices receive a uniform arrangement of the categories $\ge2$ and independent targets in the corresponding bands; this template is the same for blue sources and for $X_1$ sources. Every candidate index creates a new unresolved child with a fresh uniform unused ID.
\item \emph{Other lists.} They use the objects of resolved vertices, independent category words indexed by query position, and the table $\Pi$ for core lookups. Target lookups consult only the stored matching.
\item \emph{Unseen IDs.} A query to an unseen ID is matched uniformly among the objects outside the stored matching; the private objects defined below are not excluded. Exterior ownership and announcements use the stored matching and the public records.
\item \emph{Stopping.} The skeleton stops after the $T$-th upper list of the fragment, at public stopping or at the horizon, or after a complete exterior answer that contacts the resolved root $o$; in each case, it stops before the following announcements.
\end{itemize}
Observe that the skeleton never tests equalities with private objects. Hence the skeleton, and in particular its discovery tree, does not depend on the private choices described next.

\paragraph{Private assignments.}
Independently for every upper list of the fragment, we choose a uniform $32$-subset of its $d_1$ candidate ranks. If the source object is blue, then the selected entries use $\pi_1,\ldots,\pi_{32}$ in increasing order, and the other candidates use independent uniform objects of $E_1$; if the source object lies in $X_1$, then all candidates use independent uniform objects of the next layer. Starting from $o$, this assigns objects, which we call \emph{private objects}, to the unresolved vertices along the discovery tree. We realize the private choices in interleaved order, at each upper list of the fragment right after its public template, so that every prefix and every first witness depend only on choices already made. This gives the same joint law as sampling the whole skeleton first, because no step of the skeleton reads a private choice. If a private assignment repeats an object, we continue artificially without repair (distinct vertices keep distinct IDs, and the core list of a repeated object is still read from $\Pi$). Before a discrepancy, distinct first exposures have distinct objects and unused words, so the laws of the category words and endpoint samples agree with the conditional input. Finally, the entry vertex of a component of private objects in $X_1$ lies in the first $T$ layers; hence, before the cutoff, no path through it reaches layer $3T$, and no unknown list of the last layer of $X_1$ affects the skeleton.

\paragraph{Common prefixes.}
We compare the two executions through the probabilities of their finite prefixes. A \emph{prefix} is a finite initial list of read choices (masks, endpoint samples, and the matches made by priorities, where each priority is integrated out after the match that it makes), and its \emph{cylinder} is the event that an execution begins with these choices, whatever the unread variables are. We compare only \emph{common prefixes}, on which the two executions have not yet disagreed and the fragment has not completed its retiring answer. On such a prefix, the two executions have the same public history, the same stored matching, and the same caches. The private objects are distinct and disjoint from the stored matching; let $b$ be their number, and let $u$ be the number of objects outside the stored matching. Since at most $T$ upper sources of the fragment have been queried, we have
\[
 b\le DT,\qquad u-b\ge n_0-(D+1)Q-DT=\Omega(N).
\]
For an exterior query to an unseen ID, the real execution chooses uniformly among the $u-b$ genuinely unmatched objects, whereas the skeleton chooses uniformly among $u$ objects. Thus, for every common match, that is, one that the real execution can make, the ratio of the two probabilities is $u/(u-b)$. (A skeleton match to a private object has no real counterpart, and it is omitted from the sums over common prefixes.) All other factors agree. Indeed, for a fixed $\Pi$, every core lookup in a common prefix has the same source and color, so the whole density $\psi_s(\Pi)$ is a common factor; category words agree, since the first queries of the real execution read distinct sources; fresh IDs have the same law, because the registries of seen IDs agree; and repeated queries contribute nothing. Exterior announcements are functions of the stored matching, unless they involve an unresolved vertex, and such an event is an upper contact that ends the local experiment. Multiplying over at most $Q$ exterior matches and using $\log(1+x)\le x$, we obtain
\begin{equation}\label{eq:appendix-prefix-ratio}
 \frac{\Pr_{\rm real}[\text{common prefix}\mid s,o]}
      {\Pr_{\rm skel}[\text{same prefix}\mid s,o]}
 \le\prod\frac{u}{u-b}\le\exp(C_\eta QT/N).
\end{equation}

Apart from stopping, the horizon, and the cutoff, agreement can end only in the following ways: an upper target of the fragment equals an old object in $U_1$; an upper target of an exterior list equals a private object; the match of an exterior query to an unseen ID is a private object (which has no real counterpart); or an upper entry of an exterior list contacts the root. Each case other than the third one is an upper contact incident to the fragment, which then retires after the current answer. The third case has no real counterpart; hence, in the real execution, agreement can end only at the final answer of the local experiment. All other repetitions of IDs are reproduced exactly by the skeleton: contacts among resolved vertices, old lower targets (including repetitions within one answer), and the answers to repeated queries, which are copied from the cache. Indeed, every seen lower vertex is resolved, and lower targets lie outside $U_1$, so they never equal a private object.

\paragraph{Rare paths.}
The skeleton construction lets us estimate blue paths without conditioning on agreement. Condition on the whole skeleton. The discovery tree of the fragment in the skeleton has at most $(D+1)Q$ vertices, and a downward path of length $h$, from a vertex to one of its descendants, is determined by its last vertex. Along a blue path, the prescribed candidate rank must be selected at each of its $h$ sources, which have distinct query positions. The private masks are independent of the skeleton, and each selection has probability $\theta$; hence
\begin{equation}\label{eq:appendix-skeleton-path}
 \Pr_{\rm skel}[\text{some length-$h$ path has all its ranks selected}\mid s,o]
 \le (D+1)Q\,\theta^h.
\end{equation}
We note that this bound requires neither injectivity of the private assignment nor any conditioning on agreement or on the absence of contacts; indeed, such conditioning could bias the masks.

It remains to transfer this bound to the real fragment. If the actual fragment has blue height at least $h+2$, then a blue discovery path of length $h$ existed before its final answer, since one answer increases the height by at most one, also when it contains the retiring contact or the cutoff. Thus the height event is contained in the event $\mathcal W_h$ that a blue path of length $h$ appears while the executions agree and before the comparison ends. We partition $\mathcal W_h$ according to its first witness into the disjoint cylinders of common prefixes. Their skeleton counterparts are disjoint, since the first witness is determined by the prefix, and they lie in the event of \eqref{eq:appendix-skeleton-path}. Summing \eqref{eq:appendix-prefix-ratio} over these cylinders gives $\Pr[\mathcal W_h\mid s,o]\le\exp(C_\eta QT/N)(D+1)Q\theta^h$, as required.
\end{proof}

\section{Details of the local reconstruction}
\label{app:local-reconstruction}

This appendix supplies the verifications used in Section~\ref{os:reconstruction}. We first show that the allocated tapes reproduce the conditional input (Lemma~\ref{os:continuation-law}), and then we show that the records reconstruct the local history (Lemma~\ref{os:decoder}). Both proofs are careful bookkeeping of simple ideas. The first is a form of the principle of deferred decisions: the allocated experiment indexes the randomness of first queries by query position instead of by source object, and since every source has only one first query, this changes no law. The second checks, operation by operation, that the decoder can compute every answer from $Z$, except where the answer depends on the masks or on an unresolved identity, and that the records cover exactly these cases.

\subsection{Posterior factorization and allocation of the tapes}
\label{app:local-allocation}

\paragraph{The posterior factorization.}
We first verify the posterior factorization \eqref{os:posterior} directly from the likelihood of a birth history. Fix an atom of the full analytical history at a birth that has positive probability, and let $\mathfrak e_s$ and $\mathfrak m_s$ be its exposed colored equalities and its matched pairs, respectively. For every value of the complete core, the array of seeds, and the ID bijection, the atom has zero likelihood unless the core extends $\mathfrak e_s$, the bijection extends $\mathfrak m_s$, and the queried seeds have their recorded values; on this support, the likelihood is constant in all remaining coordinates. Indeed, a core observation imposes its equality, and every other elementary random choice has its fixed probability given its source, category, and list index. Moreover, a prefix of a priority has a skip set determined by the earlier matching, and decisions, contacts, counters, ownership changes, and announcements are functions of the exposed data. In particular, an incoming edge or an announcement does not inspect the unused seed of its target. Normalizing proves \eqref{os:posterior}. Since this is an exact conditional identity, no approximation error is divided by a small birth probability. (A priority that was already partly scanned at the birth keeps its inspected prefix and the conditionally uniform completion of its suffix.)

\paragraph{The allocated tapes.}
Fix a birth state $(s,o)$ of a fragment of $\mathcal F$ and a level $j$, and use the terminology and the stopping rule of Section~\ref{os:reconstruction}. We sample $\Pi\sim\psi_s$ and put it in $Z$, and every core lookup uses this table, which is indexed by sources. For every future query position $t$, we put in $Z$ a uniform $d_{j+1}$-set $G_t\subseteq[D]$, a uniform arrangement of the categories larger than $j+1$ on its complement, and a uniform ordered list $L_t$ of length $d_j$ with the prescribed counts of the categories at most $j$. Let $\Theta_t$ be an independent uniform $d_j$-subset of the ranks of $G_t$ (in increasing order of list index). At a $U_j$ source of the fragment, the selected entries receive $L_t$ in increasing order of list index, the other candidates receive category $j+1$, and the noncandidates receive the common arrangement of the categories larger than $j+1$. This is a uniform category word, since one may first choose the list indices of the low, middle, and high categories and then order the low and high lists. At a blue source, an entry of category~$0$ reads the core entry of its color, that is, of its rank among the entries of category~$0$ in the list, which need not equal its list index or its candidate rank. At an $X_{j+1}$ source of the fragment, the candidate indices receive an independent uniform arrangement of the categories at most $j+1$ from $Z$, and the answer ignores $\Theta_t$. Exterior and lower sources have independent whole category words in $Z$, and exterior blue sources also read $\Pi$.

The endpoint samples are allocated as follows. For every position, hypothetical source object, mode, list index, and category, $Z$ contains an independent sample from the destination prescribed by Definition~\ref{def:no}. At a noncandidate index, there is one common category and one common target sample, shared by all hypothetical upper sources of the fragment. This sharing is valid because every nonempty source in $U_{j+1}$ sends category $c>j+1$ to $E_c$; distinct actual entries still use independent samples. The allocation is made before the source choices are known, and it covers all modes and objects, whether or not a copy is used; the actual source and owner determine which copy is read. Thus $Z$ is an advance allocation of randomness, it is independent of $\Theta$, and it does not record which copies were used.

\begin{proof}[Proof of Lemma~\ref{os:continuation-law}]
Let $V_0$ be the sources queried at the birth and $\mathfrak m_0$ the matched pairs. For an unqueried source $v$, let $\Xi_v=(C_v,Y_v)$ be its seed (its category word and the samples of its nonblue endpoints, as in Section~\ref{os:fragments}; a point mass at the sources in the last layer of $X_{k+1}$), with law $\nu_v$; the targets of entries of category~$0$ at blue sources come from $\Pi$ and are not part of the seed. By \eqref{os:posterior} the birth law is
\[
 \psi_s(d\Pi)\,\Unif_{\mathfrak m_0}(d\ell)\prod_{v\notin V_0}\nu_v(d\Xi_v),
\]
where $\Unif_{\mathfrak m}$ is uniform on bijections extending $\mathfrak m$.

\emph{Seeds read from a block.} For a position $t$, let $\mathsf b_t$ be its whole block of list randomness, including $\Theta_t$, the common template, and all counterfactual copies (but not the core and not the priorities), with law $\Dist(\mathsf b_t)$, and let $g_{v,m}(\mathsf b_t)$ be the seed read by source $v$ in mode $m$. We claim that $(g_{v,m})_*\Dist(\mathsf b_t)=\nu_v$ for every admissible pair $(v,m)$, that is, every mode $m$ that is possible for the region of $v$. For a $U_j$ source of the fragment, a specified category word has probability
\[
 \left[\binom D{d_{j+1}}\binom{d_{j+1}}{d_j}\frac{d_j!}{\prod_{i\le j}(d_i-d_{i-1})!}\frac{(D-d_{j+1})!}{\prod_{i\ge j+2}(d_i-d_{i-1})!}\right]^{-1}
 =\frac{\prod_{i=0}^{k+1}(d_i-d_{i-1})!}{D!},
\]
where the second binomial coefficient accounts for the middle category, whose multiplicity is $d_{j+1}-d_j$. Given the word, the endpoint samples that the source reads are independent with the prescribed destinations, and entries of category~$0$ at blue sources have a dummy coordinate in the seed. The $X_{j+1}$ sources of the fragment combine their own arrangement of the categories at most $j+1$ with the common arrangement of the categories larger than $j+1$, which gives the same law of the word. Exterior and lower sources use whole independent copies. Observe that sharing a template among counterfactual sources at one position is harmless, since only one of them is read. Given the history of operations before an answer, the next unused block still has law $\Dist(\mathsf b_t)$, and a repeated query consumes no block.

\emph{Kernels.} For $\nu_v(x)>0$ define $\mathsf K_{v,m}(x,\mathsf b)=\Pr[\mathsf b_t=\mathsf b]\ind\{g_{v,m}(\mathsf b)=x\}/\nu_v(x)$; it is normalized by the claim, and
\begin{equation}\label{os:seed-density-identity}
 \nu_v(x)\mathsf K_{v,m}(x,\mathsf b)=\Pr[\mathsf b_t=\mathsf b]\ind\{g_{v,m}(\mathsf b)=x\},
\end{equation}
which keeps the whole block, including unused copies. For priorities, if $r_{\mathfrak m}$ IDs are unmatched, let $\mu$ be the uniform law of a full ID priority $p$, let $f_{\mathfrak m}(p)$ be the first unmatched ID in $p$, and let $\mathsf J_{\mathfrak m}(y,p)=r_{\mathfrak m}\mu(p)\ind\{f_{\mathfrak m}(p)=y\}$. For a fresh object $x$,
\begin{equation}\label{os:priority-density-identity}
 \Unif_{\mathfrak m}(\ell)\,\mathsf J_{\mathfrak m}(\ell(x),p)=\mu(p)\,\Unif_{\mathfrak m\cup\{(x,f_{\mathfrak m}(p))\}}(\ell),
\end{equation}
both nonzero sides being $\mu(p)/(r_{\mathfrak m}-1)!$; exchanging the roles of IDs and objects gives the identity for a query to an unseen ID.

\emph{Coupling.} We start from an input (core, seeds, and ID bijection) drawn from the birth law and augment it causally, that is, step by step as the execution proceeds. For each request for a fresh match, we draw the whole priority from $\mathsf J_{\mathfrak m}$, prescribing the match given by $\ell$; for each first query, we draw $\mathsf b_t$ from $\mathsf K_{v,m}(\Xi_v,\cdot)$. The next queried ID is a function of the public history, and its preparation (matching, starting an initial or dependent owner, and determining the region, layer, and mode) reads no unused block; in particular, exterior scans and their (E) records precede the block. Distinct IDs have distinct objects, and exposed sources are recognized by their caches; hence every first query has a new source and advances the position index once, also at an empty list. The normalized kernels leave the law of the input unchanged, and partly inspected priorities are completed under their conditional laws, which contributes a common factor.

\emph{Density identity.} Fix an atom of the expanded execution at its stop, which specifies all read choices, the matching, the assignment of sources to positions, and the stop tag. Its sources $v_1,\ldots,v_t$ and their modes are functions of earlier data, and the sources are distinct. Applying \eqref{os:seed-density-identity} at each position cancels each consumed factor $\nu_{v_i}$ exactly once and leaves $\prod_{v\text{ unread}}\nu_v$. If $r_0$ IDs are unmatched initially and $r_1$ at the end, then the priority factors telescope to $\frac1{r_0!}\prod_{r_1<i\le r_0}i=\frac1{r_1!}$, and the remaining indicator says that $\ell$ extends the final matching. What remains is the density of the allocated execution, times independent seeds at unread sources and a uniform completion of the bijection, with $\Pi$ unchanged; integrating against $\psi_s$ preserves the equality. The private state at the stop (the assignment of sources to positions, the ownership and discovery records, the caches, the inspected prefixes of the priorities, the records, and the stop tag) is a deterministic function of the expanded choices. Repeated queries read caches, repeated endpoints within an answer use the pair installed at their first occurrence, and both executions stop after the retiring answer and before its announcements. Since these atoms partition the finite execution space, summing the identities proves the lemma at every birth state of positive probability.
\end{proof}

\subsection{Verification of the decoder}
\label{app:decoder}

\begin{proof}[Proof of Lemma~\ref{os:decoder}]
Fix the birth state and the level. The decoder runs the algorithm itself and computes every answer from $Z$, except where an answer depends on the masks or on an unresolved identity; the records cover exactly these cases, and an invariant makes this precise. The decoder maintains three data structures: the public registry (discovery parents, ownership, and cached lists), a partial injective map from IDs to objects, and a queue of entry vertices of components in $X_{j+1}$ whose exposed parts remain to be replayed. At every operation boundary, we maintain the following invariant. The public bookkeeping and the inspected prefixes of the priorities agree with the actual run; every unresolved vertex is owned and lies in $U_{j+1}$; all exterior and lower vertices are resolved; a $U_j$ vertex may be resolved individually, but its lists are never replayed from $Z$ alone; in each component in $X_{j+1}$, either all exposed vertices are resolved or none of them is; and the replay queue is empty. This invariant holds at the birth. The query positions and the tags of the records in $\mathcal M$ tell the decoder which records to process at each query. We check the invariant for each kind of operation in turn.

\emph{Queries to unseen IDs.} Before examining the answer, the decoder processes the (E) records of the query, installs their pairs, and drains the replay queue. It then scans the priority. Every used object that precedes the first genuinely unmatched object is either already known or supplied by these records, and so the first eligible object is the true match. Observe that no answer record and no future priority suffix is read.

\emph{Answers with records of types (A)--(D).} The decoder first registers the complete public pattern in list order, updating the seen status, the discovery parents, the ownership, and the cache. For each fresh ID, it advances the ID priority of that entry past the IDs seen immediately before the entry, which reproduces the inspected prefix. Only then does it install the supplied pairs; in this way, an ID that is introduced by one entry and repeated at a later entry is treated as fresh at its first occurrence. A fresh $X_{j+1}$ target identified by (A) is the entry vertex of a new component in $X_{j+1}$ and its only exposed vertex, so the component is resolved; an old target in $X_{j+1}$ that is unresolved triggers (C), which supplies the entry vertex of its component. Next, if the current list is an upper list of the fragment, its noncandidate entries are evaluated from the common template in $Z$ (they are lower, and so they need neither the source object nor the mask) and bound, in list order, to the registered pattern. This resolves all lower targets, even when the source is an unresolved $X_{j+1}$ vertex that is resolved only during replay. Recorded lists whose source is exterior, lower, or a resolved $X_{j+1}$ vertex are evaluated from $Z$ at their stored positions and bound to their registered patterns, which also resolves fresh targets that are not listed in the record. (We note that the pattern, and not the partial map, determines which appearances are fresh.) A list of an unresolved $X_{j+1}$ source is evaluated during the replay triggered by (C), and a $U_j$ list of the fragment is never evaluated from $Z$ alone, since (A) supplies its mask and its target pairs. After a final (D) answer, no further invariant is needed. None of these steps adds a record.

\emph{Replay.} Starting at an entry vertex, the decoder visits the exposed part of its component. Its lists ignore $\Theta$, and their source objects and stored positions determine their raw endpoints from $Z$; these endpoints are bound to the IDs of the stored patterns. (For a past contact, the computed endpoint is bound to the recorded ID, without consulting the possibly incomplete map.) Suppose that replay reaches an unresolved seen target outside its own subtree. Then this target is not in $U_j$, by monotonicity, and it is not lower, because lower vertices are resolved; hence it lies in another component in $X_{j+1}$, and the occurrence is a contact in $U_{j+1}$. This contact is the current one, since an earlier contact would already have resolved the replayed component by (C), and so (C) has just installed the entry pair of the other component. Each exposed list is replayed at most once, and no unexposed list or future priority location is read.

\emph{Answers without records.} For a $U_j$ source of the fragment or an unresolved $X_{j+1}$ source of the fragment (for an unresolved source, the decoder need not know which of the two it is), a noncandidate entry is computed from its common category and target in $Z$, and a candidate entry returns the first unused ID of its shared priority (otherwise (A) or (C) would have been triggered); the new upper vertex stays unresolved. No unresolved source lies in the last layer of $X_{j+1}$. Indeed, an entry vertex lies in the first $T$ layers, and reaching layer $3T$ along edges inside $X_{j+1}$ takes $2T$ first queries in the component, which is resolved by (C) at its $T$-th query. Exterior sources, resolved $X_{j+1}$ sources, and lower sources have lists that are computable from $Z$. A hit of such a list on an unresolved upper vertex is covered by (B) or (C), or it is impossible by monotonicity, and lower lists cannot return to $U_j$. Answers to repeated queries are simply copied.

\emph{Announcements and stopping.} Announcements about other owners are computed from their resolved vertices. Any contact that could make such an announcement depend on the unresolved part of the distinguished fragment also retires the distinguished fragment, and the experiment stops before the announcements of that answer; record (D) supplies the retiring answer and the stop tag. Public stopping and the horizon are recognizable. Hence the invariant is maintained after every operation, and \eqref{os:decoder-identity} follows.
\end{proof}

\section{Bounding both directional degrees}
\label{app:indegree}

In the distributions of Definition~\ref{def:no}, the targets of steps and jumps are independent and uniform, and so the maximum indegree is not bounded by a constant. In this appendix, we bound it as outlined in Section~\ref{sec:construction-overview}: every masking object receives a fixed number of incoming slots, and the entries are routed to distinct random slots instead of independent random targets. A single target is then still uniform on the set that Definition~\ref{def:no} prescribes; only the joint law of the targets changes. We verify that the proof of Sections~\ref{rt:section} and~\ref{os:section} carries over. Only the constants $C_\eta$ change, together with a negligible factor in the height estimate, where $QT/N$ becomes $QT/W=N^{-2\beta+o(1)}$. The algorithm still queries outgoing lists only, and all parameters of Sections~\ref{os:params} and~\ref{os:analysis-params} are unchanged. We put
\begin{equation}\label{in:capacity}
 R=(3k+4)D.
\end{equation}

\begin{theorem}[Capacity-constrained distributions]
\label{in:main}
The distributions $\hDY$ and $\hDN$ of Definition~\ref{in:def} below have the following properties.
\begin{enumerate}[label=(\roman*)]
\item Every input has maximum outdegree at most $D$ and maximum indegree at most $2R$, and so does the simple graph obtained by replacing each list by its set of targets. Every input of $\hDY$ is acyclic. The core of $\hDN$ has the law of Section~\ref{os:sampler}, and every directed cycle of an input of $\hDN$ lies in its core.
\item Theorem~\ref{os:structural} holds for $\hDN$, with the same parameters.
\item Proposition~\ref{rt:transcript-tv} holds for $\hDY$ and $\hDN$.
\end{enumerate}
\end{theorem}

We do not derive Theorem~\ref{in:main} by comparing the new distributions with $\DY$ and $\DN$. Such a comparison would have to account for collisions within a pool (see Section~\ref{rt:likelihood}): each of them changes the probability of a transcript by a factor of about $1-1/R$, and their number may be of order $Q^2/W\gg1$. Instead, we repeat the two halves of the proof. The transcript formula acquires a new common factor, which again does not depend on the core (Section~\ref{in:likelihood}), and the exploration analysis acquires one new type of record, because the slot taken by an entry depends on the slots taken by masked entries (Section~\ref{in:exploration}).

\subsection{Incoming slots}
\label{in:construction}

\paragraph{Tokens and pools.}
The injections below must be independent of the category words. We therefore name every entry by its category and its rank within that category, so that the set of entries routed into a given pool is fixed in advance. For a nonempty list of an object $v$, a category $c$, and $1\le\iota\le d_c-d_{c-1}$, the \emph{token} $(v,c,\iota)$ is the $\iota$-th entry of category $c$ in the list of $v$, in list order. Since every category word contains exactly $d_c-d_{c-1}$ letters $c$, the set of tokens of $v$ is fixed in advance, and the category word determines only their list indices. (For a blue token of category $0$, $\iota$ is its color.) The \emph{pool} of a token is the set in which Definition~\ref{def:no} chooses its uniform target: the layer $X_{i,t+1}$ for a step from $X_{i,t}$ with $t<3T$; the band $E_{i+1}$ for a step from $X_{i,3T}$; the band $E_c$ for a jump of category $c$; and, under the YES rule, the band $E_1$ for a core entry of the last blue layer. The other core entries have no pool. Thus a pool is either a \emph{layer pool} $X_{i,t}$ with $t\ge2$ or a \emph{band pool} $E_i$; for $t\le T$, the layer pool $X_{i,t}$ is contained in the band pool $E_i$, but the two are different pools. Let $\mathcal T_P$ be the set of tokens with pool $P$; this is a deterministic set. In every pool $P$ that contains it, an object has $R$ \emph{incoming slots}, and we write $\mathcal S_P=P\times[R]$ and $L_P=|\mathcal S_P|=R|P|$.

A layer pool $X_{i,t}$ receives tokens only from the $W$ objects of $X_{i,t-1}$, so that $|\mathcal T_P|\le DW\le L_P$. A band pool receives at most $Dn_0=(3k+4)DN=L_P$ tokens, since this is the total number of entries. Hence $|\mathcal T_P|\le L_P$ for every pool, under both rules, and so the injections in the following definition exist.

\begin{definition}[Capacity-constrained distributions]
\label{in:def}
The distributions $\hDN$ and $\hDY$ are defined as $\DN$ and $\DY$ in Definition~\ref{def:no}, with the same objects, cores, category words, and ID bijection, except for the targets of the tokens that have a pool. Independently for every pool $P$, and independently of all other randomness, let $J_P:\mathcal T_P\to\mathcal S_P$ be a uniformly random injection.
A token $\tau$ with $J_P(\tau)=(w,a)$ targets $w$, and the pair $(w,a)$ is called the \emph{slot} of $\tau$.
\end{definition}

A query returns only targets, and slots are hidden from the algorithm. Every target lies in the set that Definition~\ref{def:no} prescribes, and the target of a single token is still uniform on its pool; only the joint law of the targets has changed. In particular, Lemma~\ref{os:properties}(i)--(ii) and Lemma~\ref{os:growth} hold verbatim for $\hDN$, and the argument after Definition~\ref{def:no} shows that every input of $\hDY$ is acyclic.

\begin{lemma}[Degrees]
\label{in:degrees}
In every input of $\hDN$ or $\hDY$, every object has at most $D$ outgoing and at most $2R$ incoming occurrences. Every directed cycle of an input of $\hDN$ lies in its core, and replacing each list by its set of targets does not increase any degree.
\end{lemma}

\begin{proof}
Outdegrees are bounded by the list length $D$. A blue object receives no token with a pool, and at most one core occurrence of each of the $32$ colors. A masking object $w\in X_{i,t}$ lies in at most two pools, namely the layer pool $X_{i,t}$ if $t\ge2$ and the band pool $E_i$ if $t\le T$. Since each $J_P$ is injective, at most $R$ tokens of each pool target $w$, and so $w$ has at most $2R$ incoming occurrences. The statement about cycles follows from Lemma~\ref{os:properties}(i), and collapsing repeated targets can only remove occurrences.
\end{proof}

\paragraph{Deferred exposure of slots.}
The unexposed part of an injection is sampled on demand. If the exposed tokens of $P$ have used a set $A_P$ of slots, then, by symmetry, a uniform injection extending the exposed values sends a further token to a uniform slot of $\mathcal S_P\setminus A_P$. As for the residual bijection in Section~\ref{os:fragments}, we implement this choice by a \emph{slot priority}, that is, by a uniformly random linear order $p=(p[1],p[2],\ldots)$ of $\mathcal S_P$: the token takes the first slot of $p$ outside $A_P$, which we call its \emph{accepted} slot, and the number $\mathrm{rej}$ of slots of $p$ before the accepted one, all of which lie in $A_P$, is its \emph{rejection count}. If $a_P=|A_P|$, and $b$ of the used slots belong to $w\in P$, then the next target equals $w$ with probability $(R-b)/(L_P-a_P)$. As long as at most $DQ$ tokens have been exposed, we have $a_P\le DQ\le L_P/2$ for large $N$, since $Q=o(W)$, and therefore
\begin{equation}\label{in:diffuse}
 \Pr[\text{the next target in $P$ equals $w$}\mid\text{exposed targets and slots}]
 \le\frac{R}{L_P-a_P}\le\frac2{|P|}.
\end{equation}
Hence Lemma~\ref{os:properties}(iii) holds for $\hDN$, with the bound $1/W$ replaced by $2/W$, at every exposure state that contains the slots of the exposed tokens and in which at most $DQ$ tokens are exposed; the bound $2/N$ for core entries is unchanged.

\subsection{Transcript probabilities}
\label{in:likelihood}

As in Section~\ref{rt:likelihood}, the probability of a marked transcript is a common factor times an embedding count, and only the factor of the nonblue targets changes. We keep the marked transcripts of Section~\ref{rt:likelihood}; slots are not marked. For a marked transcript $\xi$ of positive $\hDN$ probability and a pool $P$, let $m_P(\xi)$ be the number of exposed tokens with pool $P$, and let $a_{P,w}(\xi)$ be the number of those that target $w\in P$. These numbers are determined by the marks. Indeed, the pool of an exposed entry is determined by its category, which is given by the marked category word, and by the region and layer of its source, which are marked unless the source is blue, in which case the pool is a band; and every target of a token with a pool is nonblue, and hence marked.

\begin{lemma}[Pool marginals]
\label{in:pool-marginal}
Let $\tau_1,\ldots,\tau_m\in\mathcal T_P$ be distinct, let $w_1,\ldots,w_m\in P$, and let $a_w=|\{s:w_s=w\}|$ for $w\in P$. Then
\begin{equation}\label{in:pool-factor}
 \Pr\bigl[J_P(\tau_s)\in\{w_s\}\times[R]\text{ for all }s\le m\bigr]
 =\frac{\prod_{w\in P}(R)_{a_w}}{(L_P)_m}.
\end{equation}
In particular, this probability does not depend on $|\mathcal T_P|$.
\end{lemma}

\begin{proof}
The restriction of a uniform injection to $m$ specified tokens is a uniform injective map from these tokens to $\mathcal S_P$, and there are $(L_P)_m$ such maps. Those with the required targets assign distinct slots of $w$ to the $a_w$ tokens that target $w$, for every $w$, which gives $\prod_w(R)_{a_w}$ maps.
\end{proof}

For $\xi$ of positive $\hDN$ probability, we replace the common factor \eqref{rt:common-factor} by
\begin{equation}\label{in:common-factor}
 \widehat c_{\mathcal A}(\xi)=
 \binom{D}{d_0,\,d_1-d_0,\,\ldots,\,d_{k+1}-d_k}^{-\ell(\xi)}
 \frac{1}{(n_0)_{|V_\xi|}}
 \prod_P\frac{\prod_{w\in P}(R)_{a_{P,w}(\xi)}}{(L_P)_{m_P(\xi)}}.
\end{equation}

\begin{lemma}[Exact likelihood with capacities]
\label{in:exact-likelihood}
Lemma~\ref{rt:exact-likelihood} holds for $\hDN$ and $\hDY$, with $\widehat c_{\mathcal A}$ in place of $c_{\mathcal A}$.
\end{lemma}

\begin{proof}
We repeat the proof of Lemma~\ref{rt:exact-likelihood}, changing only the factor of the nonblue targets. Fix an injective map $f:B_\xi\to B$. Together with the marks, $f$ determines the source object, and hence the token, of every exposed entry. The ID bijection, the category words, the core indicators, the empty lists of the last layer, and the consistency indicator contribute as before. The injections are independent of each other and of all other randomness, so, by Lemma~\ref{in:pool-marginal}, the nonblue targets contribute the product over $P$ in \eqref{in:common-factor}. This product depends only on the numbers $m_P(\xi)$ and $a_{P,w}(\xi)$, and not on $f$: another choice of $f$ changes the tokens, but neither their number in each pool nor their targets. Summing over $f$ gives \eqref{rt:no-fixed-core}.

For \eqref{rt:yes-subcount}, we restrict the sum to embeddings into $G_\circ$ that avoid blue layer $W$. Then every queried blue source uses the same colored transition in $\hDY$ as in $G_\circ$, no exposed entry is a core entry of the last blue layer, and the exposed tokens of every pool, with their targets, are the same as in the NO computation. The YES pool $E_1$ contains, in addition, the tokens of the last blue layer, but these are not exposed, and by Lemma~\ref{in:pool-marginal} they do not change the factor. Here we use that both distributions give every object the same number $R$ of slots.
\end{proof}

Lemma~\ref{rt:pointwise} uses Lemma~\ref{rt:exact-likelihood} only through the identities \eqref{rt:no-fixed-core} and \eqref{rt:yes-subcount}, in which the common factor cancels. Hence Lemma~\ref{rt:pointwise} holds for $\hDY$ and $\hDN$, and so does Proposition~\ref{rt:transcript-tv}, as soon as Theorem~\ref{os:structural} holds for $\hDN$. The rest of this appendix proves the latter.

\subsection{Exploration with slots}
\label{in:exploration}

Sections~\ref{os:generations} and~\ref{os:chronology} use the input distribution only through Lemmas~\ref{os:properties}, \ref{os:growth}, \ref{os:blue-size}, and~\ref{os:local}. We have already noted that Lemma~\ref{os:properties}(i)--(ii) and Lemma~\ref{os:growth} hold for $\hDN$, and that Lemma~\ref{os:properties}(iii) holds with $2/W$ in place of $1/W$. The weaker bound affects only the constants in the record count \eqref{os:record-count} and in the retirement bound \eqref{os:retirement-rec}. The witness bound \eqref{os:clan-witness} and the bound \eqref{os:cycle-closing} on cycle closure concern occurrences between blue vertices, that is, core entries, for which $2/N\le1/W$ is unchanged. It remains to prove Lemmas~\ref{os:blue-size} and~\ref{os:local} for $\hDN$. Their proofs use the independence of the targets in the posterior factorization \eqref{os:posterior}, in the allocation of the tapes (Lemma~\ref{os:continuation-law}), in the reconstruction (Lemma~\ref{os:decoder}), and in the erased process of Appendix~\ref{app:blue-erasure}. We now modify these four steps.

\paragraph{The analytical state and the posterior.}
We add to the analytical state of Section~\ref{os:fragments} the slot of every exposed token and the inspected prefixes of slot priorities. This changes neither the exploration nor the fragments, since no decision of the algorithm or of the fragment protocol depends on slots; it only refines the conditioning in (C1) and (C2), and all bounds below hold at every value of the refined state. For a birth state $s$ and a pool $P$, let $J_{P,s}$ be the partial injection formed by the exposed tokens of $P$ and their slots. Then \eqref{os:posterior} is replaced by
\begin{equation}\label{in:posterior}
 \psi_s(\Pi)\otimes\Unif(\text{residual bijection})
 \otimes\bigotimes_{v\text{ unqueried}}\Dist(C_v)
 \otimes\bigotimes_P\Unif\bigl(\text{injections }\mathcal T_P\to\mathcal S_P\text{ extending }J_{P,s}\bigr).
\end{equation}
The proof is the same as that of \eqref{os:posterior} and of its verification in Appendix~\ref{app:local-allocation}. Before any exposure, these factors are independent. An exposed entry with a pool, together with its slot and the word of its source, which has been read, fixes one value of one injection; it does not involve any unread word, and the token sets do not depend on unread words. All other observations are as before, and the skip predicates of all priority scans, including slot priorities, involve only exposed data. Hence the likelihood of an atom is constant in all unexposed coordinates on its support.

\paragraph{Allocation of the slot priorities.}
In the conditional experiment of Section~\ref{os:reconstruction}, we replace the samples of nonblue endpoints in the tape $Z$ by slot priorities: for every future query position $t$, every list index $i\in[D]$, and every pool $P$, the tape $Z$ contains an independent uniform slot priority $p_{t,i,P}$ of $\mathcal S_P$. The entry at position $t$ and list index $i$, if its token has pool $P$, reads $p_{t,i,P}$ and takes its accepted slot with respect to the current set $A_P$ of used slots. The copies are allocated before the sources are known, and hypothetical sources at one position whose tokens at index $i$ have the same pool share one priority; since one source occupies each position, each priority is read at most once. At a noncandidate index, every hypothetical upper source of the fragment has the same category $c>j+1$ and pool $E_c$, and so all of them read one common category and one common priority, which replaces the common target sample. The tape $Z$ remains independent of $\Theta$. We note that $Z$ contains the priorities, and not the injections that result from them: the sets of used slots, and hence the accepted slots, depend on the actual sources and their categories, and they are part of the evolving state rather than of the tape.

\begin{lemma}[Exact law of the continuation with slots]
\label{in:continuation-law}
Lemma~\ref{os:continuation-law} holds for $\hDN$, where the tape $Z$ contains the slot priorities instead of the endpoint samples, and where the variables of the adjacency lists are the category words and the completed injections $J_P$.
\end{lemma}

\begin{proof}
We follow the proof in Appendix~\ref{app:local-allocation}. The seed of an unqueried source now consists of its category word alone, and \eqref{os:seed-density-identity} applies to it as before. For the slot priorities, we use the following analogue of \eqref{os:priority-density-identity}. Let $f$ be a partial injection from $\mathcal T_P$ to $\mathcal S_P$ with $m$ free slots and $\ell$ unmatched tokens, let $\Unif_f$ be the uniform law of the injections extending $f$, let $\mu$ be the uniform law of a slot priority $p$, and let $g_f(p)$ be the first free slot of $p$. Since $g_f(p)$ is uniform on the free slots, $\mathsf K_f(x,p)=m\,\mu(p)\ind\{g_f(p)=x\}$ is a normalized kernel, and for every injection $J$ extending $f$ and every unmatched token $\tau$,
\begin{equation}\label{in:slot-density-identity}
 \Unif_f(J)\,\mathsf K_f(J(\tau),p)=\mu(p)\,\Unif_{f\cup\{\tau\mapsto g_f(p)\}}(J),
\end{equation}
both nonzero sides being $\mu(p)/(m-1)_{\ell-1}$. In the coupling, we start from an input (core, category words, injections, and ID bijection) drawn from the birth law \eqref{in:posterior}; at each exposure of a token $\tau$ of $P$, we draw the priority that it reads from $\mathsf K_f(J_P(\tau),\cdot)$, where $f$ is the current partial injection, and unread priorities keep their prior law. Applying \eqref{in:slot-density-identity} at each exposure, in chronological order, telescopes to the density of the allocated execution times a uniform completion of each injection, just as the ID priorities telescope in the proof of Lemma~\ref{os:continuation-law}. The rest of that proof is unchanged.
\end{proof}

\paragraph{Slot records.}
Fix a level $j$ and a birth state $(s,o)$, and use the terminology of Section~\ref{os:reconstruction}. The accepted slot of an entry depends on the set of used slots, and this set depends on the masks: the decoder cannot evaluate a candidate entry of a $U_j$ list of the fragment, since its category, and hence its pool, depends on the mask, and so it does not know which slot this entry has used. If a later entry rejects this slot, then the decoder, for which the slot is free, would accept it. A new type of record supplies the rejection count in this case, just as records of type (E) supply the objects that a scan skips. In addition to the data structures of Appendix~\ref{app:decoder}, the decoder maintains, for every pool $P$, a set $K_P$ of slots that it knows to be used, together with the accepted slots of the entries that it has evaluated. Let $A_P$ be the actual set of used slots. The decoder maintains
\begin{equation}\label{in:known-slots}
 K_P\subseteq A_P,
\end{equation}
and at the birth, all used slots are part of $s$, so that $K_P=A_P$. The decoder evaluates the entries of every list that it computes from $Z$, that is, of every exterior, lower, or resolved $X_{j+1}$ list, the noncandidate entries of every upper list of the fragment, and, during replay, the entries of the replayed $X_{j+1}$ lists; after each evaluation, it inserts the accepted slot into $K_P$. Thus the only entries with a pool that the decoder does not evaluate are candidate entries of the $U_j$ lists of the fragment, and candidate entries of unresolved $X_{j+1}$ lists that have not yet been replayed. Their pools lie in $U_{j+1}$: a candidate of a $U_j$ source has category at most $j+1$, and an unresolved $X_{j+1}$ source does not lie in the last layer $X_{j+1,3T}$, since its component is resolved at the cutoff in (C) before it reaches that layer. Hence, at every time,
\begin{equation}\label{in:unknown-slots}
 \sum_P|A_P\setminus K_P|\le D\,\mathrm{work}_{j+1}(\phi),
\end{equation}
and $K_P=A_P$ for every pool $P$ outside $U_{j+1}$. We add one type of record to (A)--(E).
\begin{enumerate}
\item[(S)] At an entry with a pool $P$ and rejection count $\mathrm{rej}\ge1$, record $\mathrm{rej}$, tagged by its position and its list index, if the entry belongs to an upper list of the fragment, or if one of its $\mathrm{rej}$ rejected slots is not in the set $K_P$ that the decoder uses for this entry, namely its set at the beginning of the answer together with the accepted slots of the earlier entries of the same list (see the proof of Lemma~\ref{in:decoder}).
\end{enumerate}
Like (A)--(E), records of type (S) serve only the analysis: the encoder writes them into $\mathcal M$, and the algorithm never sees them. In particular, they are not disclosures, and they do not affect the fragment protocol. A record of type (S) specifies three numbers of size at most $DQ$, so that the alphabet $\Sigma$ still has size $N^{C_\eta}$, and one answer produces at most $D$ such records.

\begin{lemma}[Exact reconstruction with slots]
\label{in:decoder}
Lemma~\ref{os:decoder} holds for $\hDN$, with the records of type (S) included in $\mathcal M$.
\end{lemma}

\begin{proof}
The decoder knows every used slot outside $U_{j+1}$; inside $U_{j+1}$, it misses only the slots of entries that it has not evaluated, and records of type (S) supply every rejection count that it cannot compute. We follow the proof in Appendix~\ref{app:decoder}, and we add the following to its invariant: at every operation boundary, $K_P\subseteq A_P$ for every pool $P$, with equality for every pool outside $U_{j+1}$, where $A_P$ is the actual set of used slots at that boundary; and every slot of $K_P$ is the accepted slot of an entry that has been exposed in the actual run. This holds at the birth.

\emph{Order of the updates.} The decoder changes $K_P$ only by inserting accepted slots. A repeated query reads no slot priority, since the priorities are indexed by query positions, and so it changes neither $A_P$ nor $K_P$. At a first query, the decoder processes the (E) records and their replays as in Appendix~\ref{app:decoder}. It then evaluates the entries of the current list in list order, except the candidate entries of a $U_j$ list of the fragment (even if (A) supplies their targets) or of an unresolved $X_{j+1}$ list, inserting each accepted slot before the next entry, and only then drains the replay queue filled by the records of this answer. Thus, for an entry of a list that is not an upper list of the fragment, $K_P$ is the decoder's set at the beginning of the answer together with the accepted slots of the earlier entries of the same list. This set is determined by the past, so the encoder can compute it, and (S) is well defined. Every entry whose pool lies outside $U_{j+1}$ is evaluated at its exposure, which gives the equality outside $U_{j+1}$.

\emph{Evaluation of an entry.} The decoder computes the pool $P$ from the source object, the mode, and the category (for a noncandidate entry of an upper list of the fragment, from the common category), and it reads $p=p_{t,i,P}$. If the entry belongs to an upper list of the fragment, then its accepted slot is $p[1+\mathrm{rej}]$, where $\mathrm{rej}$ is given by the record of type (S) at this entry, and $\mathrm{rej}=0$ if there is no such record. Otherwise, the accepted slot is $p[1+\mathrm{rej}]$ if there is a record, and the first slot of $p$ outside $K_P$ if there is none. In both cases, this is the actual accepted slot. Indeed, in the second case without a record, the slots $p[1],\ldots,p[\mathrm{rej}]$ lie in $K_P$, whereas $p[1+\mathrm{rej}]$ is free, and hence not in $K_P$, since $K_P$ is contained in the set of slots used before this entry. The object of the accepted slot is the target, which is bound to the stored public pattern as in Appendix~\ref{app:decoder}, and the accepted slot is then inserted into $K_P$, before the next entry is processed. This also covers slots used by earlier entries of the same answer: all entries of an answer belong to one list, and either the list is an upper list of the fragment, whose rule does not involve $K_P$, or all its entries are evaluated in list order. Thus \eqref{in:known-slots} is maintained.

\emph{Replay.} The rule for upper lists of the fragment does not refer to the set of used slots at the time of the query, and this makes historical replay possible. When a component in $X_{j+1}$ is resolved, the decoder replays its exposed lists at their stored positions, and it computes their accepted slots from the allocated priorities and the records of type (S) at these positions, without reconstructing the historical sets $A_P$. (This is why (S) records every rejection in an upper list of the fragment, and not only the rejections of slots that are unknown to the decoder.) The replayed lists were exposed at the current or earlier positions, so their slots were actually used, and \eqref{in:known-slots} is maintained. The answers of the $U_j$ lists of the fragment, and of unresolved $X_{j+1}$ lists, without records of types (A)--(C) are reconstructed as in Appendix~\ref{app:decoder}: their noncandidate entries are evaluated by the rule above, and their candidate entries return fresh IDs from their shared priorities, so that their accepted slots are not needed. The rest of the case analysis in Appendix~\ref{app:decoder} is unchanged.
\end{proof}

\paragraph{Expected number of slot records.}
Records of type (S) are rare, because a fresh slot priority rarely begins with a used slot. We bound their number following (C1). Consider first an entry with a pool $P$ in an upper list of the fragment. Its priority has not been read before, and so, conditionally on the past, $p[1]$ is a uniform slot of $\mathcal S_P$, which is used with probability $a_P/L_P\le DQ/(RW)\le Q/W$. A list has at most $D$ entries, so the expected number of these records, summed over $\mathcal F$, is at most $D(Q/W)\E\,\mathrm{work}_{j+1}$. Consider next an entry with a pool $P$ in another list, and let $b_P=|A_P\setminus K_P|$ before it is processed, where $K_P$ is the set that the decoder uses for this entry; by the proof of Lemma~\ref{in:decoder}, the sets $A_P$ and $K_P$ are determined by the past. Each of the $b_P$ used slots that are unknown to the decoder precedes all $L_P-a_P$ free slots in the fresh priority with probability $1/(L_P-a_P+1)$, and so the entry produces a record with conditional probability at most
\[
 \frac{b_P}{L_P-a_P+1}\le\frac{2b_P}{RW}.
\]
By \eqref{in:unknown-slots}, we have $b_P\le D\,\mathrm{work}_{j+1}(\phi)$, where $\mathrm{work}_{j+1}(\phi)$ now denotes the final value for the distinguished fragment~$\phi$. Summing over the at most $DQ$ entries of the experiment, and then over $\mathcal F$ by the tower rule, bounds the expected number of these records by $2D^2(Q/(RW))\E\,\mathrm{work}_{j+1}\le C_\eta(Q/W)\E\,\mathrm{work}_{j+1}$. The bounds for (A)--(E) in Section~\ref{os:reconstruction} change only by the factor $2$ in \eqref{in:diffuse}. Hence \eqref{os:record-count} holds for $\hDN$, after increasing the constant in $\Delta$; we note that $\Delta=N^{-\beta+o(1)}$ and $N^\zeta\Delta=o(1)$ are unchanged.

The remaining steps of the proof of Lemma~\ref{os:local}, namely the union bound \eqref{os:cut-expectation} on the path cut, use only Lemmas~\ref{os:continuation-law} and~\ref{os:decoder}, the independence of $\Theta$ and $Z$, and the record count; by Lemmas~\ref{in:continuation-law} and~\ref{in:decoder}, they apply verbatim. The retirement bound \eqref{os:retirement-rec} uses only Lemma~\ref{os:properties}(iii), and so it holds with a larger constant in $\Delta$. Hence Lemma~\ref{os:local} holds for $\hDN$.

\paragraph{The height estimate.}
Finally, we adapt the erased process of Appendix~\ref{app:blue-erasure}, which must exclude used slots, and not only used objects.

\begin{lemma}[Conditional estimate of the blue height with slots]
\label{in:blue-height}
Assume $Q=o(W)$ and $QT/W=o(1)$. Then Lemma~\ref{lem:conditional-blue-height} holds for $\hDN$, with the factor $\exp(C_\eta QT/N)$ replaced by $\exp(C_\eta QT/W)$.
\end{lemma}

\begin{proof}
The skeleton of Appendix~\ref{app:blue-erasure} ignores the private objects. Now it must also ignore the slots that the private assignments use, and each of its exposures into a pool costs a further factor close to~$1$ in the comparison. We follow the proof in Appendix~\ref{app:blue-erasure}, where \emph{upper} means $U_1$, and we modify the skeleton and the private assignments as follows. The skeleton stores the slots used at the birth and the slots of its own exposures. An exposure of the skeleton with a pool takes the first slot of its priority that is not stored; thus it does not exclude the slots of private assignments, just as the skeleton does not exclude private objects when it matches an unseen ID. The private assignments of the candidates of upper lists of the fragment choose their targets in $E_1$ and in the layers of $X_1$ by taking the first slot of their priorities that is neither stored nor used by an earlier private assignment. The skeleton still reads no private choice, so it remains independent of the private masks. On a common prefix, the real set of used slots of a pool is the disjoint union of its stored slots and its private slots.

Consider a common prefix and an exposure of the skeleton into a pool $P$ in which $a$ slots are stored and $b$ slots are private. Since at most $T$ upper lists of the fragment have been queried, we have $b\le DT$, and $a\le DQ$. For a \emph{common} slot, that is, a slot that is neither stored nor private, the real and the skeleton probabilities of accepting it are $1/(L_P-a-b)$ and $1/(L_P-a)$, respectively, after the priority is integrated out. Since $L_P-a-b\ge RW-DQ-DT\ge RW/2$ for large $N$, their ratio is
\[
 \frac{L_P-a}{L_P-a-b}\le1+\frac{2DT}{RW}\le\exp\Bigl(\frac{2T}{W}\Bigr).
\]
A skeleton exposure that accepts a private slot has no real counterpart, and it is omitted from the sums over common prefixes. Private assignments exclude the same slots in both executions, so their factors agree. Multiplying over the at most $DQ$ exposures of the skeleton with a pool gives the factor $\exp(2DQT/W)$, and together with the factor of \eqref{eq:appendix-prefix-ratio}, the ratio of the prefix probabilities is at most $\exp(C_\eta QT/N+2DQT/W)\le\exp(C_\eta QT/W)$.

Agreement can end in one new way: a skeleton exposure may accept a private slot. Like a skeleton match of an unseen ID to a private object, this has no real counterpart. If, instead, an exposure of the skeleton accepts a free slot of a private object, then this object lies in $U_1$, and hence, by monotonicity, so does the source; in the real execution, this is an upper contact incident to the fragment, which retires after the current answer, as in the case of an upper target of an exterior list that equals a private object. Hence, in the real execution, agreement still ends only at the final answer of the local experiment. The bound \eqref{eq:appendix-skeleton-path} on rare paths and the partition by first witnesses are unchanged, and the lemma follows.
\end{proof}

By \eqref{os:scales} and \eqref{os:small-scales}, we have $QT/W=N^{-2\beta+o(1)}=o(1)$, so that the factor $\exp(C_\eta QT/W)$ is at most $2$ for large $N$, and the proof of Lemma~\ref{os:blue-size} applies verbatim to $\hDN$.

\begin{proof}[Proof of Theorem~\ref{in:main}]
Part (i) is Lemma~\ref{in:degrees}, together with the remarks after Definition~\ref{in:def}; the core is the NO core of Definition~\ref{def:no}. For (ii), Lemmas~\ref{os:blue-size} and~\ref{os:local} hold for $\hDN$ by Lemma~\ref{in:blue-height} and by the preceding paragraphs, and Lemma~\ref{os:properties} holds with the constants discussed at the beginning of Section~\ref{in:exploration}. The proofs in Sections~\ref{os:generations} and~\ref{os:chronology} use nothing else about the distribution, and so they give Theorem~\ref{os:structural} for $\hDN$, with the same parameters and a failure probability of the form \eqref{os:structural-failure}, up to the values of the constants. Part (iii) follows from (ii) and Lemma~\ref{in:exact-likelihood}, as explained after that lemma.
\end{proof}
\addtocontents{toc}{\protect\setcounter{tocdepth}{2}}

\section{A linear lower bound for the BFS approach}
\label{app:bfs}

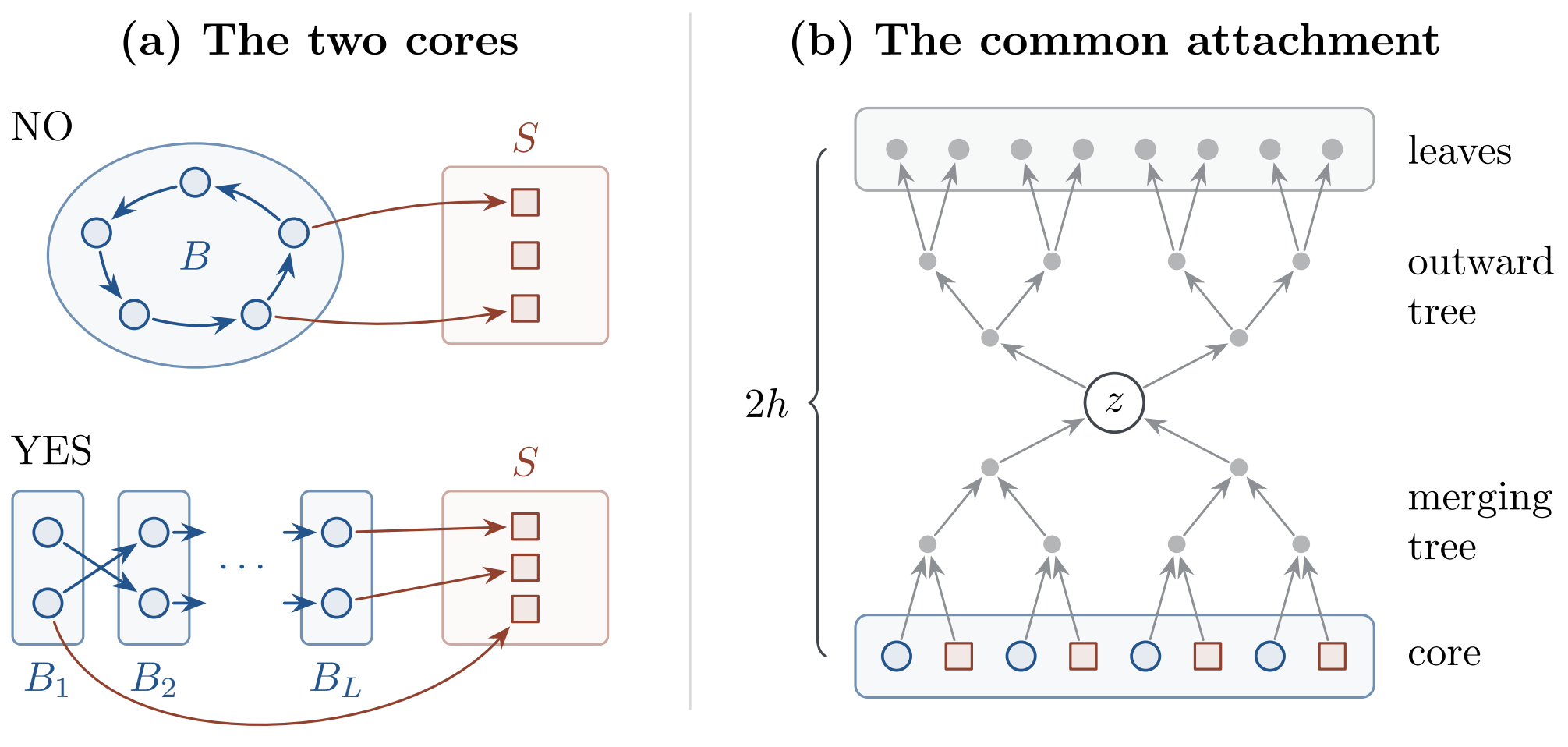
\begin{figure}[!t]
\centering
\begin{tikzpicture}[x=1cm,y=1cm,line cap=round,line join=round,
  bv/.style={acy/bluedot,minimum size=2.4mm},
  sk/.style={rectangle,draw=acyRed,fill=acyRed!12,line width=.6pt,minimum size=2.2mm,inner sep=0pt},
  tv/.style={circle,fill=acyInk!40,inner sep=0pt,minimum size=1.5mm},
  act/.style={acy/coreedge,line width=.7pt,shorten >=1pt,shorten <=1pt},
  ina/.style={acy/rededge,line width=.6pt,shorten >=1pt,shorten <=1pt},
  tre/.style={acy/edge,line width=.55pt,draw=acyInk!60,shorten >=.5pt,shorten <=.5pt},
  lab/.style={acy/note,font=\small}]
\node[acy/title] at (2.6,5.55) {(a) The two cores};
\node[lab,anchor=west] at (-.15,4.85) {NO};
 \draw[acy/bluearea] (1.55,3.75) ellipse [x radius=1.25,y radius=.95];
 \node[lab,text=acyBlue] at (1.55,3.75) {$B$};
 \foreach \name/\ang in {n1/90,n2/162,n3/234,n4/306,n5/18}
   \node[bv] (\name) at ($(1.55,3.75)+(\ang:.88 and .62)$) {};
 \foreach \from/\to in {n1/n2,n2/n3,n3/n4,n4/n5,n5/n1}
   \draw[act] (\from) to[bend right=14] (\to);
 \draw[acy/redarea] (3.65,3.0) rectangle (5.05,4.5);
 \node[lab,text=acyRed] at (4.35,4.75) {$S$};
 \foreach \name/\yy in {s1/3.3,s2/3.75,s3/4.2}
   \node[sk] (\name) at (4.35,\yy) {};
 \draw[ina] (n5) to[bend left=8] (s3);
 \draw[ina] (n4) to[bend right=8] (s1);
\node[lab,anchor=west] at (-.15,2.1) {YES};
 \foreach \tag/\xx/\lab in {y1/.3/{B_1},y2/1.2/{B_2},yL/2.75/{B_L}} {
   \draw[acy/bluearea,rounded corners=2pt] (\xx-.3,.45) rectangle (\xx+.3,1.75);
   \node[lab,text=acyBlue] at (\xx,.15) {$\lab$};
   \node[bv] (\tag a) at (\xx,1.4) {};
   \node[bv] (\tag b) at (\xx,.8) {};
 }
 \node[lab,text=acyBlue] at (1.975,1.1) {$\cdots$};
 \draw[act] (y1a) -- (y2b);
 \draw[act] (y1b) -- (y2a);
 \draw[act] (y2a) -- (1.68,1.4);
 \draw[act] (y2b) -- (1.68,.8);
 \draw[act] (2.27,1.4) -- (yLa);
 \draw[act] (2.27,.8) -- (yLb);
 \draw[acy/redarea] (3.65,.45) rectangle (5.05,1.75);
 \node[lab,text=acyRed] at (4.35,2.0) {$S$};
 \foreach \name/\yy in {t1/.75,t2/1.1,t3/1.45}
   \node[sk] (\name) at (4.35,\yy) {};
 \draw[ina] (yLa) -- (t3);
 \draw[ina] (yLb) -- (t2);
 \draw[ina] (y1b) .. controls (.75,-.55) and (3.3,-.45) .. (t1);
\draw[draw=acyInk!20,line width=.5pt] (5.75,-.1)--(5.75,5.8);
\begin{scope}[xshift=7.15cm]
  \node[acy/title] at (2.2,5.55) {(b) The common attachment};
\draw[acy/bluearea,rounded corners=3pt] (0,0) rectangle (4.4,.7);
  \node[lab,anchor=west] at (4.55,.35) {core};
  \draw[draw=acyInk!45,fill=acyInk!4,rounded corners=3pt,line width=.6pt] (0,4.3) rectangle (4.4,5.0);
  \node[lab,anchor=west] at (4.55,4.65) {leaves};
  \foreach \i in {0,...,7} {
    \pgfmathsetmacro\xx{.35+.528*\i}
    \ifodd\i \node[sk] (c\i) at (\xx,.35) {}; \else \node[bv] (c\i) at (\xx,.35) {}; \fi
    \node[tv,minimum size=1.8mm] (l\i) at (\xx,4.65) {};
  }
\foreach \i in {0,...,3} {
    \pgfmathsetmacro\xx{.35+.528*(2*\i)+.264}
    \node[tv] (m\i) at (\xx,1.3) {};
    \node[tv] (o\i) at (\xx,3.7) {};
  }
  \foreach \i in {0,1} {
    \pgfmathsetmacro\xx{.35+.528*(4*\i)+.792}
    \node[tv] (mm\i) at (\xx,1.95) {};
    \node[tv] (oo\i) at (\xx,3.05) {};
  }
  \node[acy/vertex,minimum size=5mm,font=\small] (z) at (2.198,2.5) {$z$};
  \foreach \i/\p in {0/0,1/0,2/1,3/1,4/2,5/2,6/3,7/3} {
    \draw[tre] (c\i) -- (m\p);
    \draw[tre] (o\p) -- (l\i);
  }
  \foreach \i/\p in {0/0,1/0,2/1,3/1} {
    \draw[tre] (m\i) -- (mm\p);
    \draw[tre] (oo\p) -- (o\i);
  }
  \foreach \i in {0,1} {
    \draw[tre] (mm\i) -- (z);
    \draw[tre] (z) -- (oo\i);
  }
  \node[lab,anchor=west,align=left] at (4.55,1.5) {merging\\tree};
  \node[lab,anchor=west,align=left] at (4.55,3.5) {outward\\tree};
  \draw[decorate,decoration={brace,amplitude=4pt,mirror},draw=acyInk,line width=.6pt]
    (-.25,4.65) -- (-.25,.35) node[midway,left=5pt,lab] {$2h$};
 \end{scope}
\end{tikzpicture}
\caption{The construction underlying Theorem \ref{thm:bfs-linear}, schematically. (a) In both cores, an active port of a blue vertex (blue) stays in $B$, and an inactive port (red) enters the sinks $S$. In the NO core, the active ports follow random permutations of $B$ and create cycles; in the YES core, they advance from $B_i$ to $B_{i+1}$, and every port of the last layer $B_L$ enters $S$. (b) The attachment, drawn for $b=2$ and $h=3$: a merging tree leads from all $M$ core vertices, including the sinks, to its root $z$, and an outward tree leads from $z$ to $M$ leaves. A search rooted in the core reaches all $M$ leaves at distance $2h$, so it must make $M$ queries before it can explore a core vertex at distance $2h+1$.}
\label{fig:bfs}
\end{figure}

We prove the obstruction to the BFS approach of Chen et al.~\cite[Section~9, item~2]{CRSS22} discussed in the introduction. The argument is independent of the near-linear lower bound in the main text: it uses its own fixed degree and distance parameters and compares two input distributions that are fixed in advance. The idea is that a BFS must explore a linear number of vertices before it can search deep in the core. We attach the same acyclic gadget to a random NO core with cycles and to a layered acyclic YES core: a merging tree that leads from every core vertex to a new vertex~$z$, followed by an outward tree from~$z$ (see Figure~\ref{fig:bfs}(b)). A search rooted in the core reaches all leaves of the outward tree, a constant fraction of all vertices, before any core vertex at a larger distance. Hence, below a linear number of queries, every search explores only a small ball of the core, and so few core queries cannot distinguish the two cores.

\paragraph{The restricted algorithm class.}
A search \emph{explores} a vertex by reading its \emph{row}, that is, its complete set of outgoing neighbors. An outgoing BFS enqueues every newly discovered outneighbor and explores vertices in nondecreasing directed distance from its root. It cannot skip an unfinished distance level to pursue a deeper branch. We allow arbitrary adaptive ordering within each level. A tester may run at most $r$ such searches, where $r$ is fixed independently of $n$ but may depend on the fixed distance and degree parameters. It may choose roots adaptively, interleave searches, allocate budgets adaptively, and make any randomized decision from their combined transcripts. Every query must be made by one of these searches, to explore a vertex. The searches share all rows returned so far: any search may read such a row without a new query, but a row that has never been returned must be queried; in particular, an unqueried sink must be queried when it is explored. Every search keeps its own discovered and explored sets and obeys its own BFS levels, including when it reads a shared row. Thus, if the total number of queries is $q$, each individual search explores at most $q$ distinct vertices. At constant degree, this convention differs by at most a constant factor from indexed access when exploring a vertex reads the entire outgoing list.

\begin{theorem}[Two-sided linear lower bound for outgoing BFS]
\label{thm:bfs-linear}
There exist absolute constants $D_{\rm BFS},\eps_{\rm BFS},c_{\rm BFS}>0$ such that the following holds. For every fixed $r$ and all sufficiently large $n$, no tester in the above $r$-search class, using at most $c_{\rm BFS}n$ queries, can be a two-sided $\eps_{\rm BFS}$-tester for acyclicity, that is, accept every acyclic graph and reject every $\eps_{\rm BFS}$-far graph, each with probability at least $\frac23$, on all simple $n$-vertex directed graphs of maximum indegree and maximum outdegree at most $D_{\rm BFS}$.

\end{theorem}

We prove Theorem~\ref{thm:bfs-linear} by constructing, for every large $n$, two distributions on $n$-vertex graphs that are fixed before the tester runs and supported on acyclic and on $\eps_{\rm BFS}$-far graphs, respectively. For every tester in the above class with at most $c_{\rm BFS}n$ queries, the complete transcripts under the two distributions are at total variation distance $o(1)$, uniformly over its query rule. The constants $D_{\rm BFS}$, $\eps_{\rm BFS}$, and $c_{\rm BFS}$, which we fix below, do not depend on $r$, whereas the threshold in $n$ may.

\paragraph{The NO and YES cores.}
Throughout this appendix, let $a=2^{20}$ and $b=a^{64}$; we prove Theorem~\ref{thm:bfs-linear} with $D_{\rm BFS}=b$, $\eps_{\rm BFS}=a/(2048b^2)$, and $c_{\rm BFS}=1/(12b)$.
For a power of two $N$, put
\[
 L=2^{\lfloor(\log_2N)/3\rfloor},\qquad W=N/L.
\]
Thus $L=\Theta(N^{1/3})$ and $W=\Theta(N^{2/3})$. As in the main text, the \emph{core} is the part of the input that carries the cycles of NO instances, although its construction here is different (see Figure~\ref{fig:bfs}(a)). A core has $2N$ vertices, which we call \emph{objects} before IDs are assigned: a set $B$ of $N$ \emph{blue} objects and a set $S$ of $N$ \emph{sinks}, which have no outgoing edges. Every blue object has $a$ ordered outgoing \emph{ports}, that is, list positions, indexed by $j\in[a]$. The \emph{class} of an object, that is, whether it is blue or a sink, and in the YES core also its layer, are hidden by an independent uniformly random assignment of IDs. We first specify the targets of all ports, then delete self-loops and collapse repeated targets to obtain a loopless simple graph.

In the NO core, independently for each $j\in[a]$, choose a uniform permutation $\sigma_j:B\to B$ and a uniform bijection $\tau_j:B\to S$; together they form \emph{generator}~$j$. For every $u\in B$ and $j\in[a]$, choose an independent fair bit $\xi_{u,j}$. Port $j$ of $u$ is \emph{active} if $\xi_{u,j}=1$; an active port targets $\sigma_j(u)$, and an inactive one targets $\tau_j(u)$. Vertices of $S$ have empty rows.

In the YES core, partition $B$ into $L$ layers $B_1,\ldots,B_L$ of width $W$. For every $i<L$ and $j\in[a]$, choose an independent uniform bijection $\sigma_{j,i}:B_i\to B_{i+1}$. Choose the bijections $\tau_j:B\to S$ and the fair bits independently as in NO; here generator~$j$ consists of the maps $\sigma_{j,i}$, $i<L$, and $\tau_j$. At $u\in B_i$, $i<L$, an active port targets $\sigma_{j,i}(u)$ and an inactive port targets $\tau_j(u)$. At $u\in B_L$, every port targets $\tau_j(u)$ regardless of its bit. In particular, the last blue layer is not a layer of empty rows. Again all vertices of $S$ are sinks.

Every YES edge advances in the order $B_1,\ldots,B_L,S$, so this core is acyclic. Both directional degrees in either simple core are at most $a$: every object has at most $a$ ports, and each generator contributes at most one incoming occurrence at any target, since its maps are injective. The random tables are sampled once; repeated queries never resample a row.

\begin{lemma}[Farness of the NO core]
\label{lem:bfs-core-far}
With probability $1-o(1)$, the simple NO core $G$ satisfies $\fas(G)\ge aN/64$.
\end{lemma}
\begin{proof}
As in the proof of Proposition~\ref{rc:farness}, it suffices to show that every half of $B$ receives many edges from the other half. For a fixed $U\subseteq B$ of size $N/2$, let $Z_U$ be the number of ports $(u,j)$ with $u\in B\setminus U$ and $\sigma_j(u)\in U$, that is, of ports from $B\setminus U$ that would target $U$ if they were active. Then $\E Z_U=aN/4$. Expose the $a$ permutations one image at a time. Conditional completions for two choices of the next image can be coupled by a transposition, which changes $Z_U$ by at most two. Hence the Doob martingale has $aN$ increments of absolute value at most two, and the bounded-difference inequality (see, e.g., \cite{McD89}) gives
\[
 \Pr[Z_U<aN/8]\le \exp(-aN/512).
\]
The union bound over at most $2^N$ sets $U$ is $o(1)$. Conditional on the permutations, each of these crossing occurrences is independently activated with probability $1/2$. Whenever $Z_U\ge aN/8$, a Chernoff bound gives
\[
 \Pr[\text{fewer than }aN/32\text{ active crossing occurrences}
       \mid \sigma_1,\ldots,\sigma_a]
 \le \exp(-aN/128).
\]
Another union bound shows that every balanced cut has at least $aN/32$ active occurrences in this direction, with probability $1-o(1)$.

Let $R$ be the number of activated blue occurrences lost by collapsing duplicate targets at their sources. Since two independent generators at a fixed source have the same blue image with probability $1/N$,
\[
 \E R\le N\binom a2\frac1{4N}=\frac14\binom a2=O_a(1).
\]
Thus $R<aN/64$ with probability $1-o(1)$ by Markov's inequality. Self-loops never cross a cut. After simplification, every balanced cut therefore has at least $aN/64$ distinct edges from its second side into its first. In any acyclic graph obtained by deleting edges, restrict a topological order to $B$ and take its first $N/2$ elements as $U$. All those original backward edges must have been deleted, proving the claim.
\end{proof}

\begin{lemma}[Comparison of core transcripts for few queries]
\label{lem:bfs-core-tv}
For every adaptive query rule making at most $s$ distinct core queries, where $s<W/2$ and $(a+1)s<N/2$, the complete transcripts $H^{\mathrm N}_s$ and $H^{\mathrm Y}_s$ of the rule on the NO core and on the YES core satisfy
\begin{equation}\label{eq:bfs-core-tv}
 \TV\bigl(\Dist(H^{\mathrm N}_s),\Dist(H^{\mathrm Y}_s)\bigr)
 \le C_a(s+1)^2\left(\frac1N+\frac1W+\frac1L\right).
\end{equation}
The estimate is uniform over the query rule and holds even for the oracle that returns ordered ports, before forgetting multiplicities and deleting loops.
\end{lemma}
\begin{proof}
Until a blue object repeats or the last YES layer is reached, both cores answer like a random out-forest: every active port leads to a fresh blue object, and every inactive port to a sink. We therefore couple each of the two core laws with a common auxiliary experiment $\mathsf F$ in which this holds exactly, and we bound the probability that a coupling fails. Fix the algorithm's coins; a repeated query returns the same row. It is enough to compare the stronger answers with ordered ports: ordinary answers, which are sets of neighbors, are deterministic functions of these answers. Write $m=(a+1)s$ for an upper bound on the total number of IDs seen through queries and answers. Classes in the coupling are private and are not supplied to the algorithm.

\emph{The common experiment.} Keep the actual $N$ sink objects, but represent blue objects by distinct abstract placeholders. When a previously unseen ID is queried and $u$ blue placeholders and $v$ sink objects have appeared, make it a blue root with probability
\[
 \frac{N-u}{2N-u-v};
\]
otherwise assign it a uniform unused sink object. Each newly queried blue placeholder uses independent fair port bits. An active port creates a fresh blue placeholder. At an inactive port $j$, choose a sink uniformly from the images not previously used by an inactive port of generator $j$. This samples the relevant part of $\tau_j$ without replacement. A sink previously seen through another generator or a root query returns its existing ID; a new object receives a uniform unused ID. Sink queries return empty rows. Sampling of fresh IDs can be implemented with a random ordering of the unused IDs, removing a directly queried ID before subsequent draws.

In $\mathsf F$, the blue placeholders and the edges of active ports form an out-forest rooted at the blue roots. It has at most $s$ components and height at most $s$, since every edge of a path requires a query to its source. The experiment itself need not be the oracle of a completed graph.

\emph{Coupling NO to $\mathsf F$.} Expose the object--ID matching lazily, assigning a uniform unused blue object to each new blue root. Until disagreement, couple the root classes, port bits, sink images, and fresh IDs identically. At an active port of generator $j$, the NO target is uniform among at least $N-s$ unused images of $\sigma_j$. Declare failure if it is an already seen blue object; otherwise assign it to the new placeholder. Consequently
\begin{equation}\label{eq:bfs-no-fresh}
 \TV\bigl(\Dist(H^{\mathrm N}_s),\Dist(H^{\mathsf F}_s)\bigr)
 \le \frac{as\,m}{N-s}.
\end{equation}
Self-loops are among the failures. Sink coincidences are coupled exactly and are not failures. The unexposed $\tau_j$ images have their correct conditional laws because all queried blue sources are distinct and these bijections are independent of the blue permutations.

\emph{Independent proposals for YES.} Generate the entire $\mathsf F$ transcript, its private classes, and its sink objects first. Independently assign each blue component root a uniform object of $B$, with replacement between components. Thus its root layer is uniform in $[L]$ and its coordinate is uniform in $[W]$. Give every other blue placeholder an independent uniform coordinate in the layer equal to its root layer plus its discovery depth. An assignment beyond the last layer is treated as a boundary failure. These assignments have no effect on the query choices made in $\mathsf F$.

A component reaches layer $L$ or beyond only if its root starts in one of the last $s+1$ layers. The probability of any such event is at most $s(s+1)/L$. For every pair of distinct blue placeholders, the probability of assignment to the same valid blue object is at most $1/W$: placeholders in the same component either lie in different layers or have independent coordinates, and different components have independent root assignments. The union bound for any repeated blue object is at most $m(m-1)/(2W)$.

\emph{Checking the YES marginal.} Reveal these independent proposals in chronological order. A proposal for a blue root is uniform over all $N$ blue objects. Use it if unused; if it is already used, declare failure and redraw uniformly among unused blue objects for the YES marginal. If $t$ blue objects are used, each unused object then has probability
\[
 \frac1N+\frac tN\frac1{N-t}=\frac1{N-t},
\]
as required. At a source in $B_i$, $i<L$, a proposed active target is uniform in $B_{i+1}$. If it is already an image of that generator at that layer, declare failure and redraw uniformly from that generator's remaining images. If $t$ images have been used, each eligible image has probability
\[
 \frac1W+\frac tW\frac1{W-t}=\frac1{W-t}.
\]
Such a failure is among the repetitions of blue objects already counted. A proposal equal to any other seen blue object is also counted as a failure, even when the bijection would permit it. A source in the last layer $B_L$ is charged as a boundary failure before its redirected row is answered. Until failure, the probabilities of the root classes, independent bits, exact sink process, and ID assignments agree in the two experiments. These corrected transitions are precisely the kernels of deferred exposure of the YES bijections and of the uniform ID matching. After the first failure, each marginal continues with its own conditional law. Completing the unused tables and ID matching therefore gives the prescribed law of the YES core, and not a graph chosen after observing the transcript. We have proved
\begin{equation}\label{eq:bfs-yes-fresh}
 \TV\bigl(\Dist(H^{\mathrm Y}_s),\Dist(H^{\mathsf F}_s)\bigr)
 \le \frac{s(s+1)}L+\frac{m(m-1)}{2W}.
\end{equation}
Combining \eqref{eq:bfs-no-fresh} and \eqref{eq:bfs-yes-fresh} proves \eqref{eq:bfs-core-tv}. Forgetting private classes and passing to sets of neighbors cannot increase total variation. Early stopping and randomized query rules are covered by the same coupling and by averaging over coins.
\end{proof}

\paragraph{A common acyclic attachment.}
Choose $M=b^h$ and use a core of size $M=2N$, so that $N=b^h/2$ is a power of two. Take a full $b$-ary merging tree of height $h$ whose leaves are \emph{all} $M$ core vertices, including the core sinks, and orient its edges toward a new root $z$. Attach a full outward $b$-ary tree of height $h$ at $z$, oriented away from $z$ (see Figure~\ref{fig:bfs}(b)). The trees share only $z$, and all their other noncore vertices are new. Fix the attachment to the core's IDs independently of the hidden blue/sink partition and of all core tables. No directed path leaving the core returns to it.

The resulting order is
\begin{equation}\label{eq:bfs-order}
 n_h=M+\frac{M-1}{b-1}+\frac{b(M-1)}{b-1}
     =M+\frac{b+1}{b-1}(M-1)\le3M=6N.
\end{equation}
For any sufficiently large desired order $n$, choose the largest $h$ with $n_h\le n$ and pad with isolated vertices. Then
\begin{equation}\label{eq:bfs-padding}
 n<n_{h+1}\le3bM=6bN,\qquad N=\Theta_b(n).
\end{equation}
Finally, reassign the IDs of the whole graph by an independent uniformly random bijection. Both directional degrees are at most $b$: a core vertex acquires one extra outgoing edge, and $z$ has $b$ incoming and $b$ outgoing edges. Every augmented YES instance is acyclic. By Lemma~\ref{lem:bfs-core-far}, an augmented NO instance $G$ satisfies, with probability $1-o(1)$,
\[
 \fas(G)\ge\frac{aN}{64}
       >\frac{an}{2048b}=\eps_{\rm BFS}D_{\rm BFS}n.
\]
Thus we obtain two input distributions, both fixed in advance, whose NO instances are $\eps_{\rm BFS}$-far with high probability. Conditioning the NO distribution on this farness event will give the distribution supported on far graphs.

\begin{lemma}[A deterministic bound on the number of core queries]
\label{lem:bfs-tree-depth}
On any graph with the above attachment, if fewer than $M$ vertices are queried, each outgoing BFS explores at most
\[
 \sum_{i=0}^{2h}a^i\le2a^{2h}
\]
core vertices. The union of $r$ searches therefore makes at most $2ra^{2h}$ distinct core queries, independently of roots, interleaving, the order within levels, and the sharing of rows.
\end{lemma}
\begin{proof}
A root outside the core has no outgoing path to it. A root inside the core reaches all $M$ leaves of the outward tree within $2h$ steps. Before exploring any vertex at distance $2h+1$, BFS must explore all vertices at smaller distance, including those $M$ leaves. Even if some rows are shared, this would require at least $M$ queries in total. Hence every explored core vertex is at directed distance at most $2h$ from the root. Any path between two core vertices stays inside the core, since the added trees give no return. The core outdegree bound $a$ now gives the displayed count. This argument uses no probabilistic assumption about the core.
\end{proof}

\begin{proof}[Proof of Theorem~\ref{thm:bfs-linear}]
Use the padded distributions just constructed. By \eqref{eq:bfs-padding}, a tester with at most $c_{\rm BFS}n$ queries makes fewer than $N/2<M$ queries. Lemma~\ref{lem:bfs-tree-depth} applies, and since $b=a^{64}$ and $M=b^h=2N$, the number of its core queries is at most
\[
 s_h=\lceil2ra^{2h}\rceil
     =\lceil2r(2N)^{1/32}\rceil=O_{a,r}(N^{1/32}).
\]
Simulate the tester using only a core oracle. Answer the exploration of a core vertex with one core query and its fixed edge to its parent in the merging tree; compute the rows of tree vertices and isolated vertices from the attachment without any core query. The simulator still charges these explorations to the tester's query budget. Returned rows are stored, so that no row is queried twice. Conditional on the tester's coins and the independently chosen attachment and final assignment of IDs, this is an adaptive query rule for the core, making at most $s_h$ queries. The independence of the attachment from the hidden core partition is essential here: its leaves are all core vertices, not just the blue ones.

For large $n$, the hypotheses of Lemma~\ref{lem:bfs-core-tv} hold. Its uniform bound, averaging, and post-processing therefore give
\begin{align}
 \TV\bigl(\Dist(H^{\mathrm N}),\Dist(H^{\mathrm Y})\bigr)
 &\le C_{a,r}N^{1/16}
       \left(N^{-1}+W^{-1}+L^{-1}\right)\notag\\
 &=O_{a,r}(N^{-13/48})=o(1).
 \label{eq:bfs-full-tv}
\end{align}
Here the complete transcript includes the chosen roots, every query and answer, the stopping time, and the final decision. Arbitrarily many tree answers within the allowed query budget are post-processing of the same small core transcript and independent auxiliary randomness. This also explains why adaptive root choices cause no difficulty.

Condition the NO distribution on being $\eps_{\rm BFS}$-far. Its probability of failing this condition is $o(1)$, so conditioning changes any transcript law by at most $o(1)$ in total variation. The conditioned NO distribution and the unconditioned YES distribution are the two distributions described after Theorem~\ref{thm:bfs-linear}, chosen before the algorithm runs. A two-sided $\eps_{\rm BFS}$-tester accepts the YES instances and rejects the far NO instances, each with probability at least $\frac23$, so its acceptance probabilities under the two distributions differ by at least $\frac13$, contrary to \eqref{eq:bfs-full-tv}. In fact, for every such rule with at most $c_{\rm BFS}n$ queries, the average of its success probabilities under the two distributions is at most $\frac12+o(1)$. This proves two-sided hardness, not just failure to reveal a cycle.
\end{proof}

\paragraph{Consequence for the proposed algorithm.}
For every fixed $C=C(\eps_{\rm BFS})$, the proposal of $C$ searches from uniform roots, each exploring at most $Cn/\log n$ vertices, makes $O(C^2n/\log n)=o(n)$ queries in total. Theorem~\ref{thm:bfs-linear} excludes a two-sided tester with these queries, whatever statistical decision it makes afterward; in particular, it excludes the proposed cycle-finding guarantee. The constants are large but fixed before $n$ grows and independent of $C$. The theorem does not apply to an unrestricted tester that follows chosen branches without completing BFS levels, to a number of searches growing with $n$, or to queries for incoming edges, which can traverse the merging tree backward.

\section{Comments about \cite{Yoshida26}}
\label{sec:Yuichi}

As mentioned in Section~\ref{sec:introduction}, a recent preprint~\cite{Yoshida26} claims an $\widetilde\Omega(n^{2/3})$ one-sided and an $\Omega(\sqrt n)$ two-sided lower bound for testing acyclicity. We could not verify the proofs of these bounds. Below we briefly explain why the proof of the one-sided bound does not hold: its NO distribution admits a simple one-sided algorithm with $O(\sqrt n)$ queries.

The NO distribution of \cite[Section 3.1 and Figure 1]{Yoshida26} modifies the construction of \cite{CRSS22} (see Figure~\ref{fig:chen-baseline} in our paper). There are $N = \frac{n}{3}$ blue vertices, whose blue edges are $(u,\pi_j(u))$, $j \in [d_B]$, $d_B = \lfloor\frac{d}{2}\rfloor$, for independent uniform permutations $\pi_j$ of blue vertices. Each blue vertex also has $d-d_B$ outneighbors in the first $T = \lfloor N^{1/3} \rfloor$ of $2T$ red layers, and its list is uniformly shuffled. A red vertex in layer $i \le T$ has $d$ outneighbors, each in a layer chosen uniformly from the window $\{i+1, \dots, i+T\}$; layers $T + 1, \dots, 2T$ consist of sinks, and no red vertex points to a blue one.

One feature of the construction in \cite{Yoshida26} is that \emph{red regions are cheap to exhaust}. Indeed, observe that a path of $\ell$ edges from a red vertex in layer $i \le T$ that avoids the sink layers makes $\ell$ forward jumps, each uniform in $\{1, \dots, T\}$, of total length less than $T$; this has probability at most $\binom{T}{\ell}T^{-\ell}\le 1/\ell!$ (up to the negligible effect of enforcing distinct outneighbors). Hence the expected number of vertices reachable from a red vertex is at most $(d+1)\sum_{\ell\ge 0} d^\ell/\ell! = (d+1) e^d = O_d(1)$. 

Therefore the following one-sided algorithm finds a directed cycle: run a standard depth-first search from a random vertex, rejecting as soon as an edge to a vertex on the DFS stack is found (which then closes a cycle). If the start vertex is blue (probability $\frac13$), no blue vertex is finished before a cycle is found, since each has a blue outneighbor; hence the DFS exhausts, at $O_d(1)$ expected cost, the red branches preceding the first blue entry of each visited blue vertex, and then follows that entry. The stack thus carries a blue walk that applies a uniformly random $\pi_j$ at each step, and for $d_B\ge 2$, after $t$ steps, the next step returns to the walk with probability at least $\frac{t}{2N}$, so a cycle is found after $O_d(\sqrt N)$ queries with constant probability. This contradicts the claim (6) in the proof of \cite[Theorem~4.10]{Yoshida26}, which asserts that every algorithm making $N^{2/3}/\log^c N$ queries finds a cycle with probability $o(1)$. This invalidates the arguments used in the proof of the $\widetilde\Omega(n^{2/3})$ one-sided lower bound for testing acyclicity in \cite{Yoshida26}.

\section*{AI methodology}
\addcontentsline{toc}{section}{AI methodology}

The human authors provided the principal ideas and designed the hard instances. Through multiple rounds of interaction, Codex (GPT-6 Astra and GPT-6.1 Sol) assisted in developing the proofs, while the human authors worked to make the arguments readable for human readers. In a later revision, Claude (Opus 5.5) assisted in polishing and refining the notation. This statement acknowledges the substantive roles of these tools without treating any of them as an author.

\bibliographystyle{alpha}
\bibliography{citations}
\addcontentsline{toc}{section}{References}

\end{document}